\documentclass[fleqn,usenatbib]{mnras}

\usepackage{pdflscape}
\usepackage{newtxtext,newtxmath}
\usepackage[T1]{fontenc}
\usepackage{xspace}

\DeclareRobustCommand{\VAN}[3]{#2}
\let\VANthebibliography\thebibliography
\def\thebibliography{\DeclareRobustCommand{\VAN}[3]{##3}\VANthebibliography}

\usepackage{graphicx}	
\usepackage{amsmath}	
\usepackage{multirow}
\usepackage{subcaption}
\usepackage{comment}

\newcommand{\angstrom}{\textup{\AA}\xspace}
\newcommand\textlcsc[1]{\textsc{\MakeLowercase{#1}}}

\newcommand{\lya}{Ly$\alpha$\xspace}
\newcommand{\hb}{H$\beta$\xspace}
\newcommand{\ha}{H$\alpha$\xspace}
\newcommand{\oiilow}{[OII]$\lambda\lambda3726,3729$\xspace}
\newcommand{\oiiia}{[OIII]$\lambda4959$\xspace}
\newcommand{\oiiib}{[OIII]$\lambda5007$\xspace}
\newcommand{\oiiiab}{[OIII]$\lambda\lambda4959,5007$\xspace}
\newcommand{\niia}{[NII]$\lambda6548$\xspace}
\newcommand{\niib}{[NII]$\lambda6584$\xspace}
\newcommand{\niiab}{[NII]$\lambda\lambda6548,6584$\xspace}

\newcommand{\rew}{$REW_{\rm Ly\alpha}$\xspace}
\newcommand{\xhi}{$X_{\rm HI}$\xspace}
\newcommand{\MUV}{$M_{\rm UV}$\xspace}
\newcommand{\fesc}{$f\rm  _{esc}^{Ly\alpha}$\xspace}
\newcommand{\xlya}{$X\rm_{Ly\alpha}$\xspace}

\title[JADES: Measuring reionisation with Ly$\alpha$]{JADES: Measuring reionisation properties using Lyman-alpha emission}

\author[G. C. Jones et al.]{Gareth C. Jones$^{1,2,3}$\thanks{E-mail: gj283@cam.ac.uk},
Andrew J. Bunker$^{1}$,
Aayush Saxena$^{1,4}$,
Santiago Arribas$^{5}$,
Rachana Bhatawdekar$^{6,7}$,\newauthor
Kristan Boyett$^{8,9,1}$,
Alex J. Cameron$^{1}$,
Stefano Carniani$^{10}$,
Stephane Charlot$^{11}$,
Emma Curtis-Lake$^{12}$,\newauthor
Kevin Hainline$^{13}$,
Benjamin D. Johnson$^{14}$,
Nimisha Kumari$^{15}$,
Michael V. Maseda$^{16}$,
Hans-Walter Rix$^{17}$,\newauthor
Brant E. Robertson$^{18}$,
Sandro Tacchella$^{2,3}$,
Hannah \"{U}bler$^{2,3}$,
Christina C. Williams$^{19}$,
Chris Willott$^{20}$,\newauthor
Joris Witstok$^{2,3}$,
Yongda Zhu$^{13}$
\\
$^{1}$ Department of Physics, University of Oxford, Denys Wilkinson Building, Keble Road, Oxford OX1 3RH, UK\\
$^{2}$ Kavli Institute for Cosmology, University of Cambridge, Madingley Road, Cambridge CB3 0HA, UK\\
$^{3}$ Cavendish Laboratory, University of Cambridge, 19 JJ Thomson Avenue, Cambridge CB3 0HE, UK\\
$^{4}$ Department of Physics and Astronomy, University College London, Gower Street, London WC1E 6BT, UK\\
$^{5}$ Centro de Astrobiolog\'{i}a (CAB), CSIC–INTA, Cra. de Ajalvir Km.~4, 28850- Torrej\'{o}n de Ardoz, Madrid, Spain\\
$^{6}$ European Space Agency (ESA), European Space Astronomy Centre (ESAC), Camino Bajo del Castillo s/n, 28692 Villanueva de la Ca\~{n}ada, Madrid, Spain\\
$^{7}$ European Space Agency, ESA/ESTEC, Keplerlaan 1, 2201 AZ Noordwijk, NL\\
$^{8}$ School of Physics, University of Melbourne, Parkville 3010, VIC, Australia\\
$^{9}$ ARC Centre of Excellence for All Sky Astrophysics in 3 Dimensions (ASTRO 3D), Australia\\
$^{10}$ Scuola Normale Superiore, Piazza dei Cavalieri 7, I-56126 Pisa, Italy\\
$^{11}$ Sorbonne Universit\'{e}, CNRS, UMR 7095, Institut d'Astrophysique de Paris, 98 bis bd Arago, 75014 Paris, France\\
$^{12}$ Centre for Astrophysics Research, Department of Physics, Astronomy and Mathematics, University of Hertfordshire, Hatfield AL10 9AB, UK\\
$^{13}$ Steward Observatory, University of Arizona, 933 N. Cherry Ave., Tucson, AZ 85721 USA\\
$^{14}$ Center for Astrophysics $|$ Harvard \& Smithsonian, 60 Garden St., Cambridge, MA 02138, USA\\
$^{15}$ AURA for European Space Agency, Space Telescope Science Institute, 3700 San Martin Drive. Baltimore, MD, 21210, USA\\
$^{16}$ Department of Astronomy, University of Wisconsin-Madison, 475 N. Charter St., Madison, WI 53706 USA\\
$^{17}$ Max-Planck-Institut f\"ur Astronomie, K\"onigstuhl 17, D-69117, Heidelberg, Germany\\
$^{18}$ Department of Astronomy and Astrophysics, University of California, Santa Cruz, 1156 High Street, Santa Cruz, CA 95064, USA\\
$^{19}$ NSF’s National Optical-Infrared Astronomy Research Laboratory, 950 North Cherry Avenue, Tucson, AZ 85719, USA\\
$^{20}$ NRC Herzberg, 5071 West Saanich Rd, Victoria, BC V9E 2E7, Canada
}

\date{Accepted XXX. Received YYY; in original form ZZZ}
\pubyear{2024}

\begin{document}
\label{firstpage}
\pagerange{\pageref{firstpage}--\pageref{lastpage}}
\maketitle

\begin{abstract}
Ly$\alpha$ is the transition to the ground state from the first excited state of hydrogen (the most common element). Resonant scattering of this line by neutral hydrogen greatly impedes its emergence from galaxies, so the fraction of galaxies emitting Ly$\alpha$ is a tracer of the neutral fraction of the intergalactic medium (IGM), and thus the history of  reionisation. In previous works, we used early JWST/NIRSpec data from the JWST Advanced Deep Extragalactic Survey (JADES) to classify and characterise Ly$\alpha$ emitting galaxies (LAEs). This survey is approaching completion, and the current sample is nearly an order of magnitude larger. From a sample of 795\,galaxies in JADES at $4.0<z<14.3$, we find evidence for Ly$\alpha$ emission in 150\,sources. We reproduce the previously found correlation between Ly$\alpha$ escape fraction (\fesc) - Ly$\alpha$ rest-frame equivalent width (\rew) and the negative correlation between Ly$\alpha$ velocity offset - \fesc. Both \fesc and \rew decrease with redshift ($z\gtrsim5.5$), indicating the progression of reionisation on a population scale. Our data are used to demonstrate an increasing IGM transmission of Ly$\alpha$ from $z\sim14-6$. We measure the completeness-corrected fraction of LAEs (\xlya) from $z=4-9.5$. An application of these \xlya values to the results of previously utilised semi-analytical models suggests a high neutral fraction at $z=7$ (${X_{\rm HI}}\sim0.8-0.9$). Using an updated fit to the intrinsic distribution of \rew results in a lower value in agreement with current works (${X_{\rm HI}}= 0.64_{-0.21}^{+0.13}$). This sample of LAEs will be paramount for unbiased population studies of galaxies in the EoR.
\end{abstract}


\section{Introduction}\label{intro}

It has been well established that early after the Big Bang ($z\sim1100$, or $t_{\rm H}=$360\,Myr), the Universe cooled enough to permit the formation of neutral hydrogen atoms (i.e., the Epoch of Recombination, e.g., \citealt{suny80,seag00}), creating the surface of last scattering (i.e., the cosmic microwave background; CMB) and marking the beginning of an epoch during which most hydrogen in the Universe was neutral (a neutral fraction of hydrogen of unity [${X_{\rm HI}}=1$]). This was followed by `Cosmic Dawn', when the first stars formed and began to ionise their surrounding gas via ultraviolet (UV) radiation ($z>10$; see review of \citealt{kles23}). The time between the formation of the first stars and when the intergalactic medium (IGM) was fully ionised (${X_{\rm HI}}\approx 0$) is the Epoch of Reionisation (EoR). The current general consensus is that the Universe was mostly ionised again by $z\sim6$ (e.g., \citealt{fan06,mcgr15,plan16}); but the discovery of neutral gas `islands' at later epochs suggests that the EoR did not conclude until slightly later (e.g., $z\sim5.2-5.3$; \citealt{kulk19,keat20a,bosm22,beck24}).

The study of the EoR is one of the major focuses of modern astrophysics, including investigations of the drivers (e.g., active galactic nuclei [AGN], small/massive galaxies, mergers; \citealt{sult18,naid20,bosm22,witt23,graz24,mada24}) and topology of reionisation (e.g., \citealt{pent14}), as well as the escape mechanisms of ionising radiation (e.g., \citealt{chis18}). Here, we focus on characterising the progression of the EoR through measurements of ${X_{\rm HI}}(z)$.

There are multiple pathways to study the evolution of \xhi, including damping wing (DW) observations of QSOs (e.g., \citealt{bana18,duro20,yang20}) and galaxies (e.g., \citealt{hsia23,umed23,faus24}), CMB studies (e.g., \citealt{plan20}), and comparisons of \lya observations to models (e.g., \citealt{maso18a,bhag23,feld24}). This latter path can further be divided into different methods, including studies of the \lya luminosity function (e.g., \citealt{konn14,inou18}), clustering of \lya-emitting galaxies (LAEs; e.g., \citealt{ouch10,ouch18,soba15}), and Lyman forest dark fractions (e.g., \citealt{keat20b,bosm22,zhu22}). Together, these studies suggest an evolution of ${X_{\rm HI}}(z\gtrsim13)=1$ to ${X_{\rm HI}}(z\sim5.3)=0$ with a midpoint of ${X_{\rm HI}}(z\sim7)=0.5$, although the exact shape of this evolution is under debate.

An additional method of characterising ${X_{\rm HI}}(z)$ is the study of the evolution of the \lya emitter fraction (\xlya). Multiple studies have compared observed and model fractions to place constraints from $z\sim2-8$ (e.g., \citealt{star10,star11,caru14,curt12,ono12,sche12,sche14,cass15,furu16,deba17,star17,goov23,fu24}). In a previous work \citep{jone24}, we utilised low spectral resolution James Webb Space Telescope (JWST; \citealt{gard23}) Near Infrared Spectrograph (NIRSpec; \citealt{jako22,boke23}) data (PRISM/CLEAR; with spectral resolving power $R\sim100$) from the first JWST Advanced Deep Extragalactic Survey (JADES; \citealt{bunk20,eise23a}) data release to estimate \xlya at $z=6$ and $z=7$. While the sample size was relatively small ($84$\,galaxies) and featured a non-standard \MUV range ($ -20.48<{M_{\rm UV}}<-16.33$), our completeness-corrected analysis resulted in a good determination of \xlya, which was used to constrain ${X_{\rm HI}}(z=7)$.

The JADES sample has since been combined with other public JWST datasets in order to further constrain ${X_{\rm Ly\alpha}}(z)$ (\citealt{naka23,napo24}), and the results are in agreement with those of \citet{jone24}. However, the diverse samples of these newer works (i.e., JADES, CEERS, and other programs) means that the selection function of each sample will be less homogeneous than a single-program dataset. In addition, neither of these works investigated the rest-frame \lya equivalent width (\rew) completeness of their dataset, which will result in skewed ${X_{\rm Ly\alpha}}(z)$ distributions. Here, we exploit the expanded JADES dataset to characterise \lya emission in the early Universe ($4.0<z<14.3$; corresponding to $\sim0.3-1.5$\,Gyr after the Big Bang).

This work is organised as follows. We discuss our sample in Section \ref{sec_sample} and our spectral fitting procedure in Section \ref{fitsec}. The correlations from this analysis are explored in Section \ref{corr_sec}. Section \ref{discsec} contains a completeness-corrected estimation of the \lya fraction, which is used to constrain \xhi and the IGM transmission of \lya. We conclude in Section \ref{conc} 

We assume a standard concordance cosmology throughout: ($\Omega_{\Lambda}$,$\Omega_m$,h)=(0.7,0.3,0.7) and use AB magnitudes.

\section{Sample}\label{sec_sample}
\subsection{Observations overview}

For this analysis, we use all observed NIRSpec spectroscopy so far from the JADES survey, which spans PID 1180 and 1181 (PI D. Eisenstein), PID 1210, 1286, and 1287 (PI N. Luetzgendorf), and PID 3215 (PIs D. Eisenstein and R. Maiolino). This survey observed galaxies in the Great Observatories Origins Deep Survey (GOODS; \citealt{dick03}) north (N) and south (S) fields with the JWST/NIRSpec Multi-Shutter Array (MSA; \citealt{ferr22}) in both low (PRISM/CLEAR; spectral resolving power $R\sim100$) and medium spectral resolution (G140M/F070LP, G235M/F170LP, G395M/F290LP; $R\sim1000$). Some survey tiers also contain high spectral resolution observations (G395H/F290LP; $R\sim2700$). 

For each JADES tier (see Table \ref{jadestable}), a large list of potential target galaxies was aggregated. Each galaxy was given a priority class (PC) dependent on e.g., redshift, HST (or JWST, if available) colours, and UV brightness (see \citealt{bunk23b,deug24} for details of PCs), which were used in the construction of MSA masks. This scheme was designed to ensure observations of both extraordinary objects (e.g., GN-z11; \citealt{bunk23a}) and a statistically significant number of representative galaxies over the probed range of redshifts. For more details, see the full description of the survey \citep{eise23a} and data release papers (\citealt{bunk23b,eise23b,deug24}). The resulting spectra were visually inspected \citep{deug24}, resulting in precise spectroscopic redshifts for each galaxy. For the two highest redshift sources in the sample, we include the updated redshifts for 183348 (JADES-GS-z14-0; $z_{\rm sys}=14.32$) and 20018044 in 1287\_DJS (JADES-GS-z14-1; $z_{\rm sys}=13.90$; \citealt{carn24}). We also adopt the updated redshift for 20013731 in 1287\_DJS ( JADES-GS-z13-1-LA; $z_{\rm sys}=13.01$) from \citet{wits24b}. Due to the spatially extended nature of \lya emission (e.g., \citealt{jung23}), we use a wide extraction aperture (5\,pixels$\sim0.5''$; e.g. \citealt{bunk23a,curt24,tang24b}).

Some targets were observed in multiple tiers due to a desire for a deeper integration or a repeated observation due to previous data being made unusable by an electrical short. To avoid including these observations, we collect all inspected galaxies that are within $0.25''$ of each other and exclude the shallower observation\footnote{While a future data release will include combined spectra from multiple survey tiers, this is not yet available.}. In all cases, the visual spectroscopic redshifts of the repeat observations agree. Since we wish to analyse the R100 data, we exclude observations where the R100 data are corrupted (e.g., due to electrical shorts), resulting in 2992 unique galaxies with good R100 data and precise redshifts. A redshift cut of $z>4$ is placed, so that we can detect \lya in the wavelength range of the R100 data. With these limits and exclusions, we find a list of 795\,unique galaxies. The distribution of sources between survey tiers is shown in Table \ref{jadestable}.

\begin{table}
\caption{JADES tier distribution of the sample analysed in this work. For each survey tier, we also list a shorthand label.}
\label{jadestable}
\centering
\begin{tabular}{ccccc|c}
\hline
PID	    &	Field	&	Tier	&	Selection	&	Label     & ${\rm N}_{z>4}$	\\ \hline
1180	&	GOODS-S	&	Medium	&	HST      	&	1180\_MHS &  114	\\ 
1180	&	GOODS-S	&	Medium	&	JWST	    &	1180\_MJS &  92	\\
1181	&	GOODS-N	&	Medium	&	HST	        &	1181\_MHN &  96	\\
1181	&	GOODS-N	&	Medium	&	JWST	    &	1181\_MJN &  126	\\
1210	&	GOODS-S	&	Deep	&	HST	        &	1210\_DHS & 66	\\
1286	&	GOODS-S	&	Medium	&	JWST	    &	1286\_MJS &  209	\\ 
1287	&	GOODS-S	&	Deep	&	JWST	    &	1287\_DJS & 36	\\
3215	&	GOODS-S	&	Deep	&	JWST	    &	3215\_DJS & 56	\\ \hline
	&		    &		    &	            &\textbf{TOTAL:}	&	 795	\\ \hline
\end{tabular}
\end{table}

\subsection{Galaxy clustering}\label{soudis}
\begin{figure*}
    \centering
    \includegraphics[width=\textwidth]{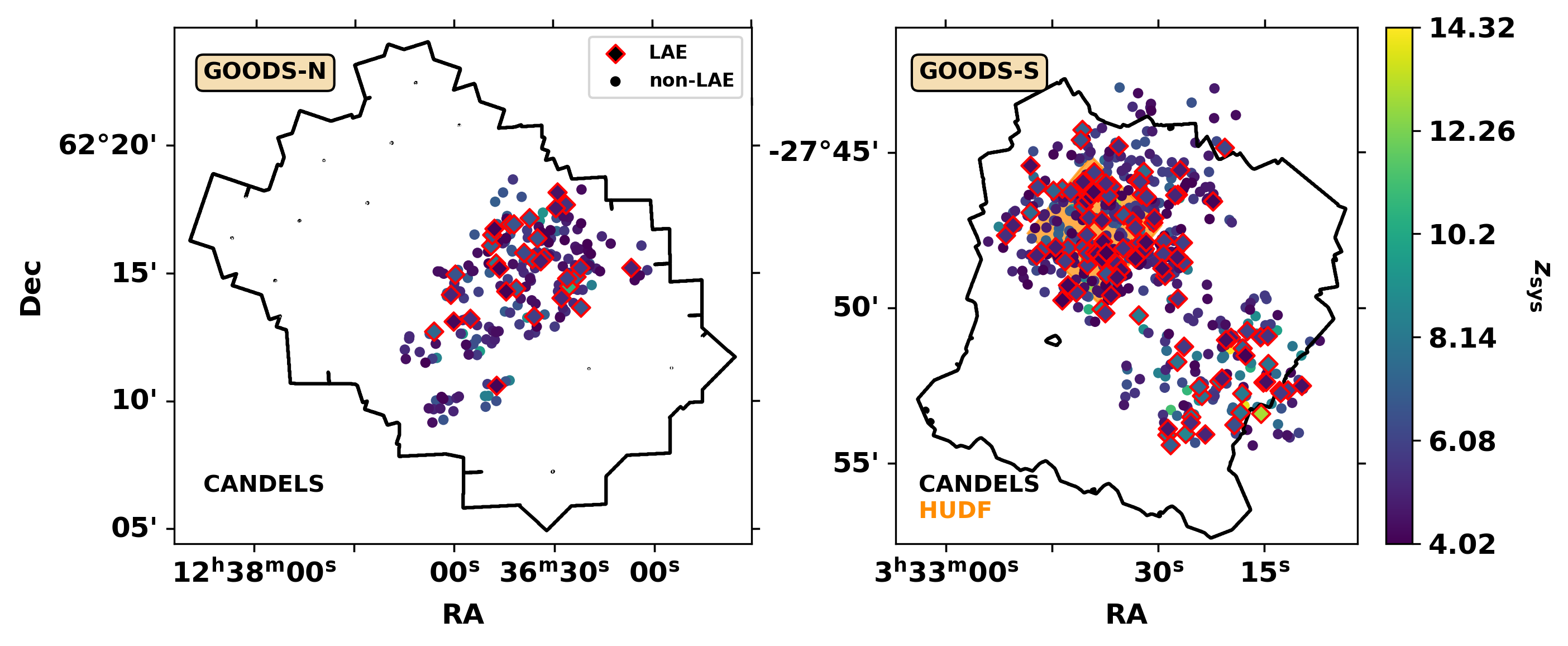}
    \caption{Spatial distribution of our sample, coloured by redshift. Sources not detected in \lya emission are represented as circles, while LAEs are red-edged diamonds (see Section \ref{fitsec} for more details). For reference, we display the footprints of the Cosmic Assembly Near-infrared Deep Extragalactic Legacy Survey (CANDELS; \citealt{grog11}) field and the Hubble Ultra Deep Field (HUDF; \citealt{beck06}).}
    \label{radecz}
\end{figure*}

The galaxies that we analyse in this work have a similar set of selection criteria and a uniform calibration pipeline, making a well-founded statistical analysis possible. The sources are well distributed across the two GOODS fields (Fig. \ref{radecz}), and some galaxies are closely clustered. However, due to the size of each field ($\sim18'$, corresponding to $\sim7.5$\,Mpc at $z=4$ or $\sim5.2$\,Mpc at $z=10$) and the number of targets, this clustering is expected. Indeed, since observation planning software (e.g., \citealt{bona23}) enables efficient observations by creating densely packed MSA slit masks without spectral overlap, we expect a number of galaxies with small projected spatial separations.

\begin{figure*}
    \centering
    \includegraphics[width=\textwidth]{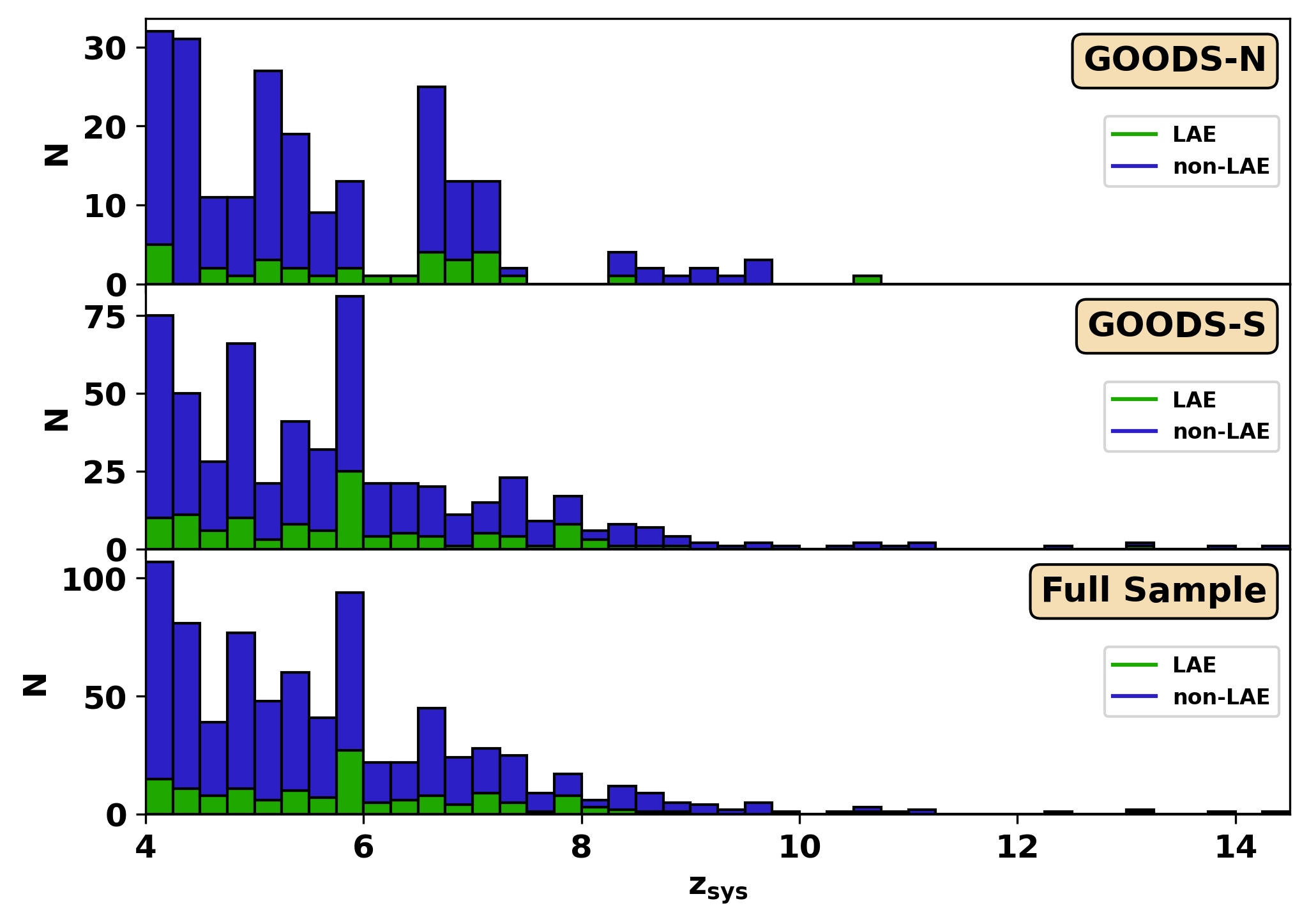}
    \caption{Redshift distribution of sources analysed in this work, coloured by \lya detection (see Section \ref{fitsec} for more details on fitting procedure).}
    \label{zbin}
\end{figure*}

Previous studies of the UV luminosity function (e.g., \citealt{donn23,hari23,robe24}) have shown that the density of galaxies for a given \MUV decreases at higher redshifts. The sample selection procedure of JADES was designed to maintain a statistical sample across a wide range of redshifts (\citealt{bunk23b,deug24}), and acts to preferentially observe more high-redshift galaxies than would be included in a flux-limited survey (Fig. \ref{zbin}).

Recently, \citet{helt23} searched JADES NIRCam \citep{riek23} data for galaxy overdensities at $4.9<z_{\rm sys}<8.9$, finding 17\,overdensities in GOODS-N and GOODS-S. By applying the same association criteria as Helton et al. (i.e., projected physical separations of $<0.1$\,Mpc and velocity offsets of $<500$\,km\,s$^{-1}$), we find that eight of our 795\,galaxies (i.e., $\sim1\%$) fall into these overdensities (three in JADES-GN-OD-7.144, one in JADES-GS-OD-6.876, two in JADES-GS-OD-7.954, and two in JADES-GS-OD-8.220). Thus, our sample is not strongly affected by high galaxy overdensities. While LAEs have been found in overdensities or close pairs (e.g., \citealt{saxe23b,wits23,witt24}), the study of LAE clustering is deferred to a future work.

\section{Spectral fitting}\label{fitsec}

Our previous work was focused on \rew, and only dealt with on the \lya line \citep{jone24}. In this work, we extend our focus to the \lya escape fraction (\fesc), which requires flux estimates of at least one Balmer line (e.g., H$\alpha$ or H$\beta$). Because of this, we extend our spectral fitting to encompass the full wavelength range covered by the PRISM/CLEAR disperser/filter combination (i.e., $0.6-5.3\,\mu$m). In addition, we include the higher resolution R1000 data, in order to verify our fits and study relationships with the velocity offset between \lya and the systemic redshift.

\subsection{Model description}\label{moddes}

The wide wavelength coverage, deep continuum sensitivity, and low spectral resolution of the R100 data mean that an appropriate model of the line and continuum emission requires careful construction. Before modelling each galaxy spectrum, we derive an estimate of \MUV by integrating the spectrum between $\lambda_{\rm rest}=1400-1500\,\angstrom$. This range is chosen to overlap with one of the windows of \citet{calz94}, and avoid contamination by possible CIV$\lambda\lambda 1548,1551$ emission (e.g., \citealt{izot24,nava24}). If this \MUV estimate has an uncertainty (based on the error spectrum) of $<0.5$\,magnitude, then we claim that \MUV is well determined. Otherwise, we determine a $3\sigma$ lower limit on \MUV based on the RMS noise level of the observed spectrum (see Appendix \ref{muvlimsec}).

The continuum at rest-frame ultraviolet (UV) wavelengths in the early Universe (i.e., at $z\gtrsim5$) is commonly fit as a power law model with a slope $\beta_{\rm UV}\sim-2$ (e.g., \citealt{yama19,cull23,topp24}). But previous works (e.g., \citealt{jone24,napo24}) suggest that the continuum just redwards of \lya ($\lambda_{\rm rest}\sim0.12-0.15\,\mu$m) is well-modelled as a power law function with a slope that may deviate from that of $\beta_{\rm UV}$. Indeed, \citet{came23} suggest that this deviation in one galaxy is a sign of two-photon nebular continuum emission (e.g., \citealt{dijk09,katz24}; but see also \citealt{li24,tacc24,terp24} for alternate interpretations), as seen in low-redshift galaxies (e.g., \citealt{hall04,john12}). For sources in the epoch of reionisation (and to some degree sources at lower redshift), reservoirs of neutral gas will create DWs (e.g. \citealt{mort11}). While the low spectral resolution of the R100 spectra results in the appearance of pseudo-DWs (e.g., \citealt{jone24}), detailed investigations into DWs at high redshift are ongoing (e.g., \citealt{fuji23,hein23,umed23}).

With this in mind, we split each spectrum into two models, with a pivot wavelength of $\lambda_{\rm rest}=0.145\,\mu$m. In the following, we refer to them as the `R100-blue' and `R100-red' models. This pivot wavelength is chosen as the middle point of the range we use to derive \MUV, and is similar to the turn-over wavelength of the nebular continuum model of \citet{came23}. The value of each model at the pivot wavelength is fixed to be the mean spectral value of the observed spectrum within $\lambda_{\rm rest}=1400-1500\,\angstrom$ (i.e., R100-red and R100-blue are required to be continuous), but the models are not assumed to be differentiable.

We first examine the R100-red model, which covers $\lambda_{\rm rest}=0.145\,\mu$m to $\lambda_{\rm obs}=5.3\,\mu$m. The continuum of this range is modelled as two power law segments: one that extends from $\lambda_{\rm obs}\ge(1+z_{\rm Ly\alpha})\times0.145\,\mu$m to the wavelength of H$\eta$ ($\lambda_{\rm obs}<(1+z_{\rm sys})0.3836\,\mu$m), and another that extends from $\lambda_{\rm obs}\ge(1+z_{\rm sys})0.3836\,\mu$m to the red limit of the spectrum ($\lambda_{\rm obs}=5.3\,\mu$m). A discontinuity between these segments is allowed, in order to capture a Balmer break or jump. While true Balmer breaks are expected to be more gradual roll-offs marked by numerous absorption lines (e.g., \citealt{bing19,furt24}), the coarse spectral resolution of our data necessitates a simple model. We note that galaxies with high nebular continuum emission may feature Balmer jumps rather than breaks and reddened rest-UV continuum slopes (e.g., \citealt{katz24,nara24,rober24}). However, the resulting rest-UV emission may still be described as a power law (e.g., \citealt{hein24}). Because our R100-red model contains two segments with separate power law slopes with no constraints on the sign of the Balmer discontinuity, we may still fit spectra of galaxies with bright nebular continuum emission.

The brightest expected emission lines (\oiilow, \hb, \oiiiab, \ha, and \niiab)\footnote{While the rich JADES dataset contains significant emission from many more lines (e.g., \citealt{came23,curti23,lase24}), we focus on the dominant emission in each spectrum.} are included via Gaussian model components at the expected wavelengths. We predict the line spread function (LSF) by first taking the fiducial resolving power curve\footnote{As recorded in the JWST documentation; \url{https://jwst-docs.stsci.edu/jwst-near-infrared-spectrograph/nirspec-instrumentation/nirspec-dispersers-and-filters}}. As noted in \citet{degr23}, this curve was derived assuming that each NIRSpec MSA slit was uniformly illuminated. Since this is not the case for each JADES galaxy, the LSF may be under-predicted. To account for this, we define the width of each Gaussian to be $\sigma_{\rm R}(\lambda)=F_{\rm R}\lambda/R(\lambda)/2.355$, where $F_{\rm R}$ represents the deviation from the fiducial LSF\footnote{While this deviation has been found to be wavelength dependent, we assume a single average value across the full wavelength range.}. Using the code PyNeb \citep{luri15}) and assuming ISM conditions of $T_{\rm e}=1.5\times10^4$\,K and $n_{\rm e}=300$\,cm$^{-3}$ (e.g., \citealt{torr24}), we derive intrinsic ratios of\oiiib/\oiiia$=2.984$ and \niib/\niia$=2.942$. \oiilow is treated as a single Gaussian line in the low-resolution R100 data.

The free parameters are thus: the systemic redshift ($z_{\rm sys}$), the power law slopes of each of the two continuum components, the normalisation of the redder power-law component, the deviation from the fiducial LSF ($F_{\rm R}$), and the integrated line fluxes of \oiilow, \hb, \oiiib, \niia, and \ha. We use LMFIT \citep{newv14} in `least\_squares' mode to find the best-fit model. Each spectrum is weighted by its inverse variance, measured from its associated error spectrum. If the initial fit is successful, then the best-fit line intensities and their uncertainties are inspected. In some cases, the first fit fails due to a non-detection of \oiilow, which is weaker than the other UV/optical lines. To remedy this, we follow failed fits with runs where the \oiilow intensity is set to 0. There are some galaxies for which we do not detect any significant emission from any of our rest-optical lines, making the measurement of $z_{\rm sys}$ from our data alone impossible. In these cases, we use the visual inspection redshift of \citet{deug24}, who used additional emission lines and inspected both the R100 and R1000 spectra. The intensities of lines that are not well detected ($<3\sigma$) are set to 0, and the fit is repeated until convergence. 

Next, we consider the R100-blue model that extends from $\lambda_{\rm obs}=0.6\,\mu$m to $\lambda_{\rm rest}=0.145\,\mu$m. An initial high-resolution model grid with bins of $0.001\,\mu$m is populated with a single power-law. A Heaviside step function with a transition at $\lambda_{\rm obs}=(1+z_{\rm Ly\alpha})\times\lambda_{\rm Ly\alpha}$ is applied to this model to represent the \lya break. Some observed spectra feature non-zero emission blue-wards of the \lya break, which may either be incomplete absorption by the intervening \lya forest (particularly at lower-$z$) or an artefact introduced during calibration. We account for this by allowing a non-zero continuum level that is constant (in units of $F_{\lambda}$) blue-wards of the \lya break. To introduce \lya emission, we add flux to the first spectral bin redwards of the \lya break. The model is then convolved with a Gaussian of width  $\sigma_{\rm R}$. 

If the R100-red model returned a well-determined $F_{\rm R}$ (i.e., $>3\sigma$), then we adopt this best-fit value for this model as well. Otherwise, we assume that $F_{\rm R}=1$. The free parameters in this model are: $z_{\rm Ly\alpha}$, the power law slope of the continuum, and the integrated line flux of \lya. Again, we use LMFIT in `least\_squares' mode to find the best-fit model and weigh each spectrum by its inverse variance. To explore the presence of \lya, we perform initial fits with a variable $F_{\rm Ly\alpha}$ (considering the line and continuum) and with $F_{\rm Ly\alpha}\equiv0$ (continuum-only). If these fits terminate successfully, then the best-fit values and reduced $\chi^2$ values are inspected. If the line and continuum fit returns a better reduced $\chi^2$, then we present the \lya properties. Otherwise, we present upper limits on \lya.

We also examine the R1000 data for each source. All available data (i.e., G140M, G235M, and G395M) are combined in order to create a composite spectrum. The wavelengths ranges around three line complexes are isolated (\lya, \hb-\oiiiab, and \niiab-\ha). Each emission line is fit using a 1-D Gaussian profile, where we assume the same \oiiiab  and \niiab ratios as for the R100 fit.

The \lya emission is modelled as a symmetric Gaussian in the R100 and R1000 data. Other works adopt a more complex asymmetric profile (e.g., \citealt{shib14}), due to the relatively high spectral resolving power (i.e., $R>1000$) of their data (e.g.; see works utilising MUSE, e.g. \citealt{keru22};  DEIMOS, e.g., \citealt{ono12}; and MOSFIRE, e.g., \citealt{oesc15}). Because \lya in our sample is shifted to $\lambda_{\rm obs}\sim0.6-1.9\,\mu$m (with a preponderance of galaxies at the lower edge), our resolving power is $R\sim30-100$ for the R100 data and $R\sim300-800$ for the R1000 data, making it difficult to resolve the true \lya profile (e.g., \citealt{saxe23a}).

Many properties of the JADES data are described in greater detail in other works, so we will not discuss them here. These include the possibility of \lya DWs (e.g., Jakobsen et al. in prep), the presence of damped \lya systems (e.g., \citealt{hain24}), UV spectral slopes (Saxena et al. in prep), and population properties derived from stacked data \citep{kuma24}.

\subsection{Further observables}\label{furobs}

The rest-frame equivalent width of each line is calculated using its integrated flux ($F_{\rm line}$), redshift ($z_{\rm sys}$), and the continuum model evaluated at the centroid wavelength ($S_{\rm C}(\lambda_{\rm line})$):
\begin{equation}\label{ew0}
{REW_{\rm Ly\alpha}} = \frac{F_{\rm line}}{(1+z_{\rm sys}) S_{\rm C}(\lambda_{\rm line})}
\end{equation}

The best-fit continuum model is used to directly determine the Balmer break by taking the ratio of the two best-fit power law components of R100-red at their overlapping point.

Using our best-fit observed \ha and \hb integrated fluxes from the R100 or R1000 data, we may directly determine the Balmer decrement (e.g. \citealt{domi13}): 
\begin{equation}
E(B-V)_{\rm BD}=\frac{2.5}{k(\lambda_{\rm H\beta})-k(\lambda_{\rm H\alpha})}\log_{10}\left(\frac{F_{\rm H\alpha,obs}/F_{\rm H\beta,obs}}{2.876}\right)
\end{equation}
where $k(\lambda)$ is the assumed dust attenuation curve \citep{calz00} and we derive an intrinsic $F_{\rm H\alpha}/F_{\rm H\beta}=2.876$ using PyNeb and assuming fiducial ISM conditions of $T_{\rm e}=1.5\times10^4$\,K and $n_{\rm e}=300$\,cm$^{-3}$ (e.g., \citealt{torr24}). This is then used to derive an intrinsic (dust-corrected) \ha integrated flux:
\begin{equation}\label{fhaint}
F_{\rm H\alpha,int}=F_{\rm H\alpha,obs}10^{k(\lambda_{\rm H\alpha})E(B-V)_{\rm BD}/2.5}
\end{equation}
Through PyNeb, our ISM condition assumptions yield intrinsic ratios of $F_{\rm Ly\alpha}/F_{\rm H\alpha}=8.789$ and $F_{\rm Ly\alpha}/F_{\rm H\beta}=24.487$, assuming case B recombination. These are combined with the result of equation \ref{fhaint} to derive the intrinsic $F_{\rm Ly\alpha}$. This is then used to derive the \lya escape fraction:
\begin{equation}
f_{\rm esc}^{\rm Ly\alpha}=F_{\rm Ly\alpha,obs}/F_{\rm Ly\alpha,int}
\end{equation}
We estimate \fesc both by de-reddening \lya and each Balmer line, and by not correcting for dust (see Section \ref{fesc_calc}).

Due to the wavelength coverage of our R100 observations (i.e., $0.60-5.30\,\mu$m), we may detect \ha for galaxies at $z\lesssim7.1$, \hb up to $z\lesssim9.9$, and \lya for $z\gtrsim3.9$. The R1000 observations have a slightly smaller wavelength coverage (i.e., $0.70-5.10\,\mu$m), so we may detect \ha for galaxies at $z\lesssim6.8$, \hb up to $z\lesssim9.5$, and \lya for $z\gtrsim4.8$.

Our R1000 data allow us to determine the \lya velocity offset with respect to the redshift based on the rest-optical lines (also derived from the R1000 data). The redshift of \lya emission is measured in two ways: from the centroid of the best-fit Gaussian model ($\Delta v_{\rm Ly\alpha,G}$), and from the wavelength corresponding to the peak flux within [-500,+1000]\,km\,s$^{-1}$ of \lya ($\Delta v_{\rm Ly\alpha,P}$). Due to the large size of our sample, these approaches are simpler than that of \citet{saxe23a}, who fit each R1000 spectrum with asymmetric and symmetric Gaussian models using an MC approach. We will use $\Delta v_{\rm Ly\alpha,P}$ in the following analyses (see Appendix \ref{delvapp} for a comparison of these velocities).

\subsection{Spectral fitting results}

As discussed in Section \ref{sec_sample}, our parent sample contained 795\,galaxies with precise spectroscopic redshifts from visual inspection of the R100 and R1000 spectra ($z=4.0-14.3$; \citealt{deug24}). Our fitting routine was applied to each galaxy, resulting in estimates on \lya, rest-optical lines, and continuum emission. 
There is evidence for \lya emission from either the R100 or R1000 spectra at $>3\sigma$ in 150\,galaxies. The best-fit \lya flux and \rew values are presented in Table \ref{lyares_table} for each such source. A set of fit examples are shown in Figure \ref{fit_ex}. The R100- and R1000-based quantities are compared in Appendix \ref{FQV}. In Section \ref{compsec}, we compare our recovered \lya properties to those of other works that studied the GOODS fields.

\begin{figure*}
\centering
\includegraphics[width=0.49\textwidth]{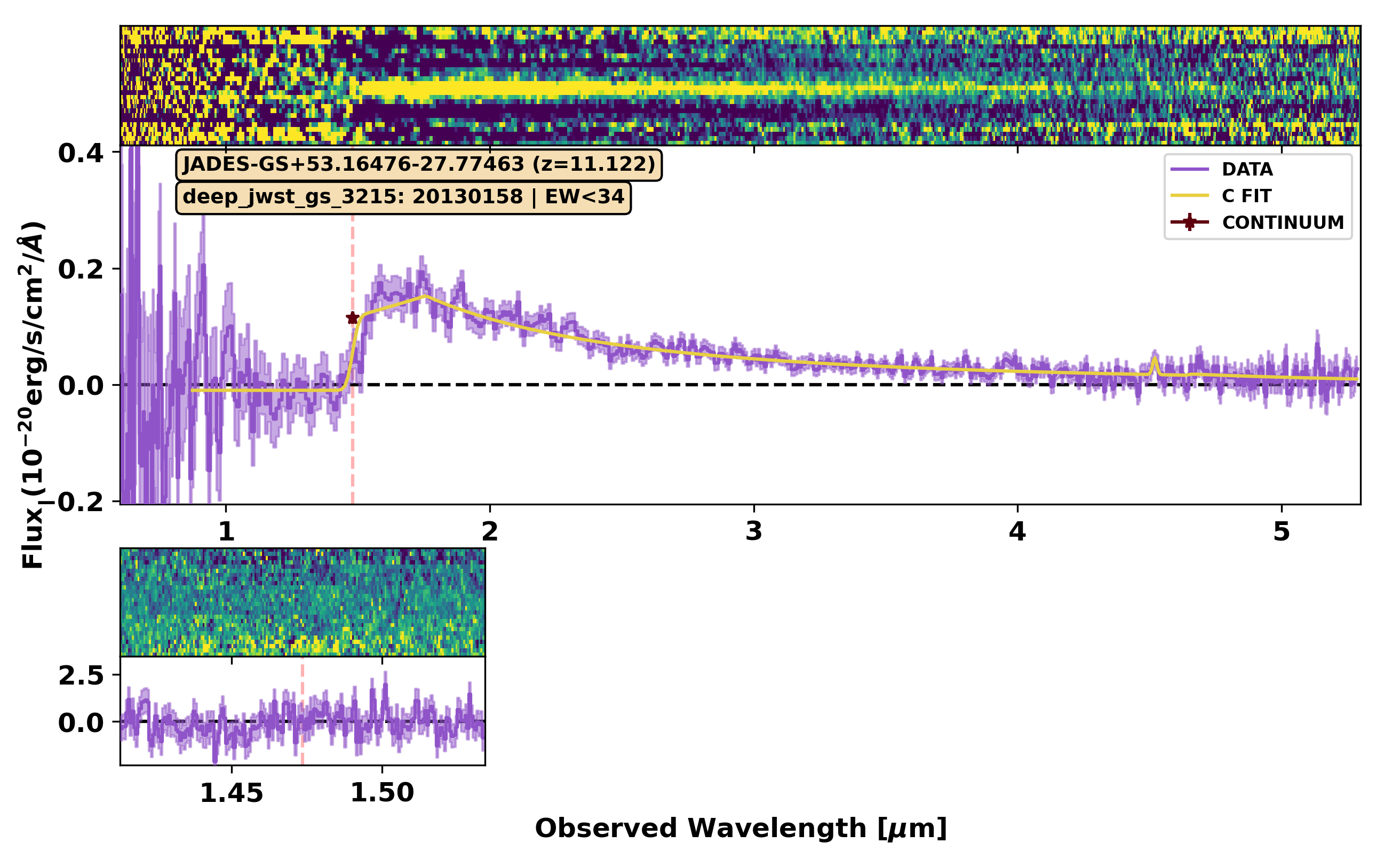}
\includegraphics[width=0.49\textwidth]{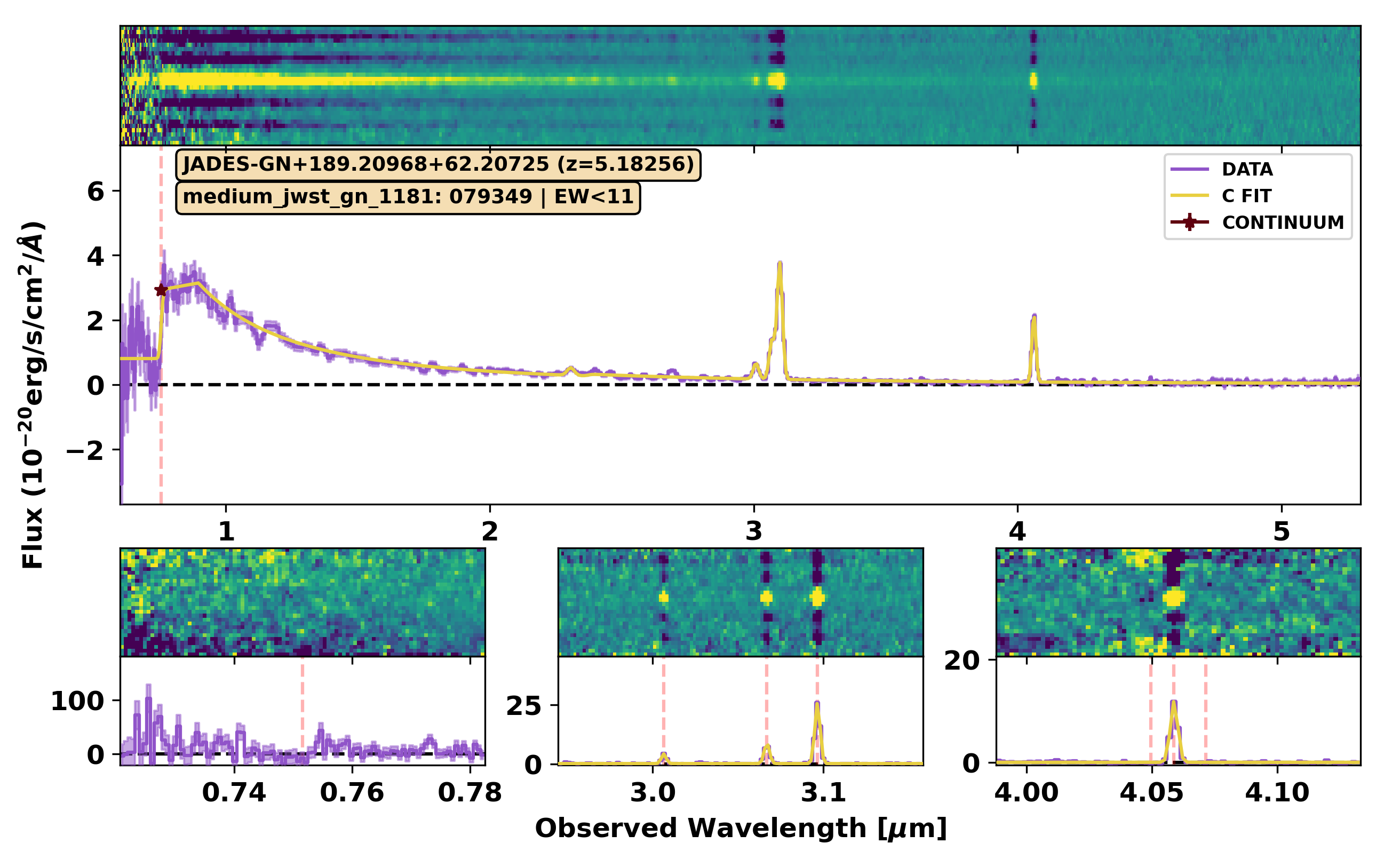}
\includegraphics[width=0.49\textwidth]{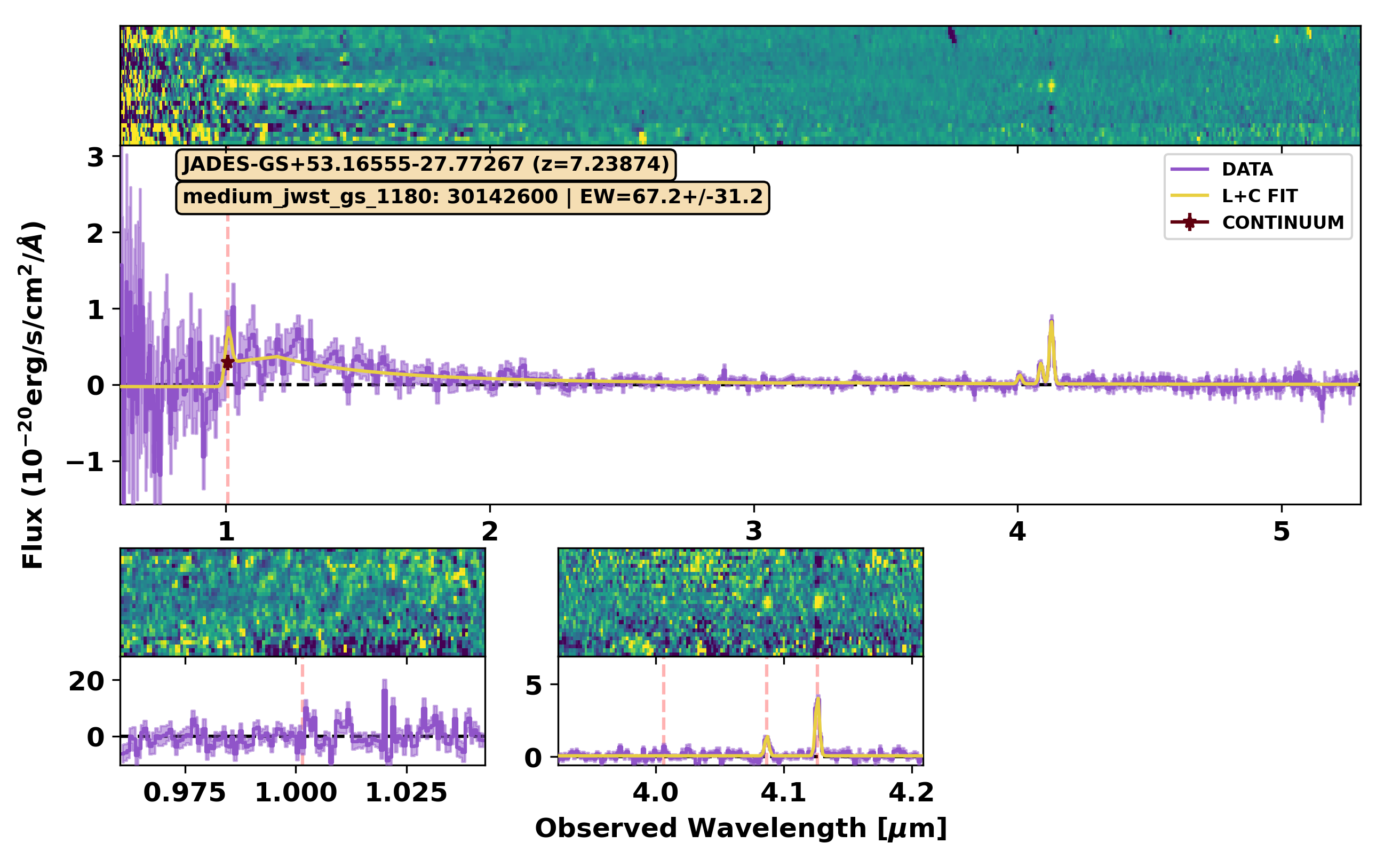}
\includegraphics[width=0.49\textwidth]{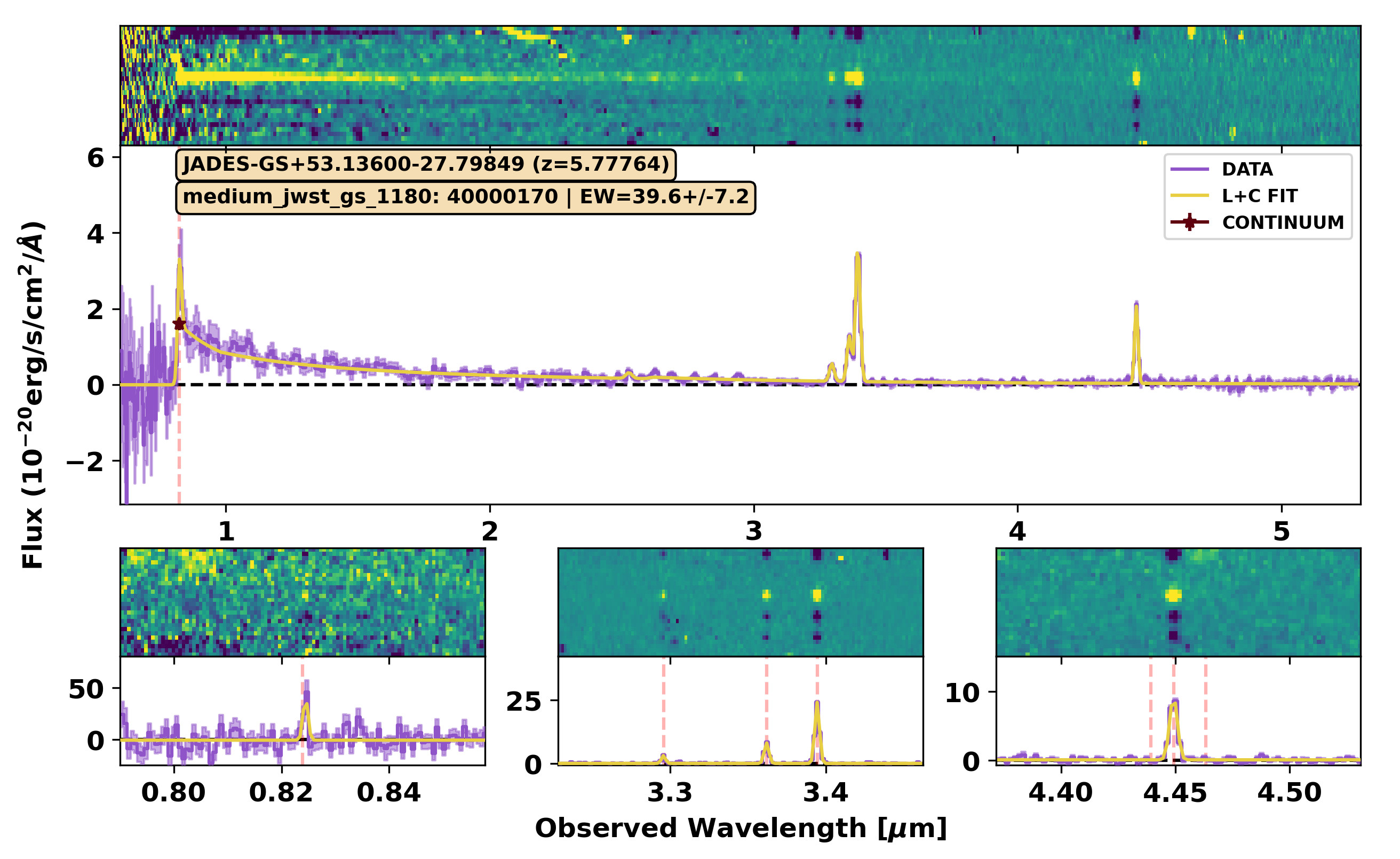}
\includegraphics[width=0.49\textwidth]{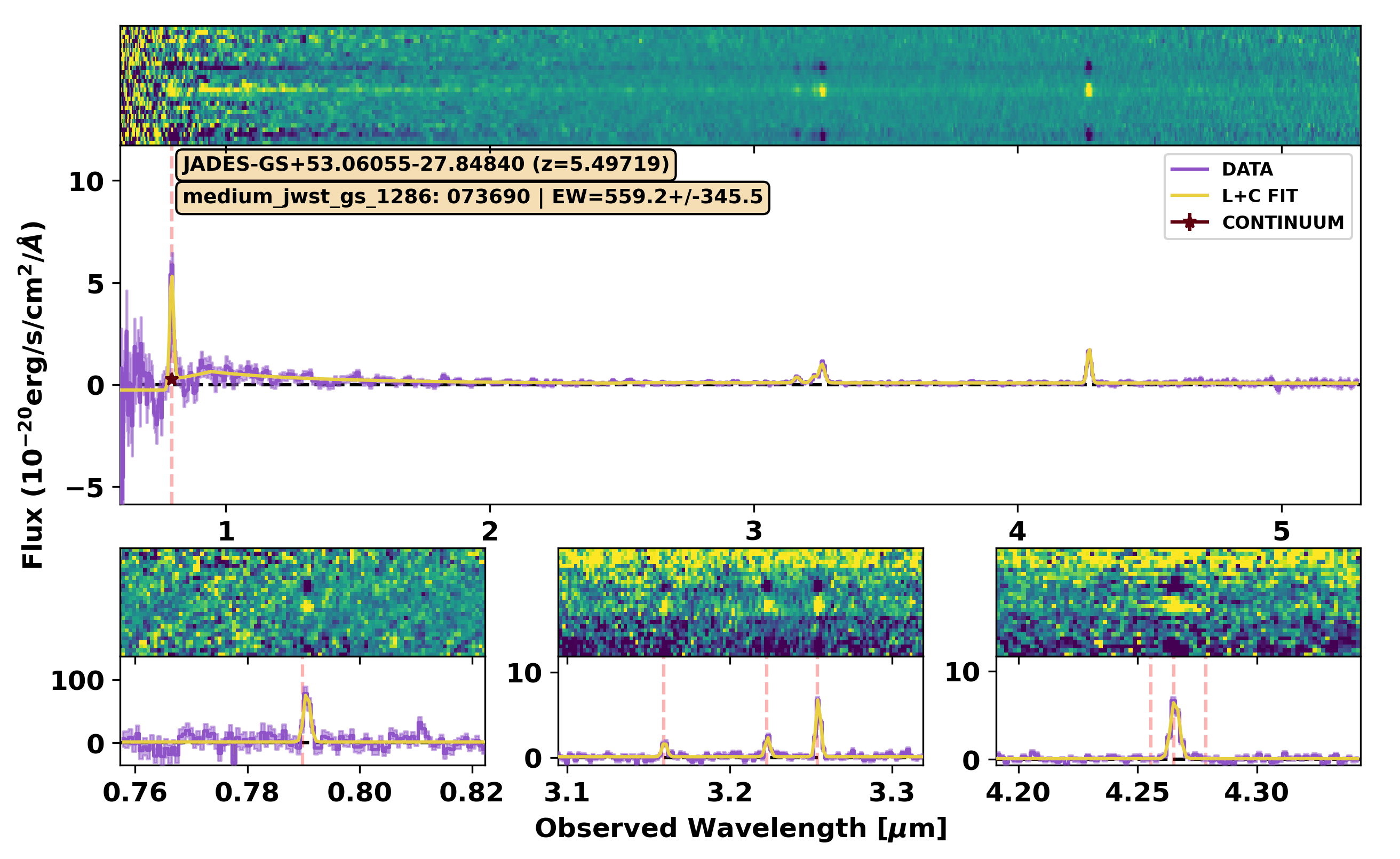}
\includegraphics[width=0.49\textwidth]{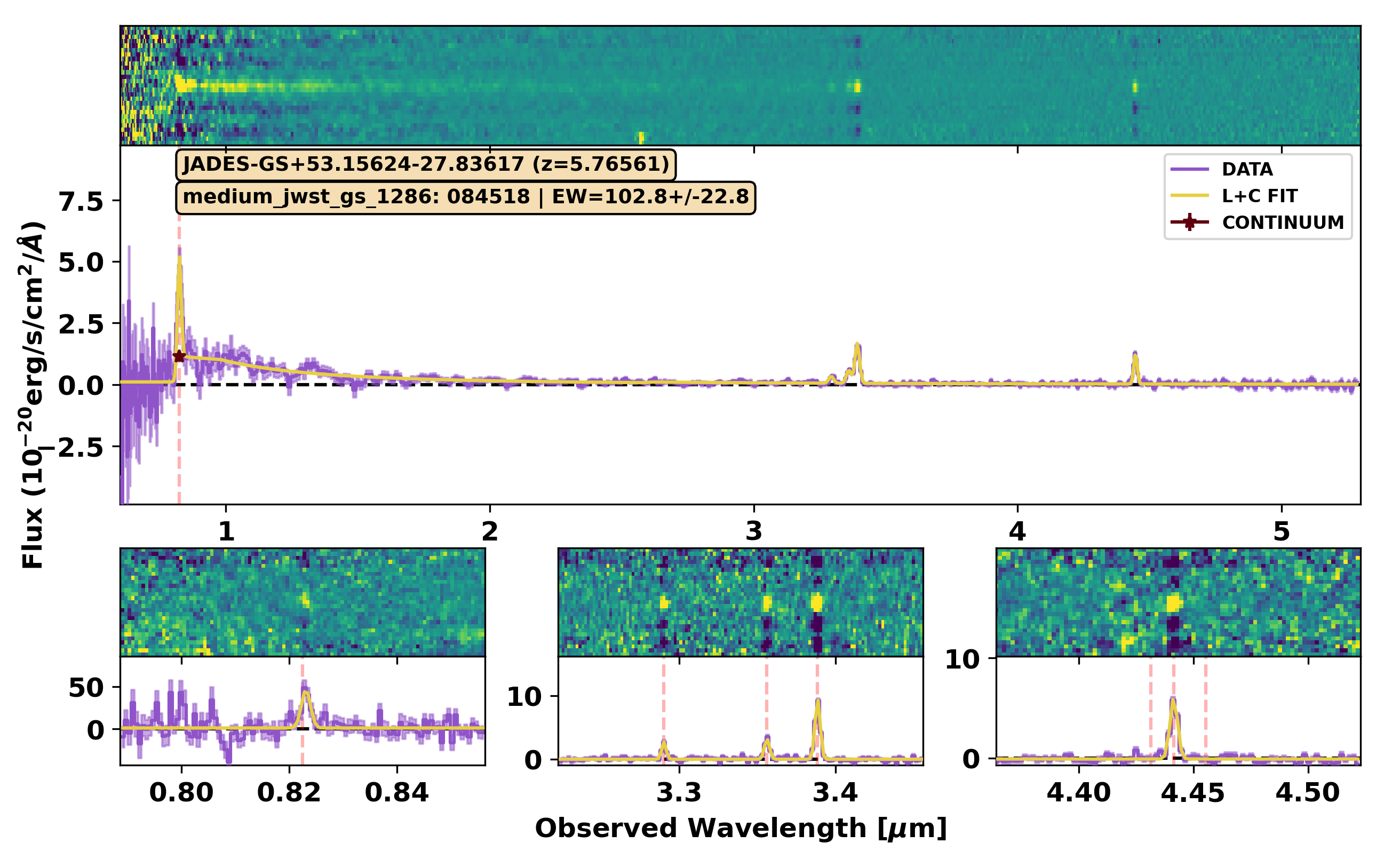}
\includegraphics[width=0.49\textwidth]{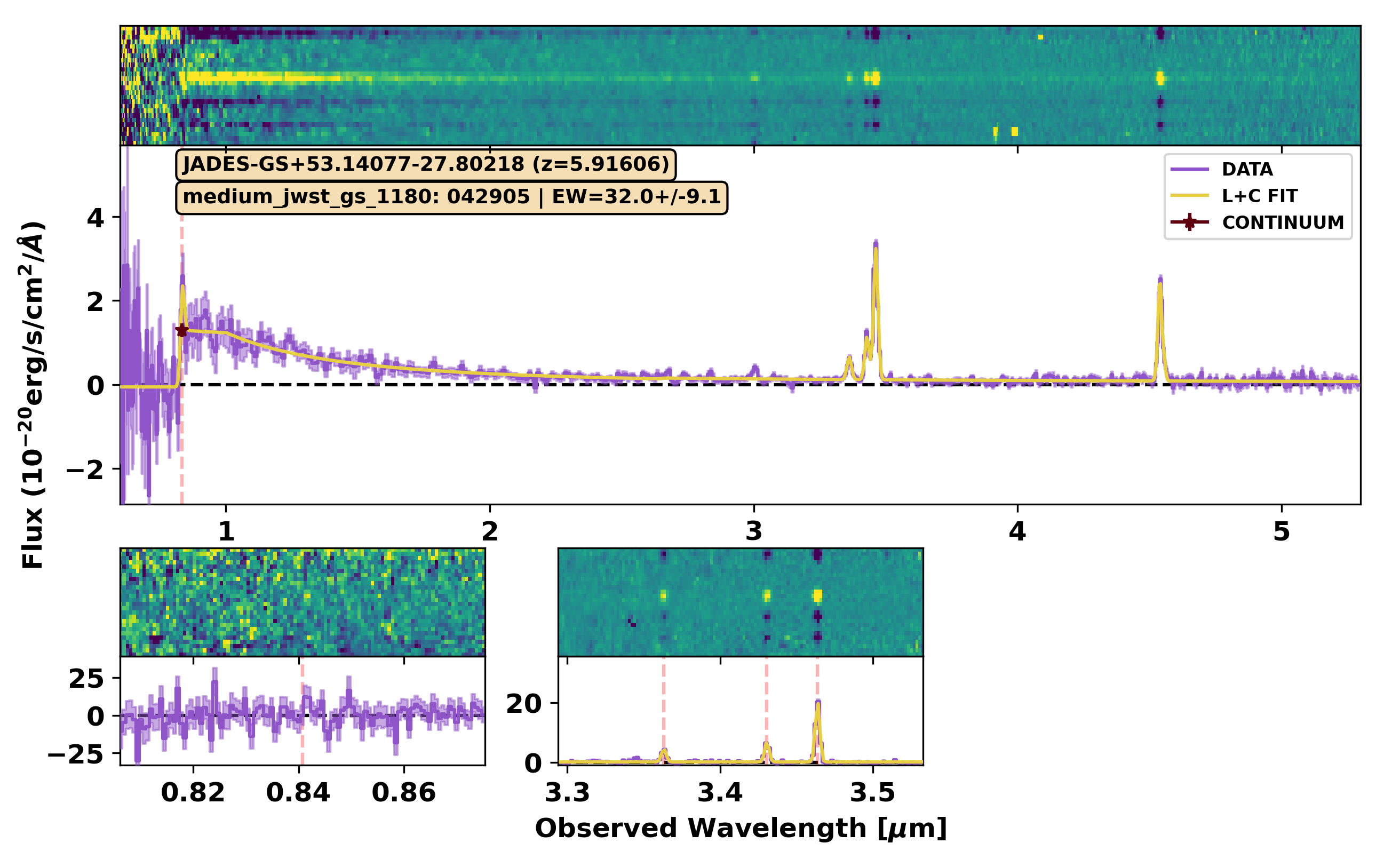}
\includegraphics[width=0.49\textwidth]{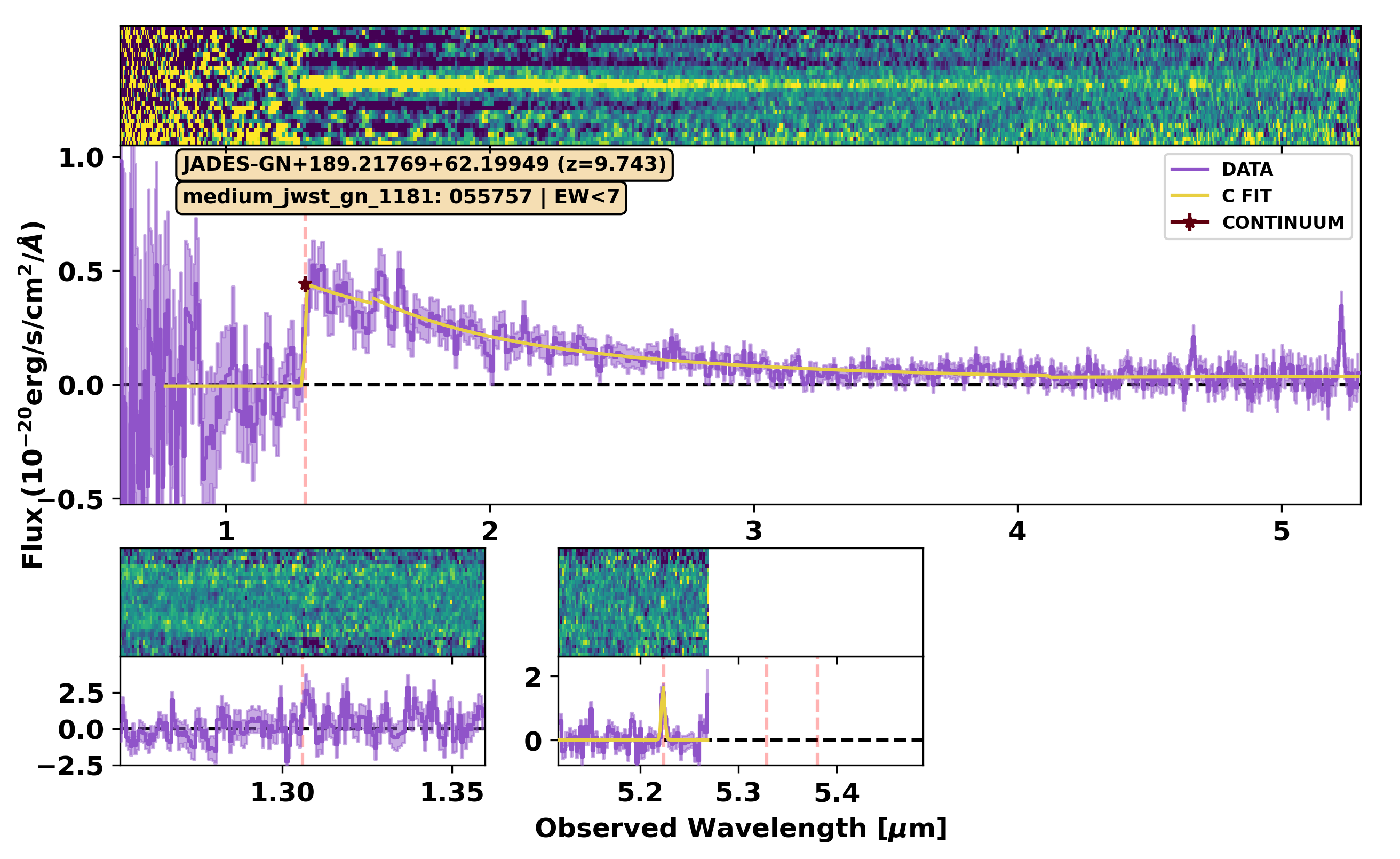}
\caption{Examples of fitting a line + continuum model to observed JADES data for the full R100 data (upper panel) and portions of the R1000 data (lower panels). In each, we show the observed 2D and 1D spectra. For the 1D spectra, we plot the observed spectrum (purple line) with an associated $1\sigma$ error (shaded region). The best-fit model, which includes the effects of the LSF, is shown by a yellow line. The continuum value at the redshifted \lya wavelength in the R100 data is represented by a brown star. The 2D corresponding spectrum of each spectrum is shown above each 1D spectrum. We show examples of sources where \lya is detected in both datasets/neither dataset/one dataset.}
\label{fit_ex}
\end{figure*}

\subsection{Comparison to previous results}\label{compsec}

In this work, we present $>100$\,galaxies at $z>4$ with evidence for \lya emission from the current JADES dataset. Because these objects are in a well-studied field, some have been previously detected in \lya emission. Additionally, a portion of these objects have been included in detailed studies (e.g., GN-z11; \citealt{bunk23a}) or analyses of populations (e.g., \citealt{tang24b}).

This work is a continuation of a previous study \citep{jone24}, which searched for \lya emission in a sample of 84\, galaxies at $5.6<z<11.9$ from R100 data in the first JADES data release (1210\_DHS, 1180\_MHS, and 1286\_MJS). Using a similar model as our R100-blue model, they found evidence for \lya emission in 17\,galaxies. All of these sources are recovered in our analysis, with \rew values in agreement (i.e., within $3\sigma$). 

Similarly, \citet{saxe23a} used R1000 data from 1210\_DHS and 1180\_MHS to find evidence of \lya emission in a sample of 17\,galaxies at $5.8<z<8.0$. For all of the galaxies in 1210\_DHS and most of the galaxies in 1180\_MHS, our results agree. However, there are some noteworthy exceptions in the medium-tier data. These may be due to different pipeline reductions (in some cases using different calibrations) or different continuum-level assumptions. As part of our effort to avoid including sources twice, we exclude one LAE in 1180\_MHS from \citet{saxe23a} in favour of a galaxy in 1286\_MJS. The two sources have a projected separation of $0.07''$($\sim0.4$\,kpc at the mean redshift of $z=6.60$) and $\Delta z=0.12$ ($\sim5000$\,km\,s$^{-1}$). This velocity offset is larger than the threshold used in most merger classification studies (e.g., \citealt{vent17,ends20,duan24}), but it is smaller than the threshold of \citet{gupt23} for a companion galaxy. An examination of the NIRCam data for these objects shows that they both lie within an extended feature\footnote{\url{https://jades.idies.jhu.edu/?ra=53.1374139&dec=-27.7652125&zoom=12}, see Appendix \ref{FV}}. Since their separation is less than the width of the MSA shutter ($0.2''$), we only include the source that was targeted using JWST-based selection and astrometry.

\citet{stan04} detected \lya emission from a source at $z\sim5.8$ (GOOD-S SBM03\#1) with Keck/DEIMOS (\rew$=30\pm10\angstrom$). This emission is coincident with an LAE in our sample (JADES-GS+53.16685-27.80413 in 3215\_DJS, \rew$=51\pm11\angstrom$). Thus, we identify this LAE as a re-detection of GOOD-S SBM03\#1 from the candidate list of \citet{stan03}.

Recently, the highest-redshift LAE in our sample (JADES-GS+53.06475-27.89024; $z=13.01$) was investigated in detail by \citet{wits24b}. By applying more detailed continuum models (e.g., two-photon continuum, absorption by damped \lya absorption systems), they find a larger \lya flux and intrinsic \rew. This highlights the need for advanced modelling for the highest-redshift sources.

\citet{wits24a} examine three JADES LAEs that also lie within our sample: JADES-GN+189.19774+62.25696 in 1181\_MHN ($z=8.2790$; called JADES-GN-z8-0-LA), JADES-GS+53.15891-27.76508 in 3215\_DJS ($z=8.4861$; JADES-GS-z8-0-LA), and JADES-GS+53.10900-27.90084 in 1287\_DJS ($z=8.7110$; JADES-GS-z8-1-LA). For each source, our \rew values are in agreement (i.e., within $3\sigma$). The first of these objects was then re-examined by \citet{nava24}, who find a similar \MUV, $\beta_{\rm UV}$, and R1000-based \lya flux and velocity offset as our model, despite using a more detailed asymmetric Gaussian model for \lya. However, their best-fit \rew value is $>7\sigma$ higher than our value due to the use of a best-fit continuum model from \textlcsc{msaexp}\footnote{\url{https://github.com/gbrammer/msaexp}} that steeply declines, resulting in a lower expected continuum level and higher \rew. \citet{tang24a} reported \lya emission from the $z=8.4861$ object from \citet{wits24a}, with comparable R1000-based flux as our value.

\citet{curt24} find tentative evidence for \lya emission (\rew$=31\pm16\,\angstrom$) in the R1000 data of a galaxy (JADES-GS-z9-0) that was observed in two tiers of JADES (1210\_DHS and 3215\_DJS). This detection was made possible by combining spectra from both programs, while our analysis of this object only used the higher-sensitivity spectra of 3215\_DJS (JADES-GS+53.11244-27.77463) and does not show evidence of \lya emission. So while higher-quality spectra may be produced by combining multiple exposures, this process lies beyond the scope of this work.

The well-studied galaxy GN-z11 also lies within our sample (JADES-GN+189.10604+62.24204 in 1181\_MJN). The JADES spectra of this source were first presented by \citet{bunk23a}, who find \rew$=18\pm2\angstrom$ using the same extraction aperture (5\,pixels) as we use in this work. Our analysis finds the same \lya flux (i.e., within $1\sigma$), but a lower \rew$=7\pm2\angstrom$. We find that the \lya break in the R100 spectrum is well fit by a Heaviside function convolved with the LSF, with no evidence of a strong DW. Because the LSF-convolved spectrum presents a lower value than the intrinsic \lya continuum level, this discrepancy in \rew is due to different assumptions on the underlying continuum level in our two works.

By combining public datasets from CEERS, JADES, GLASS, and UNCOVER, \citet{tang24b} present a set of 210\,galaxies at $z>6.5$. Of these, 110\,galaxies are from JADES, and 14 are reported as LAEs. Our independent analysis detects \lya in 13 of these objects, including the three new LAEs presented in \citet{tang24b} but excluding JADES-28342 (GN+189.22436+62.27561)\footnote{The \lya emission of this source was only detected in the R1000 data of \citet{tang24b}, but our analysis pipeline did not return an acceptable R1000 spectrum.}. The majority of our \rew values agree (i.e., within $3\sigma$), with the exception of two sources where our \rew values are higher (JADES-GN+189.14579+62.27332 and JADES-GS+53.14555-27.78380) and one source where our \rew value is lower (JADES-GS+53.13347-27.76037).
Finally, we note that our \rew values were measured with the JWST/NIRSpec MSA, which uses small observational slits ($0.20''\times0.46''$, where $0.1''$$\lesssim$$0.7$\,kpc at $z$$\geq$$4$). This is vital, as some studies have reported mismatched JWST/NIRSpec MSA and ground-based estimates of \lya flux (e.g., \citealt{jian24,jung23}), which are hypothesised to be due to the small area of the MSA slit, UV-\lya offset, or the existence of \lya halos. Simulations have also confirmed that these effects may result in inaccurate estimates of \lya flux and equivalent width from slit-based observations \citep{bhag24}. Future comparison of our values to ground-based observations should take this effect into account.

\section{Sample correlations}\label{corr_sec}

\subsection{\rew-\MUV distribution}\label{rewmuv}

\begin{figure*}
\centering
\includegraphics[width=\textwidth]{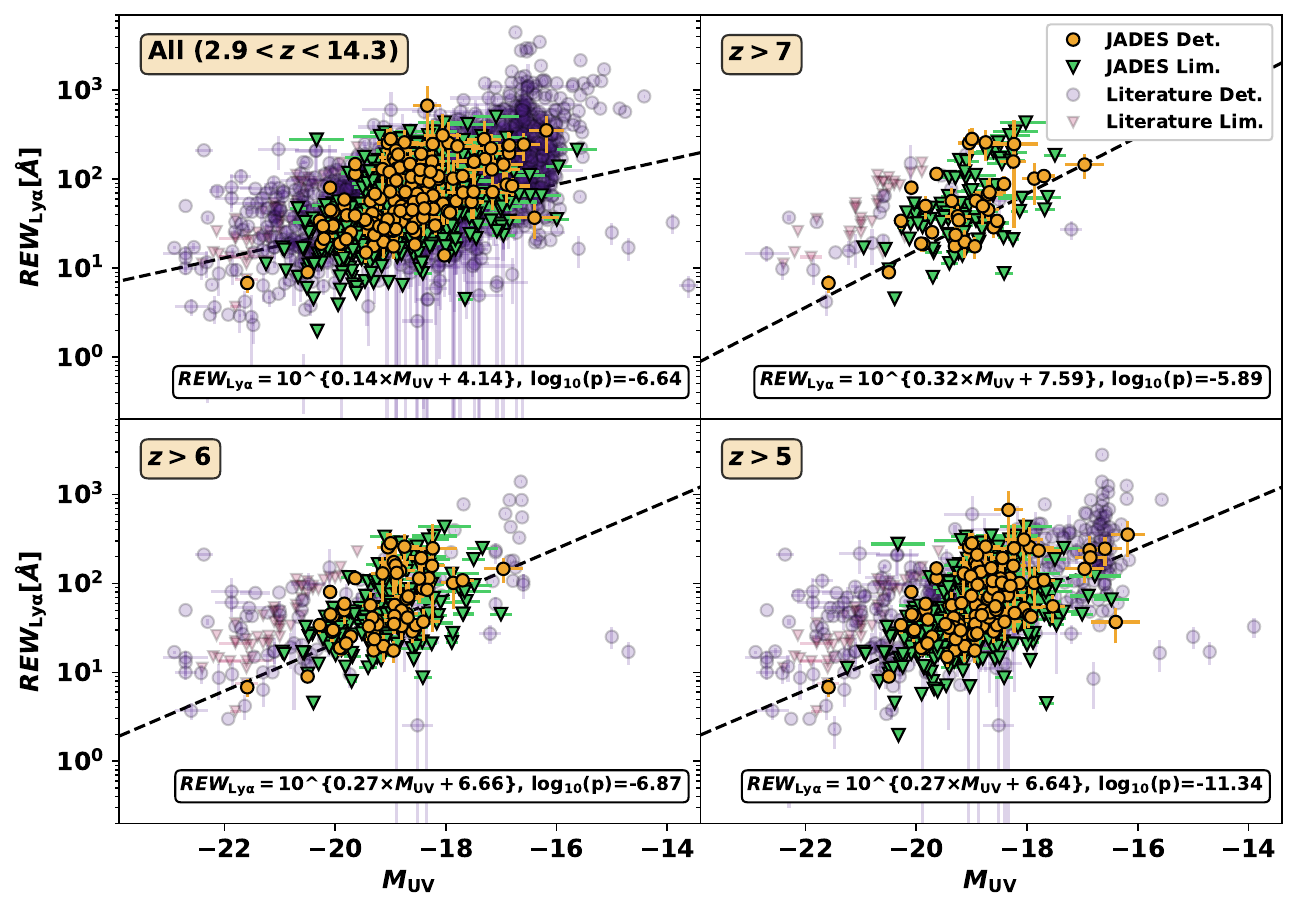}
\caption{Distribution of rest-frame \lya equivalent widths as a function of \MUV for our sample (orange circles for detections and green triangles for $3\sigma$ upper limits) and from literature (purple circles for detections and red triangles for $3\sigma$ upper limits). Only sources with robust \MUV estimates are shown. An illustrative fit to the JADES detections is shown by the black dashed line, and the best-fit parameters of this fit are included to the lower right of each panel. The literature sample (spanning a redshift range of $2.9<z<8.7$) is taken from a number of works (\citealt{cuby03,vanz11,ono12,sche12,will13,song16,shib18,pent18,hoag19,full20,tilv20,ends22,jung22,keru22,tang23}) as compiled by \citet{jone24}.}
\label{ewmuv}
\end{figure*}

The resulting distribution in \rew and \MUV is displayed in Figure \ref{ewmuv}. As found in other studies (e.g., \citealt{naka23, fu24, napo24}), there is a correlation between these values that is present at all redshift bins, implying that UV-faint galaxies feature higher \rew. Previous studies have suggested that this might be due to sensitivity effects (e.g., \citealt{jone24}), as the low \lya flux of galaxies in the lower right quadrant would require a deep blind survey to detect. But UV-bright, high-\rew galaxies (upper right quadrant), which would be easily detected, are not found. This suggests that the correlation is physical.

\subsection{${M_{\rm UV}}$-$z$ distribution}\label{muv_sec}

\begin{figure*}
\centering
\includegraphics[width=\textwidth]{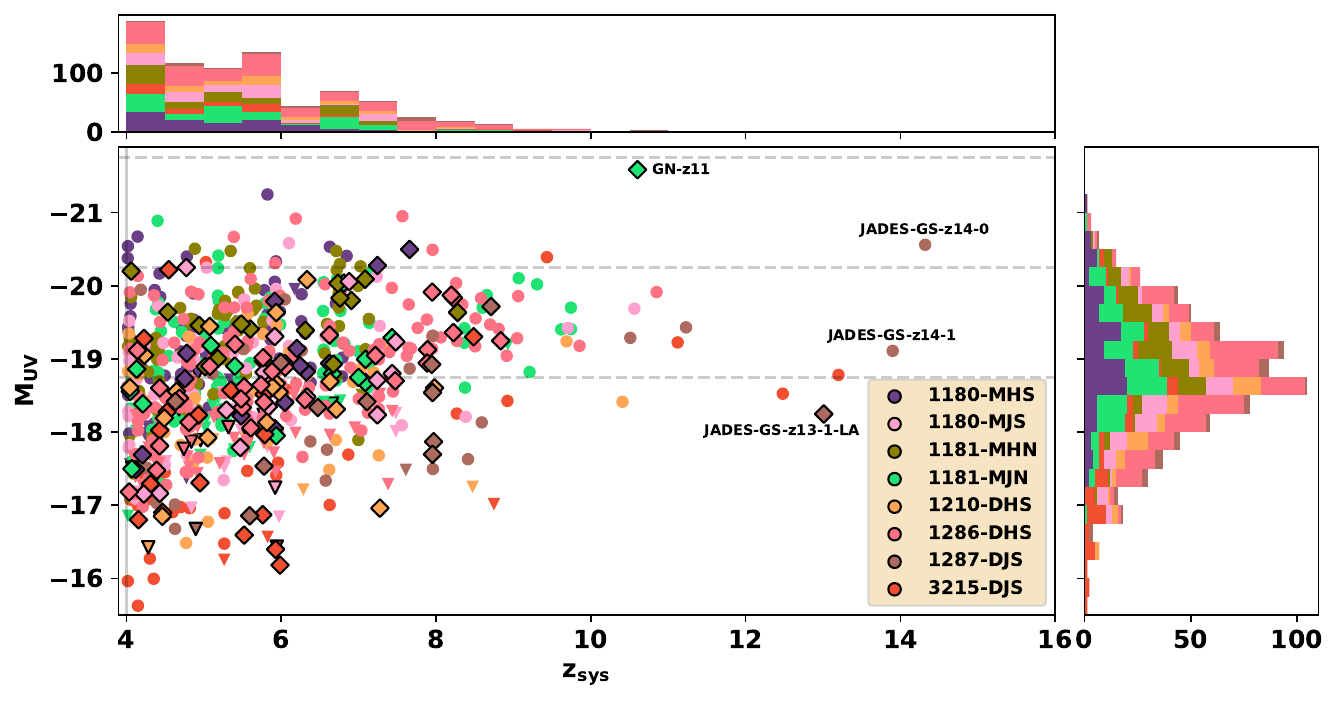}
\caption{$M_{\rm UV}$ (from NIRSpec R100 spectra) versus systemic redshift (based on rest-frame optical lines) for our sample. Galaxies observed in different tiers are coloured differently. Sources detected in \lya emission (in R100 and/or R1000) are shown as diamonds with black outlines. Horizontal dashed lines show $M_{\rm UV}$ values of -21.75, -20.25, and -18.75, while the vertical grey line shows our lower redshift cutoff ($z_{\rm sys}>4.0$). The locations of several well-studied objects are marked: GNz-11 \citep{bunk23a}, JADES-GS-z13-1-LA \citep{wits24b}, JADES-GS-z14-0, and JADES-GS-z14-1 \citep{carn24}. For sources where \MUV is not robustly measured from the observed R100 spectrum, the $3\sigma$ lower limit on \MUV is shown by a downwards-facing triangle.}
\label{muvz}
\end{figure*}

We also present the \MUV-$z$ distribution of our sample in Figure \ref{muvz}. While we exclude sources where \MUV was not significantly measured from the R100 spectra, the resulting distribution features a wide range of \MUV values (i.e., from $\sim-15.5$ to $\sim-21.75$ with a mean of $\sim-18.75$) and redshifts ($z\sim4.0-14.5$). We note the presence of four extraordinary objects: the UV-bright $z\sim10.6$ source GNz-11 (lime green point at centre top of plot; \citealt{bunk23a}), a verified LAE at $z\sim13$ (JADES-GS-z13-1-LA; \citealt{wits24b}), and two of the highest-redshift spectroscopically confirmed galaxies to date (JADES-GS-z14-0 and JADES-GS-z14-1; \citealt{carn24}), which do not exhibit \lya emission. 

Compared to the distribution from the previous work analysing \lya in JADES \citep{jone24}, we can immediately notice some improvements. First, our sample size is $\sim10\times$ the size of the previous work, due to the inclusion of data from additional JADES tiers and a wider redshift limit ($z>4$ rather than $z>5.6$). This results in a more symmetric distribution of \MUV values around \MUV$\sim-19$ and a larger number of sources in each redshift bin. The LAEs (black-outlined markers) are not clustered in a specific region of the distribution, but include UV-faint and UV-bright galaxies at nearly all redshifts.

\subsection{Ly$\alpha$ escape fraction correlations}\label{fesc_corr}

We may now consider the larger subsample of all galaxies with measures of \lya and \hb in R1000, in order to examine the full relation between \fesc and \rew for our sample (Figure \ref{fescrew}; see Appendix \ref{fesc_calc} for additional details on our \fesc calculation). The positive correlation, which has been previously evidenced through observations and simulations (e.g., \citealt{soba19,cass20,roy23,begl24,chou24}) is strengthened by the present work with the addition of more galaxies. In addition, it is clear that there is a gradient in redshift, with higher-redshift sources showing lower \rew values (e.g., \citealt{saxe23a}). 

\begin{figure}
    \centering
    \includegraphics[width=0.5\textwidth]{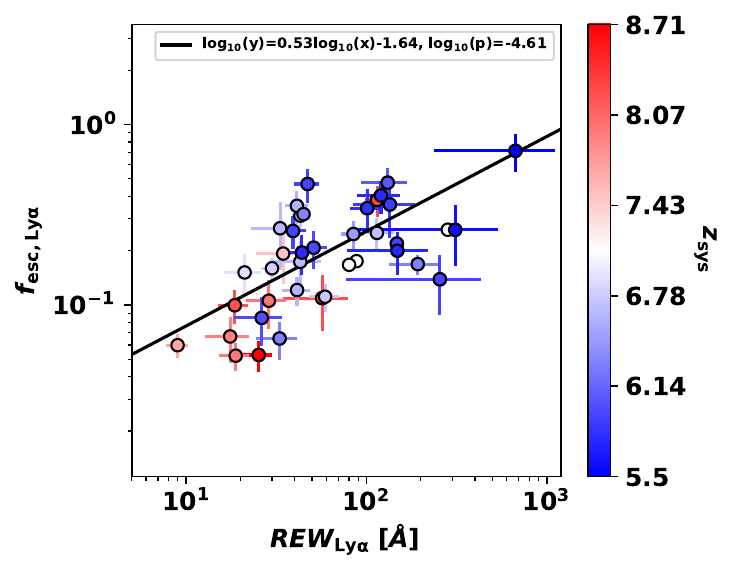}
    \caption{\lya escape fraction (derived by comparing the observed and intrinsic \lya/\hb flux ratio for the R1000 data, using no dust correction) as a function of \rew (derived using the \lya flux and $z_{\rm sys}$ from R1000 data and the continuum value from R100 data). Each point is coloured by redshift. More details about the use of these values are given in Appendix \ref{fesc_calc}.}
    \label{fescrew}
\end{figure}

\begin{figure*}
    \centering
    \includegraphics[width=\textwidth]{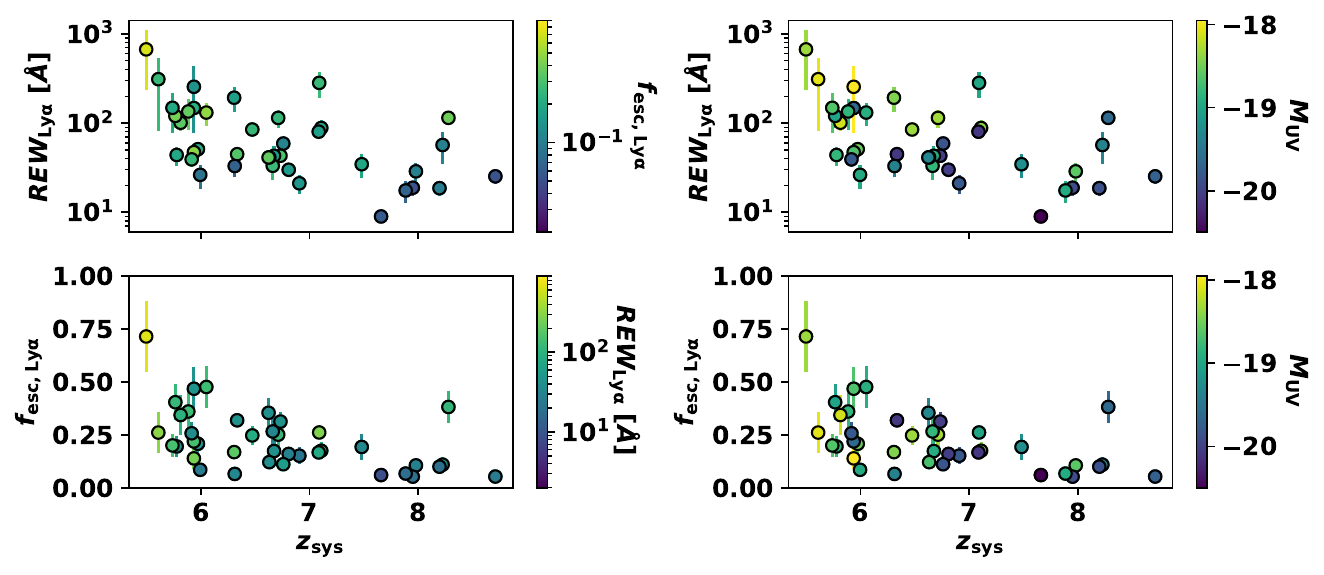}
    \caption{Redshift evolution of \rew (top row) and \lya escape fraction (lower row). In the left column, \rew and \lya escape fraction points are coloured by each other. The right-hand panels instead colour each point by \MUV.}
    \label{rewz}
\end{figure*}

This is shown more clearly in Figure \ref{rewz}, where we display \rew and \fesc as functions of redshift. Both quantities decrease with increasing redshift. One may interpret the increasing \lya escape fraction with cosmic time as a direct sign of the evolution of reionisation between $t_{\rm H}\sim0.4-1.2$\,Gyr ($z=10-5$), but it is important to rule out the possibility of selection biases. First, we consider the possibility that due to sensitivity effects, we may be biased towards more extreme systems at high redshift. But since the highest-redshift sources have low \rew, this is not the case.  Alternatively, \citet{saxe23a} find that this evolution may be caused by a relation between \rew and \MUV, with UV-fainter galaxies exhibiting higher \rew. This relation was examined in Section \ref{rewmuv}, where we are not able to discern if it is true or caused by selection effects. Our sample does not have a strong dependence of \MUV on redshift (see Figure \ref{muvz}), and there is no clear gradient in REW$_{\rm Ly\alpha}(z)$ or \fesc with respect to \MUV (right panels of Figure \ref{rewz}), so this is unlikely.

\begin{figure}
    \centering
    \includegraphics[width=0.5\textwidth]{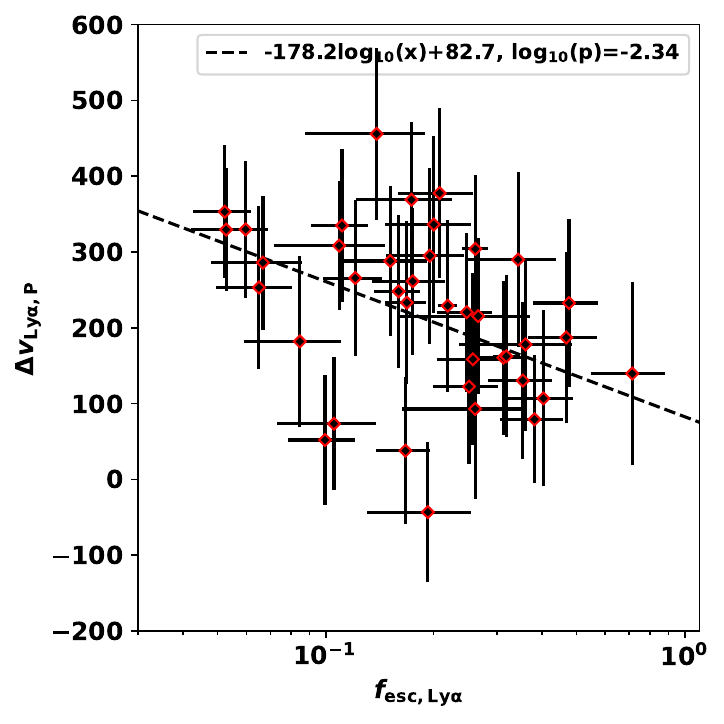}
    \caption{Correlation between $\Delta v_{\rm Ly\alpha}$ and \lya escape fraction We plot the best-fit linear relation and list the relation and p-value in the legend.}
    \label{delv_corr}
\end{figure}

Next, we examine the distribution of \fesc as a function of the Ly$\alpha$ velocity offset $\rm \Delta v_{Ly\alpha}$. Past works (e.g., \citealt{tang23,saxe23a}) found a negative correlation between $\rm \Delta v_{Ly\alpha}$ and \fesc, which is reproduced in our data (Figure \ref{delv_corr}). This implies that galaxies with high \lya escape feature \lya emission near the systemic redshift, which may be caused by a large ionised bubble (e.g., \citealt{wits23}). An ionised bubble would enable \lya to emerge largely unattenuated even at the core of the line. But in systems with neutral gas around the galaxy (i.e., small bubbles), this emission from the line core would be depleted through resonant scattering, with only photons in the red wings of the line emerging (and consequently a suppressed flux and lower \fesc).

\subsection{Dust properties}

\begin{figure}
\centering
\includegraphics[width=0.5\textwidth]{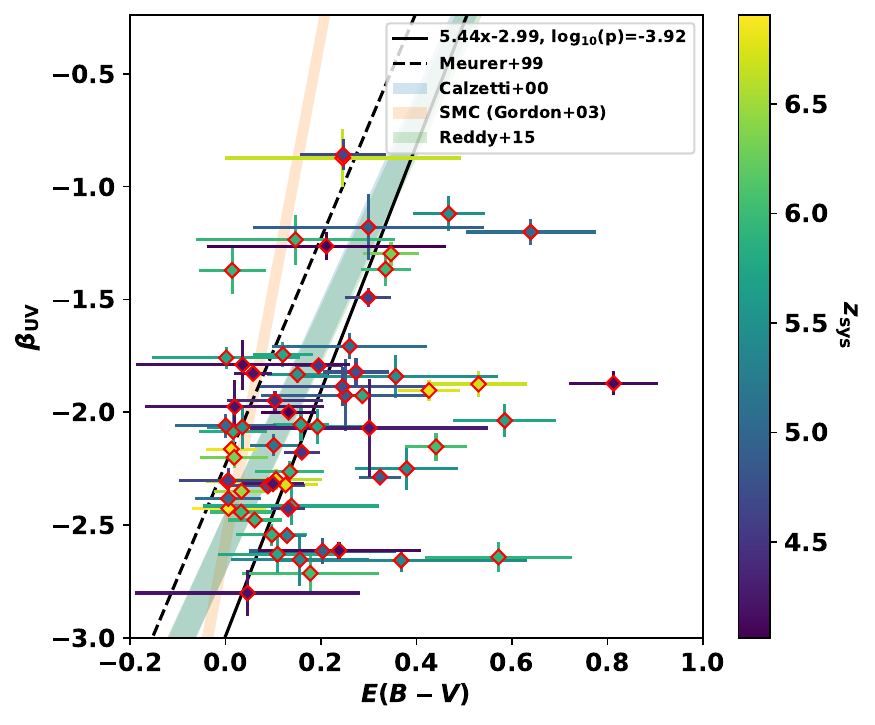}
\caption{UV spectral slope as a function of B-V colour excess (based on Balmer decrement) for LAEs, where both values are measured from the R100 spectra. Each point is coloured by spectroscopic redshift (also visually validated;  \citealt{deug24}). For comparison, we plot the relations from \citet{redd18} for multiple dust attenuation laws (\citealt{meur99,calz00,gord03,redd15}). Note that the relations of \citet{calz00} and \citet{redd15} are very similar, and overlap.}.
\label{betaebv}
\end{figure}

While not the primary focus of this work, we may also inspect the dust properties of the LAEs in our sample. To do this, we compare the rest-UV slope $\beta_{\rm UV}$ and B-V colour excess $E(B-V)$ from the R100 fits (Figure \ref{betaebv}). The latter is calculated using the observed Balmer decrement (see Section \ref{furobs}) and represents the reddening of the nebular lines, which may differ from the reddening of the stellar continuum. Our values are in agreement with the stacking analysis of \citet{kuma24}.

Redder UV slopes can be associated with increased dust extinction (e.g., \citealt{bhat21}), decreased LyC escape (e.g., \citealt{chis22}), higher Balmer break (e.g., \citealt{lang24}), and generally increased dust content (e.g., \citealt{aust24}). On the other hand, $E(B-V)$ is a direct measure of dust attenuation (e.g., \citealt{domi13,matt23}). Thus, it is expected that the two parameters should be correlated (e.g., \citealt{chis22}). Indeed, we find a positive correlation (Figure \ref{betaebv}). 

To put this result in context, we also plot the relations from \citet{redd18} between UV spectral slope and $E(B-V)$ using several dust attenuation laws. The \citet{calz00}, \citet{gord03}, and \citet{redd15} relations are calculated using two models: ``Binary Population and Spectral Synthesis'' (BPASS; \citealt{eldr12,stan16}) with low metallicity ($0.14$ solar) and those of \citet[][BC03]{bruz03} with $1.4$ solar metallicity. The span of these models is shown by a shaded region for each attenuation law. We also include the correlation of \citet{meur99}. 

The relation that we find for our LAEs features a similar slope as those of \citet{meur99}, \citet{calz00}, and \citet{redd15}, but with a lower value of $\beta$ when $E(B-V)=0$. On the other hand, the SMC curve of \citet{gord03} differs significantly. Because the calculation of our $E(B-V)$ values included the assumption of a \citet{calz00} attenuation law, this agreement is not surprising. The resulting shift between our correlations and the others could be caused by differences in galaxy properties (e.g., \citet{redd15} use galaxies at $1.5\leq z \leq 2.5$ while our sample is $4\leq z \leq 14$) or model assumptions. Further studies will shine more light on the dust properties of high-redshift LAEs. 

\section{Discussion}\label{discsec}

\subsection{IGM transmission}

Following other recent studies (\citealt{maso18a,naka23,tang24b}), we may use our full distribution of \rew values and upper limits to constrain the redshift evolution of a physical tracer of reionisation. While other works examine the neutral hydrogen fraction directly, we will examine the IGM transmission of \lya ($T_{\rm IGM}$). 

This is possible because of a few basic assumptions that are made in each of the other works. First, we assume that the continuum emission underlying \lya emission is not extincted. This allows us to calculate an observed $REW_{\rm Ly\alpha,obs.}=T_{\rm IGM}REW_{\rm Ly\alpha,emit.}$. Next, we choose the redshift range of $4.9<z<6.5$ (hereafter denoted as $z\sim5$) as reference\footnote{This epoch contains the end of the EoR ($T_{\rm IGM}=1.0$; e.g., \citealt{bosm22}), but also contains a time range where $T_{\rm IGM}<1.0$. Because of this, we use the \rew distribution as a reference to see how $T_{\rm IGM}$ evolves with $z$ rather than exploring the absolute value of $T_{\rm IGM}$.} and assume that the distribution of $REW_{\rm Ly\alpha,emit.}$ values does not change between $5<z<14$. Finally, we assume that this $REW_{\rm Ly\alpha,obs.}$ distribution only changes because of an evolving $T_{\rm IGM}$. 

\subsubsection{\rew distribution}

To begin, we follow a method similar to that of \citet{tang24a} to derive the distribution of $REW_{\rm Ly\alpha,emit.}$ values at $z\sim5$. This is done by isolating all galaxies in our sample that fall into the redshift range and have a well-determined \MUV value (see Appendix \ref{muvlimsec}). For all such galaxies with \lya detections, we calculate
the probability distribution implied by the \rew value:
\begin{equation}
P_{i,det}(REW_{\rm Ly\alpha}) = \frac{1}{\sqrt{2\pi}\sigma_i} \exp \left[\frac{-(REW_{\rm Ly\alpha,i}-REW_{\rm Ly\alpha})^2}{2\sigma_{\rm i}^2}\right]
\end{equation}
where $REW_{\rm Ly\alpha,i}$ is the measured \rew value and $\sigma_i$ is the associated uncertainty. For all galaxies that meet the $z$ and \MUV constraints but are non-detected in \lya emission, we find the probability distribution implied by the upper limit on \rew:
\begin{equation}
P_{i,lim}(REW_{\rm Ly\alpha}) = \frac{1}{\sqrt{2\pi}\sigma_i} \exp \left[\frac{-REW_{\rm Ly\alpha}^2}{2\sigma_{\rm i}^2}\right]
\end{equation}
All $P_{i}(REW_{\rm Ly\alpha})$ values are summed and the resulting distribution
is normalised. To ease further computation, we fit the distribution with a log-normal model:
\begin{equation}\label{lneq}
P(REW_{\rm Ly\alpha}) = \frac{1}{\sqrt{2\pi}\sigma REW_{\rm Ly\alpha}} exp \left[\frac{-(\ln(REW_{\rm Ly\alpha})-\mu)^2}{2\sigma^2}\right]
\end{equation}
Using the \textlcsc{optimize.curve\_fit} task of SciPy \citep{virt20}, we find best-fit values of $\mu=2.44\pm0.01$ and $1.64\pm0.01$. These are comparable to those of \citet{tang24b}, who used a similar redshift range\footnote{ The different levels of uncertainty originate from different methods of fitting.}: $\mu=2.38^{+0.28}_{-0.31}$ and $\sigma=1.64^{+0.23}_{-0.19}$.

\subsubsection{IGM transmission calculation}

With a \rew distribution in hand, we employ the Bayesian framework of \citet{maso18a} to constrain $\rm T_{IGM}$. In the case of LAEs, we have measured \rew values ($REW_{\rm i}$) and uncertainties ($\sigma_{\rm i}$), which we may use to calculate the likelihood implied by our measurement ($REW_{\rm i}\pm \sigma_{\rm i}$):
\begin{multline}
P(REW_{\rm i})_{\rm det} = \int_{0}^{\infty}\frac{1}{\sqrt{2\pi} \sigma_{\rm i}} e^{-(REW-REW_{\rm i})^2/(2\sigma_{\rm i}^2)} \\ \times P(REW/T_{\rm IGM})\,dREW
\end{multline}
For galaxies that are not detected in \lya emission, we may use our observational $1\sigma$ limits on \rew ($\sigma_{\rm i}$) to find the \rew likelihood:
\begin{multline}
P(REW_{\rm i})_{\rm lim} = \int_{0}^{\infty}\frac{1}{2}{\rm erfc}\left[\frac{REW-3\sigma_{\rm i}}{\sqrt{2}\sigma_{\rm i}}\right]  \\ \times P(REW/T_{\rm IGM})\,dREW
\end{multline}
All of these distributions are then combined to create a probability distribution for $\rm T_{IGM}$:
\begin{equation}
P(T_{\rm IGM}) = \prod_{\rm i}^N P(REW_{\rm i})
\end{equation}
where $P(EW_{\rm i})$ is the \rew probability distribution for each galaxy.

\begin{figure}
\centering
\includegraphics[width=0.5\textwidth]{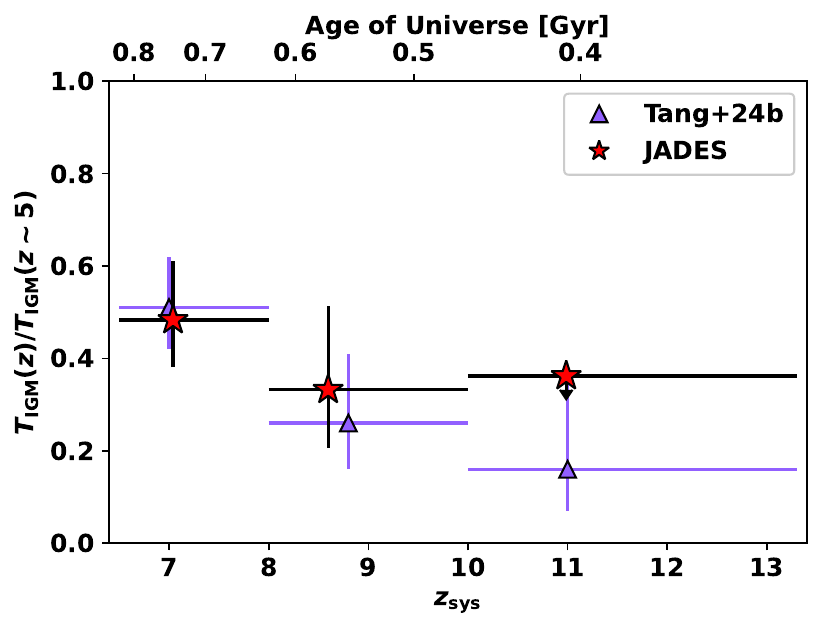}
\caption{Redshift evolution of IGM transmission of \lya (with respect to its value at $z\sim5$), as measured by \citet{tang24b} and through our analysis. Both studies show a strong increase in transmission with cosmic time.}
\label{tigmfig}
\end{figure}

Our sample is separated into three redshift bins ($6.5<z<8.0$, $8.0<z<10.0$, $10.0<z<13.3$), and we exclude galaxies where \MUV was not well determined from the R100 data. For each bin, we calculate $P(T_{\rm IGM})$ for $\rm T_{\rm IGM}=[0.01,0.02,0.03,\ldots,1.0]$ and calculate the 16th, 50th, and 84th percentiles, which we present in Figure \ref{tigmfig}. This analysis results in similar constraints on $\rm T_{IGM}$ as \citet{tang24b}.

\subsection{\lya fraction}\label{lyafrac_sec}
The large size of our sample allows for new constraints on the \lya emitter fraction ($X_{\rm Lya}$). This value represents the fraction of galaxies in an \MUV and redshift bin that are detected in \lya emission with an \rew value greater than a limit. The standard \rew limits are $25\angstrom$, $50\angstrom$, and $75\angstrom$, although some studies use slightly different values (e.g., $10\angstrom$, $55\angstrom$, \citealt{star11,ono12,naka23}). Here, we determine $X_{\rm Lya}(z)$ of our sample, after accounting for incompleteness at low \rew values.

\subsubsection{Completeness analysis}
As in \citet{jone24}, the galaxies in our sample span a wide range of redshifts and \MUV values. In order to use our measurements of \rew to place constraints on the characteristics of galaxy populations, we must examine the completeness of our sample at low values of \rew, where a weak \lya feature may be washed out by the strong spectral break at this wavelength at the low spectral resolution of the R100 data. But since \rew is a function of both the underlying continuum strength (i.e., $M_{\rm UV}$) and the \lya flux (in the case of a detection) or the error spectrum (for nondetections), we will use a series of models that take this complexity into account to determine the completeness.

There are a few key points that must be included. First, the sensitivity to \lya is dependent on both the observed wavelength of the line and the redshift, as the error spectrum for any given NIRSpec observation is not flat. More importantly, the scaling of these error spectra vary from tier to tier, with nearly an order of magnitude difference between the error spectrum of 1180\_MHS and 3215\_DJS at $\lambda_{\rm obs}\sim1\,\mu$m. In addition, our sample covers a range of $\Delta {M_{\rm UV}}>5$\,mag, strongly affecting our ability to detect low \rew emission.

An added complexity arises from the fact that we wish to examine the completeness of our sample for all \rew values beyond a lower limit, rather than the completeness at a single value. Thus, if we wish to evaluate the completeness value for a single galaxy at a given limit (e.g. $>25\angstrom$), we must create a series of models with a range of \rew values and calculate the fraction of models with an intrinsic \rew larger than the limit that are well-fit. Assuming a single value (e.g., the \rew limit; \citealt{jone24}) will result in an unrealistically low completeness value. Similarly, assuming a uniform \rew distribution is non-physical, as there are fewer extreme LAEs (e.g., \citealt{tang24a}). We choose to use a physically motivated distribution of \rew models.

\begin{figure}
\centering
\includegraphics[width=0.5\textwidth]{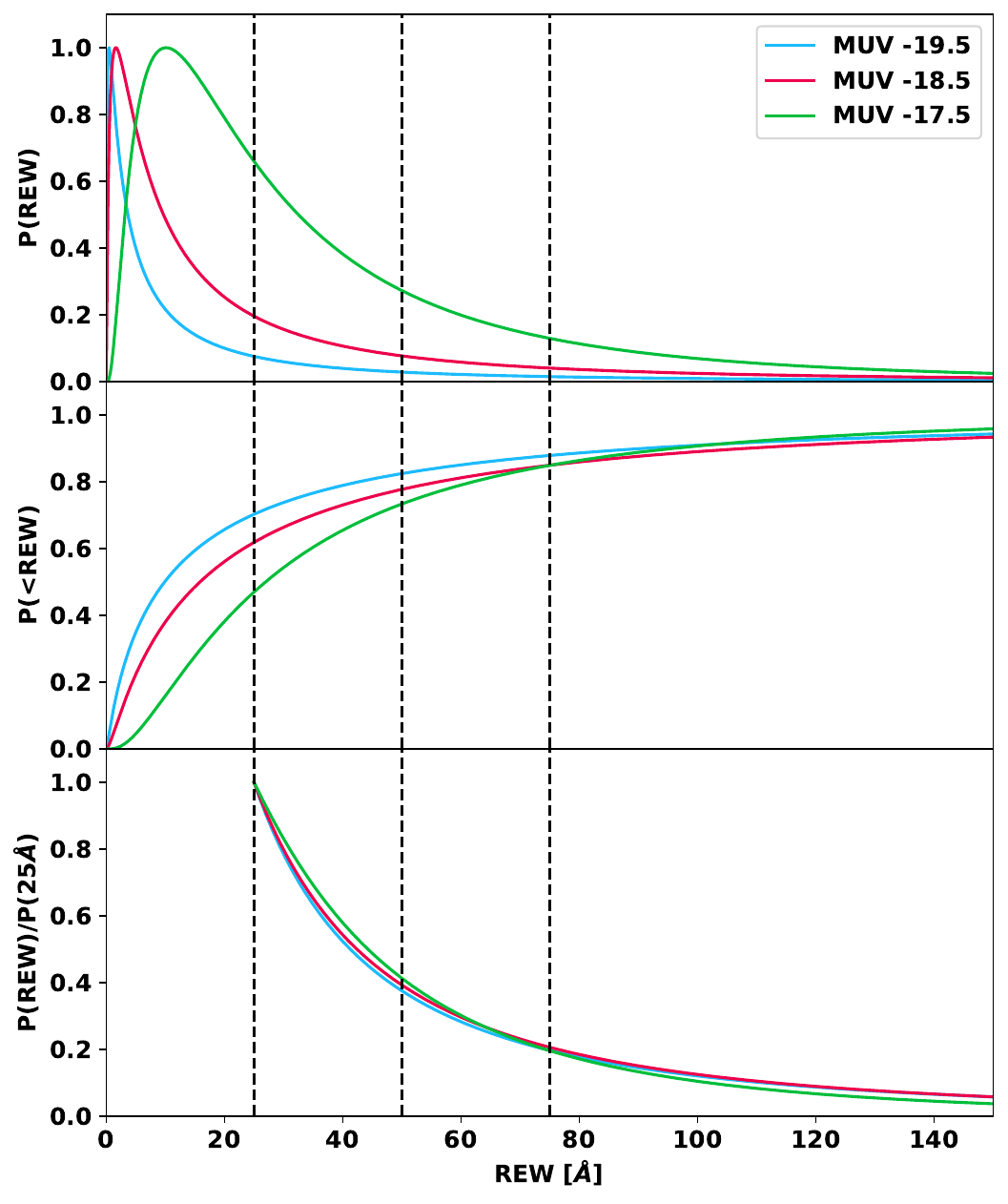}
\caption{Expected \rew distribution of galaxies at $z=5-6$, based on models of \citet{tang24a}. The top row shows the normalised distribution of galaxies as a function of \rew, while the middle row shows the CDF. The bottom row shows the same distributions as the top row, but normalised by their value at \rew$=25\angstrom$. For each panel, we show three \MUV values (\MUV$=[-19.5,-18.5,-17.5]$) with different colours, and mark three \rew values of interest (\rew$=[25,50,75]\angstrom$).}
\label{tangrew}
\end{figure}

The redshift range $z=5-6$, which lies below the expected mid-point of reionisation, can be used to get a handle on the intrinsic distribution of \rew at high redshift without much impact from IGM absorption. The \rew distribution of galaxies in this epoch was recently determined by \citet{tang24a} for three bins of \MUV=[-17.5,-18.5,-19.5]. Because their analysis already accounts for completeness, it is suitable for this analysis. This distribution (reproduced in the top panel of Fig. \ref{tangrew}) shows a moderate dependence on \MUV, with the UV-faint population containing a larger proportion of higher-\rew galaxies. This is shown in an alternate way in the middle panel, which shows the cumulative distribution function (CDF) of these distributions. The CDF of the UV-bright population rises to unity quickly, representing a wealth of low-\rew galaxies. Since we only wish to examine galaxies with \rew$>25\angstrom$, we examine the \rew distribution as normalised by the $25\angstrom$ value (bottom panel of Fig. \ref{tangrew}). These normalised distributions are quite similar, with differences of $\lesssim10\%$. Based on this similarity across \MUV values, we adopt the \MUV$=-18.5$ distribution when constructing our models.

To begin, we derive mean error spectra for each tier. For each observed galaxy, we create 30 high-resolution mock spectra using the R100-blue model from Section \ref{moddes} with different realisations of the appropriate noise spectrum. Each model has the same \MUV and redshift as the observed galaxy, but with a variable \rew (sampled from the \rew distribution described previously, with limits of $25\angstrom\leq$\rew$\leq 500\angstrom$), $\beta$ (sampled from a uniform distribution between [-2.5,2.5])\footnote{Note that this slope is different from $\beta_{UV}$, see Section \ref{moddes}.}, and deviation from the fiducial LSF ($F_{\rm R}$; sampled from a uniform distribution between [0.5,1.0]). For this analysis, we exclude all galaxies from each tier whose \MUV value cannot be measured directly from the R100 data due to high noise levels, as their \rew is poorly constrained. Since we are able to constrain \rew across a wide range of \MUV values (Figure \ref{ewmuv}), this does not strongly affect our analysis.

Each model is fit using the R100-blue model of Section \ref{moddes}, and we record the best-fit \rew ($REW_{\rm obs}$) and the associated uncertainty ($\delta REW_{\rm obs}$). The completeness for an observed galaxy at the given \rew limit (hereafter $C_{\rm j}(>REW_{\rm lim})$) is derived by dividing the number of models that meet the \rew limit with successful \lya detections ($REW_{\rm obs}>3\delta REW_{\rm obs}$ and $<3\sigma$ difference between $REW_{\rm obs}$ and $REW_{\rm input}$) by the number of such simulations. As an example, consider the galaxy with ID 1655 in 1181\_MHN ($z_{\rm sys}= 4.474$), which yields $C_{\rm j}(>25\angstrom)\sim 60\%$, $C_{\rm j}(>50\angstrom)\sim 89\%$, and $C_{\rm j}(>75\angstrom)\sim 93\%$.

\begin{figure*}
    \centering
    \includegraphics[width=\textwidth]{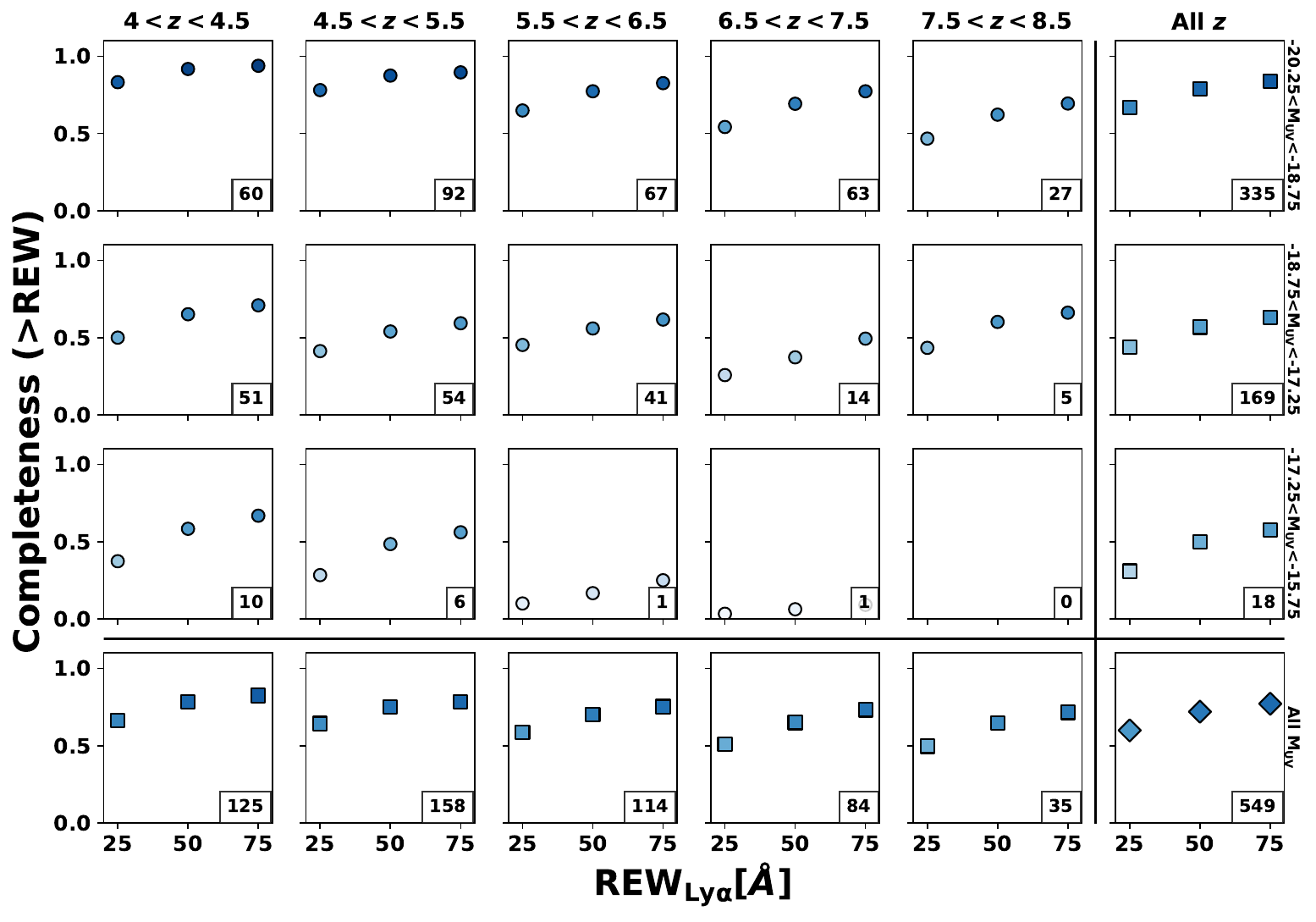}
    \caption{Completeness of galaxies in our sample for three \rew limits, divided into bins of redshift (columns, increasing from left to right) and \MUV (rows, decreasing in brightness from top to bottom). Darker points indicate higher completeness. The lower row and rightmost column show the completeness with no limits on \MUV and redshift, respectively. Thus, the lower right panel is the completeness of the full sample. The number of galaxies in each bin is shown the lower right of each panel.}
    \label{comp_panels}
\end{figure*}

This analysis allows us to examine the completeness of our sample and technique as a function of galaxy properties. We calculate the average completeness for our sample in bins of redshift and \MUV, excluding galaxies for which our completeness analysis returned $C_{\rm j}(>25\angstrom)=0$. As shown in Figure \ref{comp_panels}, the completeness increases with \rew for nearly all bins. The completeness decreases strongly with \MUV, with $C\lesssim50\%$ for the UV-faint bin. Since the majority of our sources lie in the $-20.25<{M_{\rm UV}}<-18.75$ bin, our average completeness is $\sim50-80\%$.

\subsubsection{Ly$\alpha$ fraction determination}

The \lya fraction for each redshift bin ($z_{\rm i}$, e.g., $4.5<z<5.5$) and \rew bin (e.g., $REW_{lim}=25\angstrom$) is then evaluated as:
\begin{equation}
X_{\rm Ly\alpha}(z_{\rm i},REW>REW_{\rm lim}) = \frac{N_{\rm det}(>REW_{\rm lim})}{\sum_j^{N_{\rm obs}}C_{\rm j}(>REW_{\rm lim})}
\end{equation}
where $N_{\rm det}(>REW_{\rm lim})$ is the subset of these galaxies with a detected \rew greater than $REW_{\rm lim}$ and $N_{\rm obs}$ is the total number of observed galaxies that meet the $z$ and \MUV requirements. Because the completeness factor is bound between zero and unity, this form of $X_{\rm Ly\alpha}$ will always be equal to or greater than the form lacking a completeness correction. For this calculation, we exclude galaxies where \MUV was not measurable from the R100 spectra, as well as galaxies with $C_{\rm j}(>REW_{\rm lim})=0$.

\begin{figure*}
\centering
\includegraphics[width=\textwidth]{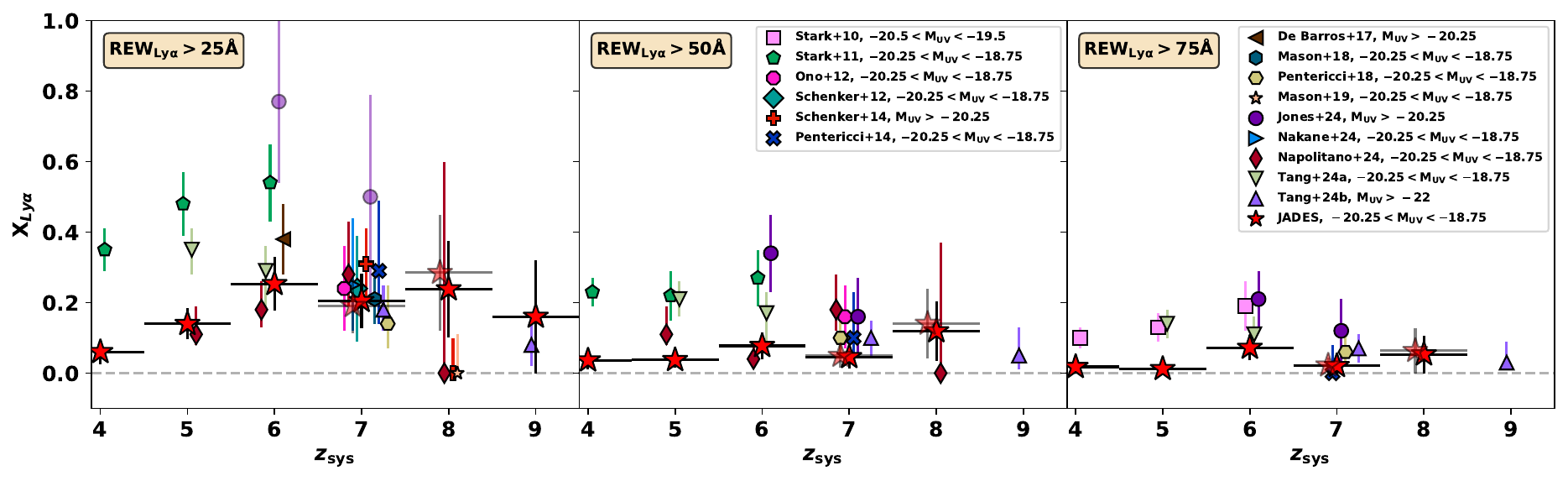}
\caption{Fraction of observed galaxies detected in \lya emission with REW$_{Ly\alpha}>25\angstrom$ (left), REW$_{Ly\alpha}>50\angstrom$ (centre), and REW$_{Ly\alpha}>75\angstrom$ (right). Our derived fractions are shown as red stars, while fractions from literature are shown by coloured markers (\citealt{star10,star11,ono12,sche12,sche14,pent14,pent18,deba17,maso18a,maso19,jone24,naka23,napo24,tang24a,tang24b}). Points are shifted in redshift for clarity. The \rew$>25\angstrom$ points of \citet{jone24} were affected by low completeness, so we display them with low opacity. For the central panel, note that \citet{star11} and \citet{ono12} used a \rew limit of $>55\angstrom$. The \lya fractions calculated by excluding the eight galaxies in possible galaxy overdensities (see Section \ref{soudis}) are depicted by low-opacity red stars.}
\label{lyafracfig}
\end{figure*}

The resulting evolution of \lya fraction with redshift is shown in Figure \ref{lyafracfig}, along with values from literature (\citealt{star10,star11,ono12,sche12,sche14,pent14,pent18,deba17,maso18a,maso19,jone24,naka23,napo24,tang24a,tang24b}). For this comparison, we exclude some \lya fractions derived only using UV-bright galaxies (\MUV$<-20.25$; \citealt{curt12,cass15,furu16,star17,yosh22,fu24}), as these are found to have systematic differences (e.g., \citealt{star11,pent14,pent18}). To avoid overcrowding of the figure, we also do not include all available studies (e.g., \citealt{mall12,treu13,caru14,tilv14,full20,kusa20,goov23}). The fractions for \citet{tang24a} and \citet{tang24b} are derived by integrating their best-fit \rew distributions. We note that \citet{napo24} presents \lya fractions derived using multiple datasets (i.e., JADES and CEERS). For each redshift bin, we include their result with the largest sample size with overdensity correction, if available. 

For most datasets, there is a clear increase in $X_{\rm Ly\alpha}$ from $z=4$ to $z=6$ and a decrease for $z>6$, due presumably to enhanced IGM absorption. However, a clear spread in values is present for each redshift bin (likely due to changes in stellar populations and ISM properties). Our $X_{\rm Ly\alpha}$ values for $4<z<6.5$ are lower than those of archival studies that used DEIMOS on Keck (i.e., \citealt{star10,star11}), but are comparable to other JWST studies (i.e., \citealt{napo24,tang24a}) as well as a study that used the FOcal Reducer/low dispersion Spectrograph 2 (FORS2) on the Very Large Telescope \citep[VLT; ][]{deba17}. 

As one of the first redshift bins containing galaxies in the EoR, the \lya fraction at $6.5<z<7.5$ has been very well explored. Our fractions are in agreement with the other results. 
We previously found that some of the \lya emitting galaxies in the $7.5<z<8.5$ bin may lie in galaxy overdensities (see Section \ref{soudis}). Since these may trace regions of increased \lya transmission (e.g., \citealt{ouch10}), we also estimate the \lya fraction when these galaxies are excluded (see faint red stars in Figure \ref{lyafrac_sec}). This results in lower $X_{\rm Ly\alpha}$ values, but not significantly (i.e., $<1\sigma$). The highest-redshift bin ($8.5<z<9.5$) has not yet been well explored, but our low fraction agrees with the findings of \citet{tang24b}.  

To summarise, our derived \lya fractions imply a similar evolution as previous studies: an increase from early times ($z\sim9.5$) to the end of the EoR (between $5.5<z<6.5$), followed by a decrease to $z\sim4$. We briefly note that proper constraints on $X_{\rm Ly\alpha}$ require knowledge of the effects of cosmic variance, selection effects, and observational biases. The JADES survey is well-suited to the discovery of LAEs, but due to its relatively small survey area (i.e., the GOODS fields) and the pre-selection of sources to be observed with the NIRSpec MSA, we may be affected by these effects. Future studies including more fields will correct this effect, but they must also take completeness into account.

\subsection{Neutral hydrogen fraction}\label{xhisec}

The evolution of the neutral hydrogen fraction is key to the study of the EoR, as it directly traces the process of reionisation. A number of studies have constrained this evolution using different techniques, including DW modelling (e.g., \citealt{umed23,duro24,spin24}), detailed reionisation simulations (e.g., \citealt{mora21,bhag23,asth24,mukh24}), and analysis of \lya fractions (e.g., \citealt{ono12,furu16,maso18b,jone24}). While a general evolution from ${X_{\rm HI}}=1$ at $z\gtrsim9$ to ${X_{\rm HI}}\sim0$ at $z\sim6$ is observed, the exact evolution of this fraction for $z\sim6-9$ is not yet well constrained. Here, we combine our \lya fractions with the model outputs of \citet{dijk11} to place constraints on the neutral fraction at $z\sim7$. 

The model originally created in \citet{dijk11} was built on the assumption that the evolution of the \lya fraction between $z=7$ and $z=6$ (a period of $\sim170$\,Myr) is predominately dictated by a changing neutral fraction (\xhi). While the intrinsic distribution of \rew may also evolve due to changes in galaxy population properties (e.g., metallicity, ISM conditions), the small timescale between these redshifts makes this assumption reasonable. In addition, they assume that the \rew distribution at $z=6$ may be described by an exponential with scale length $REW_{\rm Ly\alpha,c}=50\angstrom$. Further studies using additional observations (e.g., \citealt{pent18,naka23}) determined that $REW_{\rm Ly\alpha,c}$ was lower ($\sim30-40\angstrom$). 

This model was recently utilised by \citet{naka23}, who assumed $N_{\rm HI}=10^{20}$\,cm$^{-2}$ and outflow speed $v_{\rm wind}=200$\,km\,s$^{-1}$ and restricted the galaxy sample to those with $-20.25<{{M_{\rm UV}}}<-18.75$. As part of their results, they include a set of \rew distributions for different values of $REW_{\rm Ly\alpha,c}$ and ${X_{\rm HI}}(z=7)$.  

\begin{figure}
\centering
\includegraphics[width=0.5\textwidth]{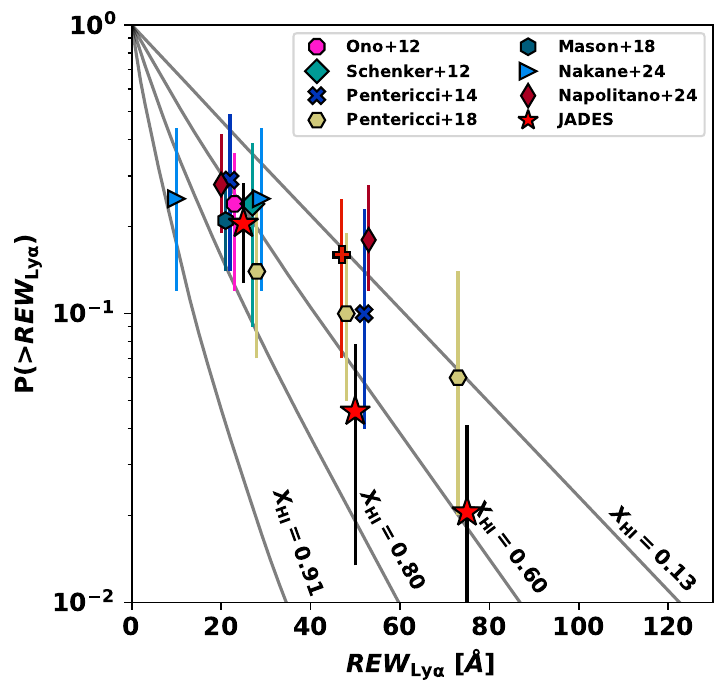}
\caption{Cumulative distribution for \rew at $z\sim7$ using galaxies with $-20.25<{M_{\rm UV}}<-18.75$. Each solid line shows the expected distribution for a model with $N_{\rm HI}=10^{20}$\,cm$^{-2}$, a wind speed of 200\,km\,s$^{-1}$, and an assumed intrinsic \rew distribution scale length of 30\,$\angstrom$, but with a different neutral fraction \citep{naka23}. Estimates from the literature (\citealt{ono12,sche12,pent14,pent18,maso18a,naka23,napo24}) are shifted by $1\angstrom$ for visibility.}
\label{xhifig}
\end{figure}

In order to investigate the neutral fraction at $z\sim7$, we isolate all galaxies between $6.5<z<7.5$ and enforce the same \MUV cut as other works ($-20.25<{M_{\rm UV}}<-18.75$). Using the \MUV-dependent parametrisation of \citet[][]{maso18a}\footnote{$REW_{\rm Ly\alpha,c}=31+12\tanh\left[4(M_{\rm UV}+20.25)\right]\angstrom$}, this \MUV range corresponds to $REW_{\rm Ly\alpha,c}\sim31-32\angstrom$. We convert the \rew PDFs of \citet{naka23} for $REW_{\rm Ly\alpha,c}=30\angstrom$ to CDFs, and compare our completeness-corrected X$_{Ly\alpha}(z=7)$ cumulative distribution with these model outputs and a set of literature values in Figure \ref{xhifig}. The values of our sample are in agreement (i.e., $\lesssim2\sigma$ discrepancy) with the other values, as previously seen in Figure \ref{lyafracfig}.

Next, we estimate \xhi using the sets of measurements in Figure \ref{xhifig}. Each $X_{\rm Ly\alpha}$ value and its uncertainties represents a probability distribution of $P(>$$REW_{\rm Ly\alpha,lim})$ for $REW_{\rm Ly\alpha,lim}\in[25,50,75]\angstrom$, while the model grid of \citet{naka23} may be used to convert $P(>$$REW_{\rm Ly\alpha,lim})$ into a distribution of \xhi for each $REW_{\rm Ly\alpha,lim}$ value. The combination of these distributions for our data results in an estimate of ${X_{\rm HI}}=0.64_{-0.21}^{+0.13}$ for our $-20.25<{M_{\rm UV}}<-18.75$ sample. If we instead use the model outputs of \citet{pent14}, which assumes $REW_{\rm Ly\alpha,c}=50\,\angstrom$, then we find a higher value (${X_{\rm HI}}=0.89_{-0.06}^{+0.04}$; see Appendix \ref{altxhi}).

\begin{figure*}
\centering
\includegraphics[width=\textwidth]{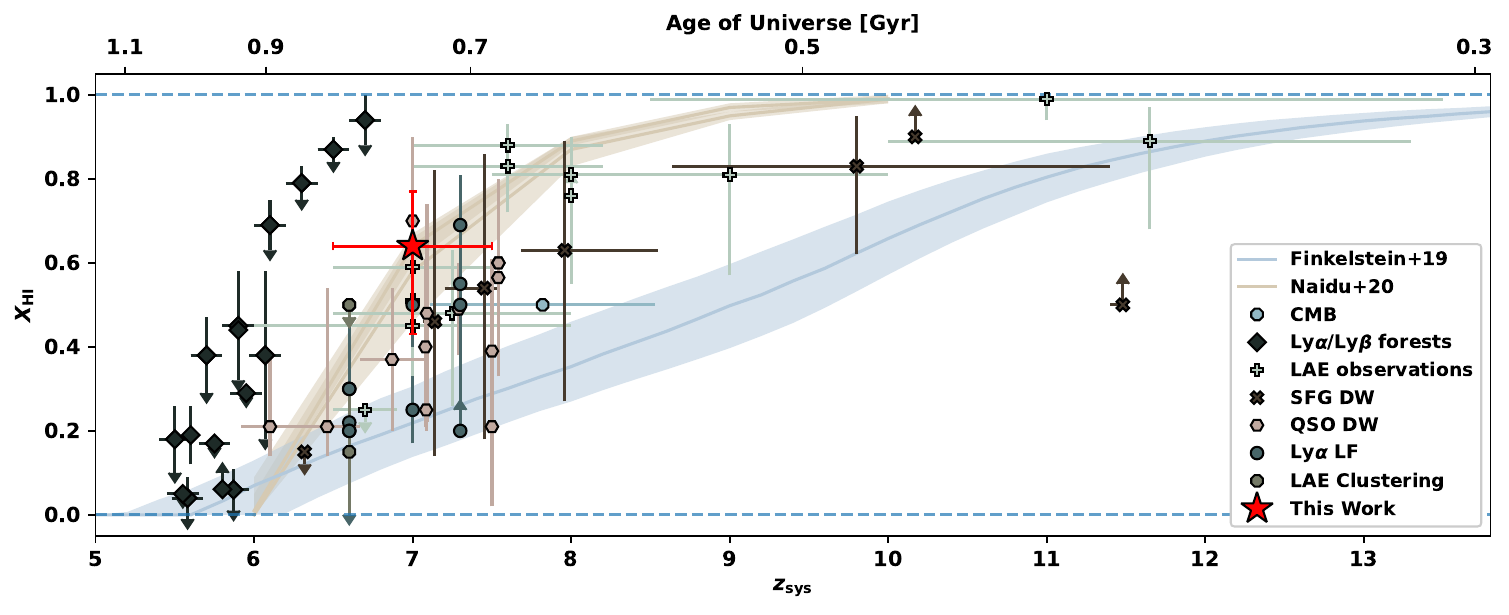}
\caption{Comparison of \xhi values derived through diverse methods. Our value, as derived through a comparison of the observed X$_{Ly\alpha}$ distribution to the model grid of \citet{pent14}, is shown as a large black star. We include the results of two reionisation models (\citealt{fink19,naid20}), a detailed CMB study \citep{plan20}, Ly$\alpha$ and Ly$\beta$ forest observations (\citealt{mcgr15,zhu22,zhu24,jin23,spin24}), LAE observations (\citealt{sche14,maso18a,maso19,hoag19,bola22,jone24,naka23,tang24b}), \lya damping wings of star forming galaxies (\citealt{curt23,hsia23,umed23,faus24}) and QSOs (\citealt{grei17,grei19,grei22,bana18,davi18,duro20,duro24,wang20,yang20}), \lya luminosity functions (\citealt{ouch10,konn14,konn18,zhen17,inou18,goto21,mora21,ning22}), and LAE clustering (\citealt{ouch10,ouch18,soba15}).}
\label{xhiz_fig}
\end{figure*}

To put this result in context, we compare our best-fit \xhi value to those of literature in Figure \ref{xhiz_fig}. While there are a multitude of estimates that have been made over the last decades, there are a few illustrative boundaries. The first is composed of the conservative upper limits at $5.5\lesssim z\lesssim6.7$ based on studies of dark pixels in \lya and Ly$\beta$ forests (e.g., \citealt{mcgr15,jin23}), which constrain the end of the EoR. We may also consider the constraints of two different models (\citealt{fink19,naid20}). The former charts the progress of reionisation if the budget of reionising photons is primarily supplied by UV-faint (\MUV$>-15$) galaxies, while reionisation in the latter model is dominated by UV-bright objects (\MUV$<-18$; `oligarchs'). Regardless of the method used, most observations result in \xhi estimates that fall between these two models. Our value of ${X_{\rm HI}}(z=7)= 0.64_{-0.21}^{+0.13}$ is in agreement with those of other studies, which predict a value of $\sim0.5$ at $z=7$ (e.g., \citealt{maso18a,grei22,naka23,duro24,tang24b}). 

\section{Conclusions}\label{conc}
Using the rich JWST/NIRSpec dataset of the full JADES survey, we have searched for \lya emission in a sample of 795\,galaxies at $4.0<z<14.3$, resulting in the detection of 150\,LAEs spanning the end of the EoR to nearly cosmic dawn ($4.0<z<13.1$). Due to the construction of the JADES survey, galaxies in our sample are distributed across the GOODS-N and GOODS-S fields, with LAEs detected over a wide range of \MUV (from -16 to -21).

The low-resolution R100 data allowed for estimates of the underlying continuum emission, while the wide wavelength coverage ($\lambda_{\rm obs}=0.6-5.3\,\mu$m) permitted the detection of rest-optical lines (e.g., \oiiiab,\ha). Most galaxies also benefit from higher-resolution R1000 data, which open a more detailed window into the fluxes of each line and the velocity offset of \lya. The resulting line and continuum properties were analysed to characterise this unique sample of galaxies. 

Similarly to previous works, our data show a positive relation between \rew and \MUV across a range of redshifts. While this correlation was proposed to be the result of sensitivity effects (i.e., a lack of galaxies with faint \lya and continuum emission; \citealt{jone24}), we still find a strong correlation in each redshift bin using our large sample that includes deep observations. Thus, the correlation is likely physical.

We calculate the \lya escape fraction of our sample using R1000 data and calculating the intrinsic \lya flux using the observed \hb flux (see Appendix \ref{fesc_calc} for discussion of this assumption). This value shows a strong positive correlation with \rew (in agreement with e.g., \citealt{roy23}). There is a strong negative correlation between \fesc and redshift ($z\sim5.5-9.5$), which may reflect IGM evolution during the EoR. 

To explore the reionising properties of individual galaxies, we examine the relation between \lya velocity offset and \fesc, which shows a negative correlation. For galaxies in the EoR, \lya near the systemic redshift will be absorbed or scattered from the line of sight, and only \lya at high relative velocities will be able to escape. As the galaxies ionise their surroundings, a lower velocity offset is required. Thus, this negative correlation also represents a direct tracer of reionisation on the galaxy scale.

All \rew measurements (both detections and upper limits) are then combined with the Bayesian framework of \citet{maso18a} to constrain the IGM transmission of \lya ($T_{\rm IGM}$) between $z=6-14$ as a function of $T_{\rm IGM}(z\sim5)$. We find a similar evolution as \citet{tang24b}: a decrease of $\sim50\%$ between $z=5-7$, and a further decrease of $\sim20\%$ between $z=7-12$.

Using the observed properties of our galaxies, and a set of mock spectra, we determine the \rew completeness of our sample and analysis technique. Instead of completeness at a given \rew value, we are interested instead in the completeness for all values above a \rew limit, and thus adopt a previously found \rew distribution \citep{tang24a} in our derivation. This analysis reveals that our completeness increases from $\sim50\%$ for \rew$>25\angstrom$ to $\sim70\%$ for \rew$>75\angstrom$ across most redshift and \MUV bins. We strongly recommend implementing completeness analyses for future works investigating \lya emission in large JWST datasets, as its exclusion introduces a non-trivial bias in the results.

A completeness correction is applied to the sample to create \lya fraction distributions: ${X_{\rm Ly\alpha}}(z)$. We find that \xlya increases between $z=4-6$ and decreases at higher redshifts, in line with other works. A non-zero ${X_{\rm Ly\alpha}}(z\sim8)$ is found, which we verify is not biased by observing galaxy overdensities.

The ${X_{\rm Ly\alpha}}(z=7)$ values are combined with the model of \citet{naka23} to place a constraint on ${X_{\rm HI}}(z=7)= 0.64_{-0.21}^{+0.13}$. Applying the same method to \lya fractions from other works results in similar \xhi values. This is placed in context with other ${X_{\rm HI}}(z\sim5.3-13.5)$ values, where it is made clear that our hydrogen neutral fraction is comparable to most values derived in other works. 

By exploiting the large dataset of JADES, we have unveiled a number of new LAEs spanning a wide range of intrinsic properties and cosmic epochs. Ongoing and future investigations will detail individual LAEs, and this sample will be combined with other large surveys to shine light on the remaining mysteries of the EoR.

\section*{Acknowledgements}
We would like to thank Laura Pentericci for useful discussions, and the anonymous referee for constructive feedback that strengthened this work.
GCJ, AJB, AS, KB, and AJC acknowledge funding from the ``FirstGalaxies Advanced Grant from the European Research Council (ERC) under the European Union’s Horizon 2020 research and innovation programme (Grant agreement No. 789056).
GCJ and JW acknowledge support by the Science and Technology Facilities Council (STFC) and by the ERC through Advanced Grant 695671 ``QUENCH.
SA acknowledges support from Grant PID2021-127718NB-I00 funded by the Spanish Ministry of Science and Innovation/State Agency of Research (MICIN/AEI/ 10.13039/501100011033).
This research is supported in part by the Australian Research Council Centre of Excellence for All Sky Astrophysics in 3 Dimensions (ASTRO 3D), through project number CE170100013.
SCa acknowledges support by European Union’s HE ERC Starting Grant No. 101040227 - WINGS.
ECL acknowledges support of an STFC Webb Fellowship (ST/W001438/1).
KH, BDJ, PR, BER, YZ acknowledges support from the NIRCam Science Team contract to the University of Arizona, NAS5-02015, and JWST Program 3215.
ST acknowledges support by the Royal Society Research Grant G125142.
H{"U} gratefully acknowledges support by the Isaac Newton Trust and by the Kavli Foundation through a Newton-Kavli Junior Fellowship.
The research of CCW is supported by NOIRLab, which is managed by the Association of Universities for Research in Astronomy (AURA) under a cooperative agreement with the National Science Foundation.

\section*{Data availability}
The data underlying this article will be shared on reasonable request to the corresponding author.

\bibliographystyle{mnras}
\bibliography{references}

\appendix

\section{\MUV limit distribution}\label{muvlimsec}

Throughout this work, the rest-UV magnitude \MUV is estimated directly from our JWST/NIRSpec R100 spectra. While these data are sensitive to the strength and shape of the rest-UV continuum (e.g., \citealt{topp24}), our sample features two types of diversity that hinder \MUV measurement: intrinsic UV brightness and observation depth. As seen in Figure \ref{muvz}, our measured \MUV values extend over a range of $\delta$\MUV$\sim5$\,magnitudes, including both UV-luminous and faint galaxies (e.g., \citealt{star17}). In addition, the JADES dataset may be separated into a deep and medium tier (see Table \ref{jadestable}), with a $\sim1$\,magnitude difference in sensitivity between the deepest and shallowest observations.

\begin{figure}
\centering
\includegraphics[width=0.5\textwidth]{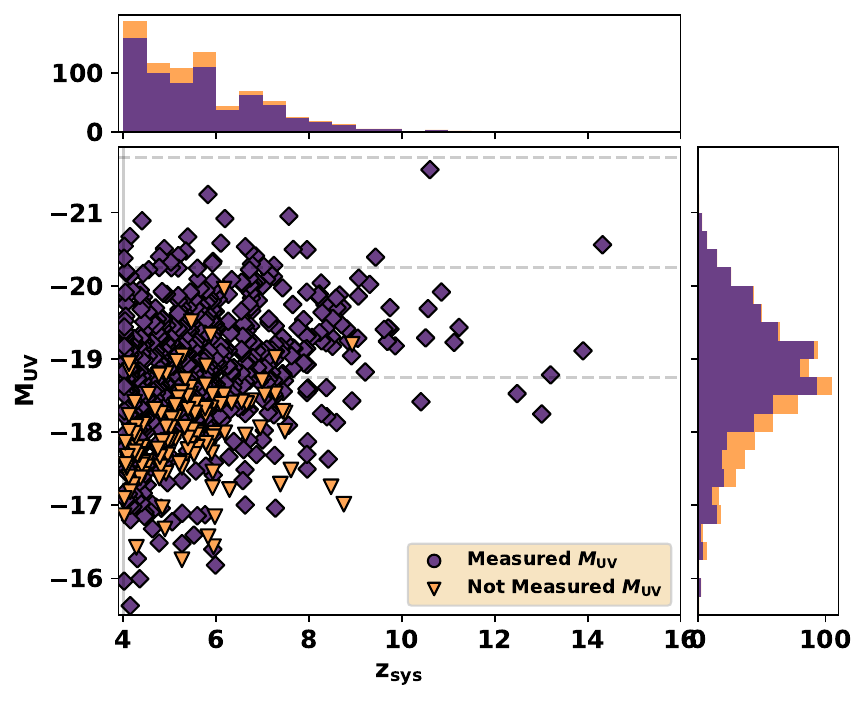}
\caption{$M_{\rm UV}$ (from NIRSpec R100 spectra) versus systemic redshift (based on rest-frame optical lines) for our sample. Sources where \MUV is robustly measured from the observed R100 spectrum are shown by purple markers, while sources where \MUV is not well determined are shown in orange.}
\label{goodbadmuv}
\end{figure}

With this sample properties in mind, we consider the possibility that our \MUV measurement technique introduces a bias towards UV-bright galaxies. The distribution of our \MUV values (separated into measurements and upper limits) as a function of systemic redshift is shown in Figure \ref{goodbadmuv}. It is clear that majority of the galaxies have reliable \MUV estimates ($\sim 85\%$ of the sample). Most of the upper limits are fainter than \MUV$=-18.75$, but we are able to measure \MUV for some galaxies below this threshold. This demonstrates that for the primary \MUV range of interest (-20.25$<M_{\rm UV}<$-18.75), we are able to measure \MUV well for most of our galaxies.

\section{Best-fit \lya properties}
\clearpage
\onecolumn
\begin{landscape}
\begin{table}
\caption{Best-fit properties of JADES LAEs. For each, we list the NIRSpec ID, position, observational tier, visual inspection redshift \citep{deug24}, $M_{\rm UV}$ values measured from R100 spectra, and \lya properties (\rew and escape fraction) from the R100 and R1000 spectra. While all galaxies with significantly detected (i.e., $>3\sigma$) \lya are listed, the low continuum detections of some sources result in $<3\sigma$ \rew values. Upper limits are given as $3\sigma$. \lya equivalent widths are calculated using the best-fit continuum value from the R100 data, while the escape fractions are calculated using the intrinsic \lya/\hb ratio with no dust correction. We present the truncated table here for illustration, but the full table (including additional line and continuum properties) is available upon request.}
\label{lyares_table}
\begin{tabular}{ccc|cc|cc|cc}
\hline
ID 			   & $z_{\rm sys}$ & Tier & M$_{\rm UV}$ & S$_{\rm C}(\lambda_{\rm Ly\alpha,obs})$ 					 	   & $F_{\rm Ly\alpha,R100}$                  & $REW_{\rm Ly\alpha,R100}$ & $F_{\rm Ly\alpha,R1000}$                  & $REW_{\rm Ly\alpha,R1000}$\\ 
\textit{JADES-}   &	       & 	  &               & $10^{-21}$\,erg\,s$^{-1}$\,cm$^{-2}$\,$\angstrom^{-1}$ &  $10^{-20}$\,erg\,s$^{-1}$\,cm$^{-2}$ & $\angstrom$            &  $10^{-20}$\,erg\,s$^{-1}$\,cm$^{-2}$ & $\angstrom$           \\ \hline 
GS+53.06475-27.89024 & 13.0100 & DJS\_1287 & $-18.25\pm0.23$ & $<0.23$ & $44\pm6$ & $<192$ & $<64$ & $<184$\\ 
GN+189.10604+62.24204 & 10.6030 & MJN\_1181 & $-21.59\pm0.03$ & $21.73\pm0.55$ & $<162$ & $<6$ & $171\pm39$ & $7\pm2$\\ 
GS+53.15862-27.83408 & 8.8365 & MJS\_1286 & $-19.25\pm0.23$ & $4.11\pm0.92$ & $<251$ & $<62$ & $155\pm28$ & $38\pm11$\\ 
GS+53.10900-27.90084 & 8.7110 & DJS\_1287 & $-19.72\pm0.04$ & $6.26\pm0.21$ & $171\pm16$ & $28\pm3$ & $153\pm28$ & $25\pm5$\\ 
GS+53.15891-27.76508 & 8.4861 & DJS\_3215 & $-19.30\pm0.05$ & $5.11\pm0.30$ & $94\pm22$ & $19\pm5$ & $86\pm14$ & $18\pm3$\\ 
GN+189.19774+62.25696 & 8.2790 & MHN\_1181 & $-19.63\pm0.09$ & $6.59\pm0.78$ & $918\pm60$ & $150\pm21$ & $698\pm55$ & $115\pm16$\\ 
GS+53.13675-27.83746 & 8.2252 & MJS\_1286 & $-19.36\pm0.15$ & $2.77\pm0.79$ & $228\pm56$ & $<105$ & $145\pm39$ & $<67$\\ 
GS+53.08932-27.87270 & 8.2242 & MJS\_1286 & $-19.83\pm0.09$ & $4.81\pm0.75$ & $222\pm55$ & $50\pm15$ & $<107$ & $<31$\\ 
GS+53.07581-27.87938 & 8.1968 & MJS\_1286 & $-19.87\pm0.09$ & $12.70\pm1.14$ & $<228$ & $<20$ & $217\pm36$ & $19\pm4$\\ 
GS+53.15682-27.76716 & 7.9799 & DHS\_1210 & $-18.60\pm0.07$ & $3.11\pm0.21$ & $65\pm16$ & $23\pm6$ & $80\pm19$ & $29\pm7$\\ 
GS+53.07670-27.88957 & 7.9690 & DJS\_1287 & $-17.87\pm0.21$ & $<0.83$ & $<113$ & $<242$ & $66\pm21$ & $<150$\\ 
GS+53.11991-27.90158 & 7.9561 & DJS\_1287 & $-18.54\pm0.10$ & $3.54\pm0.26$ & $108\pm14$ & $34\pm5$ & $<61$ & $<5$\\ 
GS+53.10561-27.89186 & 7.9548 & DJS\_1287 & $-17.70\pm0.20$ & $2.25\pm0.32$ & $222\pm15$ & $108\pm18$ & $<130$ & $<71$\\ 
GS+53.09943-27.88038 & 7.9508 & MJS\_1286 & $-19.91\pm0.07$ & $11.72\pm0.91$ & $257\pm55$ & $25\pm6$ & $199\pm34$ & $19\pm4$\\ 
GS+53.11378-27.86238 & 7.9451 & DJS\_1287 & $-18.92\pm0.07$ & $5.21\pm0.35$ & $266\pm18$ & $57\pm5$ & $<59$ & $<13$\\ 
GS+53.05373-27.87789 & 7.8906 & MJS\_1286 & $-19.12\pm0.18$ & $5.57\pm0.93$ & $<285$ & $<69$ & $97\pm29$ & $<20$\\ 
GS+53.06029-27.86354 & 7.8854 & MJS\_1286 & $-18.94\pm0.13$ & $5.10\pm0.59$ & $<185$ & $<44$ & $79\pm19$ & $18\pm5$\\ 
GS+53.13347-27.76037 & 7.6590 & MHS\_1180 & $-20.50\pm0.03$ & $22.73\pm0.86$ & $352\pm66$ & $18\pm3$ & $177\pm24$ & $9\pm1$\\ 
GS+53.20042-27.78210 & 7.4809 & MJS\_1180 & $-19.23\pm0.14$ & $8.83\pm1.20$ & $<230$ & $<36$ & $257\pm68$ & $35\pm10$\\ 
GS+53.10105-27.87581 & 7.4729 & MJS\_1286 & $-18.70\pm0.20$ & $3.52\pm0.79$ & $<204$ & $<77$ & $135\pm18$ & $45\pm12$\\ 
GS+53.18148-27.76950 & 7.4326 & MJS\_1180 & $-18.80\pm0.20$ & $4.87\pm1.21$ & $<293$ & $<82$ & $204\pm62$ & $<59$\\ 
GN+189.27524+62.21244 & 7.4318 & MJN\_1181 & $-19.29\pm0.14$ & $7.81\pm1.20$ & $<249$ & $<41$ & $155\pm43$ & $24\pm7$\\ 
GS+53.16746-27.77201 & 7.2752 & DHS\_1210 & $-16.96\pm0.36$ & $1.78\pm0.50$ & $134\pm20$ & $90\pm29$ & $214\pm28$ & $146\pm45$\\ 
GS+53.16959-27.73805 & 7.2430 & MHS\_1180 & $-20.28\pm0.08$ & $17.81\pm1.66$ & $500\pm89$ & $34\pm7$ & $<170$ & $<15$\\ 
GS+53.18674-27.77064 & 7.2425 & MJS\_1180 & $-18.23\pm0.44$ & $<2.52$ & $204\pm56$ & $<659$ & $<241$ & $<45$\\ 
GS+53.16555-27.77267 & 7.2387 & MJS\_1180 & $-18.71\pm0.27$ & $<3.01$ & $162\pm52$ & $<94$ & $<116$ & $<174$\\ 
GS+53.07543-27.85520 & 7.2175 & MJS\_1286 & $-19.05\pm0.20$ & $<2.63$ & $440\pm65$ & $<338$& - & -\\ 
GS+53.11776-27.90701 & 7.1087 & DJS\_1287 & $-18.42\pm0.11$ & $3.11\pm0.43$ & $102\pm23$ & $40\pm11$ & $223\pm39$ & $88\pm20$\\ 
GN+189.20377+62.26843 & 7.0897 & MJN\_1181 & $-18.99\pm0.16$ & $4.19\pm1.32$ & $936\pm68$ & $278\pm92$ & $954\pm41$ & $283\pm90$\\ 
GN+189.09630+62.24797 & 7.0874 & MJN\_1181 & $-18.68\pm0.25$ & $<3.58$ & $218\pm64$ & $<96$ & $<137$ & $<52$\\ 
GN+189.17975+62.28239 & 7.0850 & MHN\_1181 & $-20.09\pm0.09$ & $12.69\pm1.64$ & $705\pm88$ & $69\pm12$ & $822\pm100$ & $80\pm14$\\ 
GN+189.17253+62.24054 & 7.0003 & MJN\_1181 & $-18.75\pm0.26$ & $<4.03$ & $747\pm75$ & $<298$& - & -\\ 
GN+189.20260+62.27551 & 6.9070 & MHN\_1181 & $-19.80\pm0.06$ & $17.77\pm0.94$ & $414\pm45$ & $30\pm4$ & $298\pm68$ & $21\pm5$\\ 
GS+53.14555-27.78380 & 6.8782 & MJS\_1180 & $-20.06\pm0.10$ & $18.96\pm1.68$ & $425\pm92$ & $29\pm7$ & $<243$ & $<136$\\ 
GN+189.15531+62.28647 & 6.8089 & MJN\_1181 & $-20.04\pm0.05$ & $19.71\pm1.16$ & $303\pm66$ & $20\pm4$ & $461\pm57$ & $30\pm4$\\ 
GN+189.15197+62.25964 & 6.7590 & MHN\_1181 & $-19.83\pm0.08$ & $12.09\pm1.21$ & $422\pm68$ & $45\pm8$ & $552\pm88$ & $59\pm11$\\ 
GN+189.17514+62.28226 & 6.7330 & MHN\_1181 & $-20.03\pm0.05$ & $25.09\pm1.04$ & $879\pm49$ & $45\pm3$ & $835\pm82$ & $43\pm5$\\ 
GS+53.15579-27.81520 & 6.7122 & DHS\_1210 & $-18.31\pm0.13$ & $3.57\pm0.63$ & $149\pm27$ & $54\pm14$ & $316\pm50$ & $114\pm27$\\ 
GN+189.09145+62.22810 & 6.6740 & MHN\_1181 & $-18.93\pm0.19$ & $9.92\pm1.84$ & $280\pm81$ & $<38$ & $327\pm78$ & $43\pm13$\\ 
GN+189.14579+62.27332 & 6.6620 & MJN\_1181 & $-18.80\pm0.18$ & $9.36\pm1.28$ & $<306$ & $<51$ & $237\pm67$ & $33\pm10$\\ 
GN+189.24892+62.24974 & 6.6554 & MJN\_1181 & $>-18.39$ & $<4.64$ & $256\pm71$ & $<1467$ & $<204$ & $<223$\\ 
GS+53.16904-27.77884 & 6.6306 & DHS\_1210 & $-18.69\pm0.05$ & $6.73\pm0.48$ & $204\pm21$ & $40\pm5$ & $212\pm34$ & $41\pm7$\\ 
GS+53.17063-27.74325 & 6.6247 & MHS\_1180 & $-18.94\pm0.28$ & $<5.05$ & $420\pm100$ & $<299$ & $<302$ & $<206$\\ 
\end{tabular}
\end{table}

\begin{table}
\contcaption{}
\label{tab:continued}
\begin{tabular}{ccc|cc|cc|cc}
\hline
ID 			   & $z_{\rm sys}$ & Tier & M$_{\rm UV}$ & S$_{\rm C}(\lambda_{\rm Ly\alpha,obs})$ 					 	   & $F_{\rm Ly\alpha,R100}$                  & $REW_{\rm Ly\alpha,R100}$ & $F_{\rm Ly\alpha,R1000}$                  & $REW_{\rm Ly\alpha,R1000}$\\ 
\textit{JADES-}   &	       & 	  &               & $10^{-21}$\,erg\,s$^{-1}$\,cm$^{-2}$\,$\angstrom^{-1}$ &  $10^{-20}$\,erg\,s$^{-1}$\,cm$^{-2}$ & $\angstrom$            &  $10^{-20}$\,erg\,s$^{-1}$\,cm$^{-2}$ & $\angstrom$           \\ \hline 
GS+53.13742-27.76521 & 6.6236 & MJS\_1286 & $-19.33\pm0.10$ & $14.98\pm1.60$ & $363\pm72$ & $32\pm7$ & $471\pm46$ & $41\pm6$\\ 
GS+53.08036-27.89598 & 6.4737 & DJS\_1287 & $-18.34\pm0.09$ & $4.37\pm0.41$ & $309\pm17$ & $95\pm10$ & $277\pm30$ & $85\pm12$\\ 
GS+53.13492-27.77271 & 6.3343 & DHS\_1210 & $-20.08\pm0.02$ & $24.76\pm0.67$ & $733\pm29$ & $40\pm2$ & $813\pm41$ & $45\pm3$\\ 
GS+53.19404-27.80293 & 6.3317 & MJS\_1180 & $-18.48\pm0.23$ & $<4.18$ & $297\pm78$ & $<159$ & $<203$ & $<83$\\ 
GS+53.17836-27.80098 & 6.3260 & MHS\_1180 & $-18.82\pm0.26$ & $<7.44$ & $596\pm106$ & $<188$ & $<262$ & $<426$\\ 
GN+189.16215+62.26381 & 6.3120 & MHN\_1181 & $-19.39\pm0.08$ & $11.73\pm1.26$ & $<194$ & $<24$ & $281\pm65$ & $33\pm8$\\ 
GS+53.16611-27.77204 & 6.3067 & MJS\_1286 & $-18.44\pm0.23$ & $3.77\pm1.09$ & $501\pm49$ & $183\pm57$ & $530\pm58$ & $193\pm60$\\ 
GS+53.16902-27.80079 & 6.2432 & MHS\_1180 & $-18.91\pm0.22$ & $<4.31$ & $350\pm72$ & $<246$ & $<267$ & $<167$\\ 
GS+53.08604-27.74760 & 6.2040 & MHS\_1180 & $-19.14\pm0.20$ & $<5.34$ & $463\pm97$ & $<270$ & $348\pm71$ & $<202$\\ 
GS+53.04881-27.87750 & 6.0507 & MJS\_1286 & $-18.89\pm0.17$ & $6.86\pm1.74$ & $487\pm71$ & $100\pm28$ & $639\pm84$ & $131\pm38$\\ 
GS+53.19588-27.76843 & 6.0480 & MHS\_1180 & $-18.41\pm0.30$ & $9.46\pm2.64$ & $<310$ & $<53$ & $247\pm66$ & $<43$\\ 
GN+189.10818+62.24715 & 6.0478 & MJN\_1181 & $-18.93\pm0.13$ & $12.32\pm1.63$ & $304\pm61$ & $35\pm8$ & $<291$ & $<57$\\ 
GS+53.07281-27.84584 & 5.9945 & DJS\_1287 & $-18.95\pm0.05$ & $8.25\pm0.60$ & $151\pm26$ & $26\pm5$ & $151\pm44$ & $26\pm8$\\ 
GS+53.11052-27.79849 & 5.9849 & DJS\_3215 & $-16.18\pm0.31$ & $<0.59$ & $128\pm8$ & $<403$ & $128\pm27$ & $<462$\\ 
GS+53.16062-27.77161 & 5.9734 & DHS\_1210 & $-18.55\pm0.06$ & $6.60\pm0.56$ & $323\pm20$ & $70\pm7$ & $236\pm40$ & $51\pm10$\\ 
GS+53.16692-27.81033 & 5.9422 & DJS\_3215 & $>-16.43$ & $<0.64$ & $73\pm11$ & $<2142$ & $<127$ & $<1296$\\ 
GS+53.11041-27.80892 & 5.9362 & DHS\_1210 & $-18.66\pm0.05$ & $9.58\pm0.47$ & $292\pm16$ & $44\pm3$ & $313\pm47$ & $47\pm7$\\ 
GN+189.14972+62.22212 & 5.9361 & MJN\_1181 & $-17.95\pm0.44$ & $<4.53$ & $673\pm64$ & $<785$ & $428\pm136$ & $<536$\\ 
GS+53.12175-27.79763 & 5.9361 & DHS\_1210 & $-19.63\pm0.02$ & $10.20\pm0.52$ & $1125\pm25$ & $159\pm9$ & $1044\pm57$ & $147\pm11$\\ 
GS+53.15217-27.76817 & 5.9318 & MJS\_1180 & $>-17.93$ & $<2.40$ & $142\pm47$ & $<2296$ & $<285$ & $<146$\\ 
GS+53.15420-27.80551 & 5.9276 & DJS\_3215 & $-16.40\pm0.44$ & $1.82\pm0.49$ & $46\pm15$ & $<47$ & $<138$ & $<112$\\ 
GS+53.15444-27.77332 & 5.9220 & MJS\_1180 & $>-17.24$ & $<4.40$ & $143\pm38$ & $<230$& - & -\\ 
GS+53.14077-27.80218 & 5.9161 & MJS\_1180 & $-19.31\pm0.11$ & $12.93\pm1.70$ & $285\pm71$ & $32\pm9$ & $<285$ & $<30$\\ 
GS+53.16280-27.76084 & 5.9155 & MHS\_1180 & $-19.79\pm0.10$ & $19.50\pm2.47$ & $1517\pm98$ & $113\pm16$ & $530\pm62$ & $39\pm7$\\ 
GS+53.16773-27.76816 & 5.9118 & MHS\_1180 & $-18.05\pm0.34$ & $6.68\pm2.16$ & $<336$ & $<73$ & $246\pm68$ & $<68$\\ 
GS+53.17655-27.77111 & 5.8889 & DHS\_1210 & $-18.62\pm0.10$ & $11.64\pm1.09$ & $700\pm37$ & $87\pm9$ & $<412$ & $<20$\\ 
GS+53.16577-27.80345 & 5.8848 & MJS\_1286 & $-18.83\pm0.15$ & $4.35\pm1.21$ & $192\pm55$ & $<81$ & $401\pm100$ & $<151$\\ 
GS+53.17986-27.80828 & 5.8348 & MJS\_1180 & $-18.35\pm0.27$ & $5.76\pm1.69$ & $289\pm55$ & $<78$ & $<269$ & $<140$\\ 
GS+53.16685-27.80413 & 5.8311 & DJS\_3215 & $-18.56\pm0.07$ & $5.34\pm0.67$ & $184\pm31$ & $51\pm11$ & $<241$ & $<34$\\ 
GS+53.11351-27.77284 & 5.8141 & DHS\_1210 & $-18.13\pm0.07$ & $5.95\pm0.52$ & $608\pm17$ & $150\pm14$ & $411\pm70$ & $101\pm19$\\ 
GS+53.12210-27.80429 & 5.7881 & DJS\_3215 & $-17.97\pm0.08$ & $6.14\pm0.52$ & $160\pm17$ & $38\pm5$ & $184\pm49$ & $44\pm12$\\ 
GS+53.05313-27.87897 & 5.7792 & MJS\_1286 & $-18.41\pm0.23$ & $12.10\pm2.01$ & $794\pm61$ & $97\pm18$& - & -\\ 
GS+53.13184-27.77377 & 5.7789 & DJS\_3215 & $-18.52\pm0.04$ & $7.84\pm0.33$ & $418\pm11$ & $79\pm4$ & $364\pm35$ & $68\pm7$\\ 
GS+53.11002-27.85416 & 5.7784 & DJS\_1287 & $-17.53\pm0.22$ & $2.96\pm0.79$ & $111\pm26$ & $<58$ & $<206$ & $<33$\\ 
GS+53.13600-27.79849 & 5.7776 & MJS\_1180 & $-18.82\pm0.17$ & $15.95\pm1.64$ & $428\pm64$ & $40\pm7$ & $475\pm105$ & $44\pm11$\\ 
GS+53.16713-27.79424 & 5.7734 & MHS\_1180 & $>-18.23$ & $6.42\pm2.06$ & $250\pm75$ & $<76$ & $<335$ & $<63$\\ 
GN+189.09179+62.25374 & 5.7719 & MHN\_1181 & $-18.97\pm0.16$ & $23.54\pm2.28$ & $432\pm81$ & $27\pm6$ & $<472$ & $<23$\\ 
GS+53.15624-27.83617 & 5.7656 & MJS\_1286 & $-19.02\pm0.16$ & $11.53\pm2.28$ & $802\pm83$ & $103\pm23$ & $941\pm170$ & $120\pm32$\\ 
GS+53.13580-27.76591 & 5.7612 & DJS\_3215 & $-16.87\pm0.21$ & $<0.79$ & $98\pm12$ & $<318$ & $<125$ & $<93$\\ 
GS+53.06316-27.87341 & 5.7390 & MJS\_1286 & $-18.64\pm0.20$ & $<4.84$ & $582\pm70$ & $<283$ & $392\pm90$ & $<210$\\ 
GS+53.17350-27.82507 & 5.6090 & MJS\_1286 & $-18.06\pm0.34$ & $<3.79$ & $402\pm59$ & $<659$ & $392\pm128$ & $<687$\\ 
GN+189.13724+62.26064 & 5.6000 & MHN\_1181 & $-19.41\pm0.11$ & $20.44\pm2.54$ & $428\pm84$ & $32\pm7$ & $<642$ & $<39$\\ 
GS+53.06512-27.84905 & 5.5928 & DJS\_1287 & $-16.86\pm0.27$ & $<1.07$ & $123\pm14$ & $<231$ & $<187$ & $<124$\\ 
GS+53.11357-27.82849 & 5.5783 & MJS\_1286 & $>-17.96$ & $<7.56$ & $759\pm73$ & $<188$ & $402\pm133$ & $<142$\\ 
GS+53.14022-27.78709 & 5.5219 & MJS\_1286 & $-18.62\pm0.22$ & $<4.98$ & $327\pm57$ & $<124$ & $<927$ & $<71$\\ 
GS+53.14565-27.80150 & 5.5217 & DJS\_3215 & $-16.59\pm0.30$ & $<1.04$ & $162\pm12$ & $<261$& - & -\\ 
GS+53.06055-27.84840 & 5.4972 & MJS\_1286 & $-18.33\pm0.26$ & $<5.04$ & $997\pm82$ & $<1037$ & $1161\pm159$ & $<1305$\\ 
\end{tabular}
\end{table}

\begin{table}
\contcaption{}
\label{tab:continued}
\begin{tabular}{ccc|cc|cc|cc}
\hline
ID 			   & $z_{\rm sys}$ & Tier & M$_{\rm UV}$ & S$_{\rm C}(\lambda_{\rm Ly\alpha,obs})$ 					 	   & $F_{\rm Ly\alpha,R100}$                  & $REW_{\rm Ly\alpha,R100}$ & $F_{\rm Ly\alpha,R1000}$                  & $REW_{\rm Ly\alpha,R1000}$\\ 
\textit{JADES-}   &	       & 	  &               & $10^{-21}$\,erg\,s$^{-1}$\,cm$^{-2}$\,$\angstrom^{-1}$ &  $10^{-20}$\,erg\,s$^{-1}$\,cm$^{-2}$ & $\angstrom$            &  $10^{-20}$\,erg\,s$^{-1}$\,cm$^{-2}$ & $\angstrom$           \\ \hline 
GN+189.10968+62.29506 & 5.4839 & MHN\_1181 & $-19.48\pm0.13$ & $15.98\pm2.60$ & $361\pm95$ & $35\pm11$ & $<875$ & $<44$\\ 
GS+53.12819-27.78769 & 5.4817 & MHS\_1180 & $-18.22\pm0.34$ & $12.44\pm3.52$ & $375\pm108$ & $<56$ & $<705$ & $<187$\\ 
GS+53.13859-27.79025 & 5.4816 & MJS\_1286 & $-18.45\pm0.35$ & $14.63\pm3.85$ & $1037\pm126$ & $102\pm29$ & $<761$ & $<208$\\ 
GS+53.16570-27.78494 & 5.4716 & MJS\_1180 & $-17.79\pm0.40$ & $<6.73$ & $<239$ & $<55$ & $1021\pm272$ & $<300$\\ 
GS+53.21484-27.79458 & 5.4040 & MJS\_1286 & $-19.20\pm0.10$ & $18.91\pm2.23$ & $358\pm72$ & $30\pm7$ & $<522$ & $<21$\\ 
GN+189.23015+62.22080 & 5.4004 & MJN\_1181 & $-18.90\pm0.24$ & $<6.66$ & $629\pm101$ & $<316$ & $<996$ & $<162$\\ 
GS+53.15584-27.76672 & 5.3500 & DJS\_3215 & $-18.57\pm0.04$ & $8.18\pm0.42$ & $102\pm15$ & $20\pm3$ & $287\pm77$ & $55\pm15$\\ 
GS+53.10590-27.89486 & 5.3183 & MJS\_1286 & $>-18.07$ & $<8.49$ & $263\pm73$ & $<269$ & $<594$ & $<5$\\ 
GS+53.14837-27.74662 & 5.2914 & MJS\_1180 & $-18.30\pm0.23$ & $6.98\pm1.60$ & $411\pm51$ & $94\pm24$ & $<618$ & $<14$\\ 
GN+189.11532+62.23410 & 5.1790 & MHN\_1181 & $-19.00\pm0.10$ & $16.74\pm1.58$ & $750\pm48$ & $73\pm8$ & $<700$ & $<46$\\ 
GN+189.25460+62.23668 & 5.0917 & MJN\_1181 & $-19.18\pm0.20$ & $28.53\pm4.96$ & $1040\pm149$ & $60\pm13$ & $<1808$ & $<1516$\\ 
GS+53.09753-27.90126 & 5.0782 & MJS\_1286 & $-18.90\pm0.12$ & $14.68\pm1.85$ & $246\pm59$ & $27\pm7$ & $<1032$ & $<248$\\ 
GS+53.11535-27.77289 & 5.0765 & DHS\_1210 & $-19.44\pm0.05$ & $20.54\pm1.27$ & $186\pm42$ & $15\pm3$ & $<894$ & $<34$\\ 
GS+53.14946-27.80979 & 5.0520 & DHS\_1210 & $-17.92\pm0.14$ & $7.98\pm0.79$ & $204\pm22$ & $42\pm6$ & $<1047$ & $<96$\\ 
GN+189.02753+62.25374 & 5.0169 & MHN\_1181 & $-18.89\pm0.13$ & $22.20\pm2.46$ & $1291\pm68$ & $96\pm12$ & $<1553$ & $<121$\\ 
GS+53.19662-27.80531 & 4.9582 & MJS\_1180 & $>-17.89$ & $<8.99$ & $212\pm62$ & $<99$ & $<2678$ & $<110$\\ 
GS+53.16091-27.80354 & 4.9510 & DJS\_3215 & $-17.31\pm0.31$ & $<2.62$ & $244\pm33$ & $<521$& - & -\\ 
GN+189.14179+62.25841 & 4.9410 & MHN\_1181 & $-19.46\pm0.12$ & $32.24\pm4.06$ & $667\pm111$ & $35\pm7$ & $<3098$ & $<88$\\ 
GS+53.12103-27.81599 & 4.9296 & DJS\_3215 & $-18.23\pm0.06$ & $11.46\pm0.58$ & $450\pm16$ & $65\pm4$ & $<1937$ & $<453$\\ 
GS+53.21033-27.78916 & 4.9220 & MHS\_1180 & $>-18.80$ & $<25.28$ & $4797\pm253$ & $<1081$ & $<4659$ & $<471$\\ 
GS+53.08250-27.84946 & 4.8960 & DJS\_1287 & $>-16.68$ & $3.52\pm1.13$ & $247\pm26$ & $<120$ & $<1087$ & $<58$\\ 
GS+53.18539-27.80073 & 4.8388 & MJS\_1286 & $>-17.87$ & $<6.55$ & $318\pm67$ & $<574$ & $<2357$ & $<229$\\ 
GS+53.13613-27.80399 & 4.8080 & MJS\_1286 & $-18.13\pm0.31$ & $10.56\pm2.84$ & $659\pm83$ & $108\pm33$ & $<3034$ & $<52$\\ 
GS+53.11237-27.75960 & 4.7791 & MHS\_1180 & $-19.08\pm0.16$ & $22.04\pm3.69$ & $397\pm114$ & $32\pm11$ & $<4404$ & $<306$\\ 
GS+53.15817-27.78648 & 4.7742 & MJS\_1180 & $-20.25\pm0.03$ & $64.73\pm2.25$ & $703\pm69$ & $19\pm2$ & $<2173$ & $<256$\\ 
GS+53.12739-27.78524 & 4.7562 & MHS\_1180 & $-18.73\pm0.23$ & $21.45\pm4.96$ & $893\pm147$ & $72\pm21$ & $<1274$ & $<107$\\ 
GS+53.08773-27.87124 & 4.7425 & MJS\_1286 & $>-17.77$ & $<6.25$ & $403\pm106$ & $<3516$ & $<2783$ & $<6590$\\ 
GS+53.16948-27.76566 & 4.6870 & MJS\_1286 & $-18.46\pm0.16$ & $18.54\pm2.35$ & $1381\pm66$ & $127\pm17$ & $<2723$ & $<151$\\ 
GN+189.12252+62.29285 & 4.6819 & MJN\_1181 & $>-18.31$ & $<12.84$ & $1452\pm145$ & $<418$ & $<4583$ & $<3733$\\ 
GS+53.13284-27.80186 & 4.6480 & DHS\_1210 & $-18.56\pm0.05$ & $15.65\pm0.78$ & $163\pm24$ & $18\pm3$ & $<2488$ & $<320$\\ 
GS+53.11958-27.89815 & 4.6346 & DJS\_1287 & $-18.42\pm0.11$ & $14.89\pm1.46$ & $503\pm58$ & $60\pm9$ & $<4778$ & $<nan$\\ 
GS+53.11392-27.80620 & 4.5472 & DJS\_3215 & $-20.22\pm0.01$ & $85.49\pm1.41$ & $1402\pm40$ & $30\pm1$ & $<1869$ & $<43$\\ 
GN+189.12052+62.30317 & 4.5350 & MHN\_1181 & $-19.64\pm0.04$ & $37.07\pm2.56$ & $792\pm82$ & $39\pm5$ & $<3070$ & $<98$\\ 
GS+53.16264-27.80368 & 4.5258 & DJS\_3215 & $-18.27\pm0.05$ & $17.34\pm0.98$ & $1165\pm27$ & $121\pm8$ & $<3099$ & $<76$\\ 
GS+53.16083-27.80455 & 4.4907 & DHS\_1210 & $-18.19\pm0.13$ & $17.19\pm1.96$ & $285\pm52$ & $30\pm7$& - & -\\ 
GS+53.16743-27.77585 & 4.4690 & DJS\_3215 & $-16.90\pm0.16$ & $1.93\pm0.63$ & $94\pm21$ & $<99$ & $<2705$ & $<621$\\ 
GS+53.14700-27.81303 & 4.4646 & DHS\_1210 & $-16.85\pm0.19$ & $<2.41$ & $288\pm27$ & $<272$ & $<1550$ & $<2974$\\ 
GS+53.06169-27.87309 & 4.4306 & MJS\_1286 & $-18.60\pm0.12$ & $24.39\pm3.04$ & $597\pm85$ & $45\pm9$ & $<2939$ & $<142$\\ 
GS+53.15294-27.82658 & 4.4300 & MJS\_1180 & $-17.16\pm0.50$ & $<9.07$ & $331\pm65$ & $<234$ & $<1856$ & $<40$\\ 
GS+53.14936-27.81704 & 4.4300 & DJS\_3215 & $-18.03\pm0.05$ & $16.41\pm0.68$ & $125\pm17$ & $14\pm2$ & $<1277$ & $<167$\\ 
GS+53.04050-27.87520 & 4.4290 & MJS\_1286 & $-17.81\pm0.23$ & $13.22\pm2.75$ & $405\pm71$ & $57\pm15$ & $<3121$ & $<166$\\ 
GS+53.20020-27.75714 & 4.3910 & MJS\_1286 & $-17.48\pm0.31$ & $<6.11$ & $384\pm63$ & $<237$ & $<2139$ & $<62$\\ 
GS+53.08528-27.85042 & 4.3693 & MJS\_1286 & $>-17.37$ & $<8.23$ & $290\pm67$ & $<33471$ & $<8687$ & $<848$\\ 
GS+53.12290-27.81225 & 4.3110 & DJS\_3215 & $-17.29\pm0.14$ & $2.16\pm0.69$ & $198\pm24$ & $<170$ & $<3173$ & $<2230$\\ 
GS+53.13228-27.79811 & 4.2830 & DHS\_1210 & $>-16.42$ & $<2.71$ & $171\pm29$ & $<833$ & $<2174$ & $<41$\\ 
GS+53.15832-27.80724 & 4.2331 & DHS\_1210 & $-19.05\pm0.05$ & $36.58\pm2.09$ & $674\pm53$ & $35\pm3$& - & -\\ 
GS+53.15765-27.79791 & 4.2279 & DJS\_3215 & $-19.28\pm0.01$ & $33.05\pm1.32$ & $1184\pm40$ & $68\pm4$ & $<3618$ & $<381$\\ 
GS+53.13850-27.80681 & 4.2246 & MJS\_1180 & $-17.15\pm0.34$ & $6.72\pm1.99$ & $251\pm57$ & $<80$ & $<8027$ & $<507$\\ 
\end{tabular}
\end{table}

\begin{table}
\contcaption{}
\label{tab:continued}
\begin{tabular}{ccc|cc|cc|cc}
\hline
ID 			   & $z_{\rm sys}$ & Tier & M$_{\rm UV}$ & S$_{\rm C}(\lambda_{\rm Ly\alpha,obs})$ 					 	   & $F_{\rm Ly\alpha,R100}$                  & $REW_{\rm Ly\alpha,R100}$ & $F_{\rm Ly\alpha,R1000}$                  & $REW_{\rm Ly\alpha,R1000}$\\ 
\textit{JADES-}   &	       & 	  &               & $10^{-21}$\,erg\,s$^{-1}$\,cm$^{-2}$\,$\angstrom^{-1}$ &  $10^{-20}$\,erg\,s$^{-1}$\,cm$^{-2}$ & $\angstrom$            &  $10^{-20}$\,erg\,s$^{-1}$\,cm$^{-2}$ & $\angstrom$           \\ \hline 
GN+189.25074+62.21889 & 4.2123 & MJN\_1181 & $-18.39\pm0.18$ & $29.33\pm5.24$ & $715\pm112$ & $47\pm11$ & $<1551$ & $<53$\\ 
GS+53.16496-27.77375 & 4.2071 & MHS\_1180 & $-17.69\pm0.29$ & $19.93\pm4.21$ & $742\pm99$ & $71\pm18$ & $<2828$ & $<62$\\ 
GS+53.16302-27.77111 & 4.1562 & DJS\_3215 & $-16.80\pm0.32$ & $9.15\pm1.90$ & $394\pm40$ & $84\pm19$ & $<3431$ & $<63$\\ 
GS+53.17842-27.82131 & 4.1420 & MJS\_1286 & $-19.12\pm0.07$ & $33.45\pm4.50$ & $482\pm156$ & $<29$ & $<4393$ & $<275$\\ 
GN+189.19740+62.17723 & 4.1330 & MJN\_1181 & $-18.86\pm0.08$ & $27.29\pm4.93$ & $2648\pm166$ & $189\pm36$ & $<10254$ & $<173$\\ 
GS+53.09292-27.77619 & 4.1190 & MHS\_1180 & $-17.49\pm0.40$ & $<14.40$ & $709\pm141$ & $<268$ & $<6750$ & $<478$\\ 
GS+53.07374-27.85905 & 4.0731 & MJS\_1286 & $>-17.52$ & $<17.84$ & $619\pm145$ & $<2170$ & $<3691$ & $<42$\\ 
GN+189.19929+62.27946 & 4.0709 & MJN\_1181 & $-17.50\pm0.31$ & $<9.23$ & $269\pm79$ & $<109$ & $<4300$ & $<110$\\ 
GN+189.19328+62.25373 & 4.0610 & MHN\_1181 & $-20.20\pm0.03$ & $98.08\pm4.71$ & $1068\pm127$ & $21\pm3$ & $<8916$ & $<395$\\ 
GN+189.18525+62.23876 & 4.0501 & MJN\_1181 & $-18.56\pm0.10$ & $18.51\pm2.83$ & $245\pm76$ & $<26$ & $<2353$ & $<245$\\ 
GS+53.15548-27.80388 & 4.0448 & DHS\_1210 & $-18.61\pm0.06$ & $36.07\pm2.12$ & $519\pm68$ & $28\pm4$ & $<2331$ & $<616$\\ 
GS+53.18149-27.82922 & 4.0370 & MJS\_1286 & $-17.18\pm0.37$ & $<7.81$ & $658\pm83$ & $<311$ & $<1779$ & $<25$\\ \hline 

\end{tabular}
\end{table}
\end{landscape}
\clearpage
\twocolumn

\section{Fit quality verification}\label{FQV}
\subsection{R100-R1000 comparison}\label{prisgrat}

\begin{figure*}
\centering
\includegraphics[width=\textwidth]{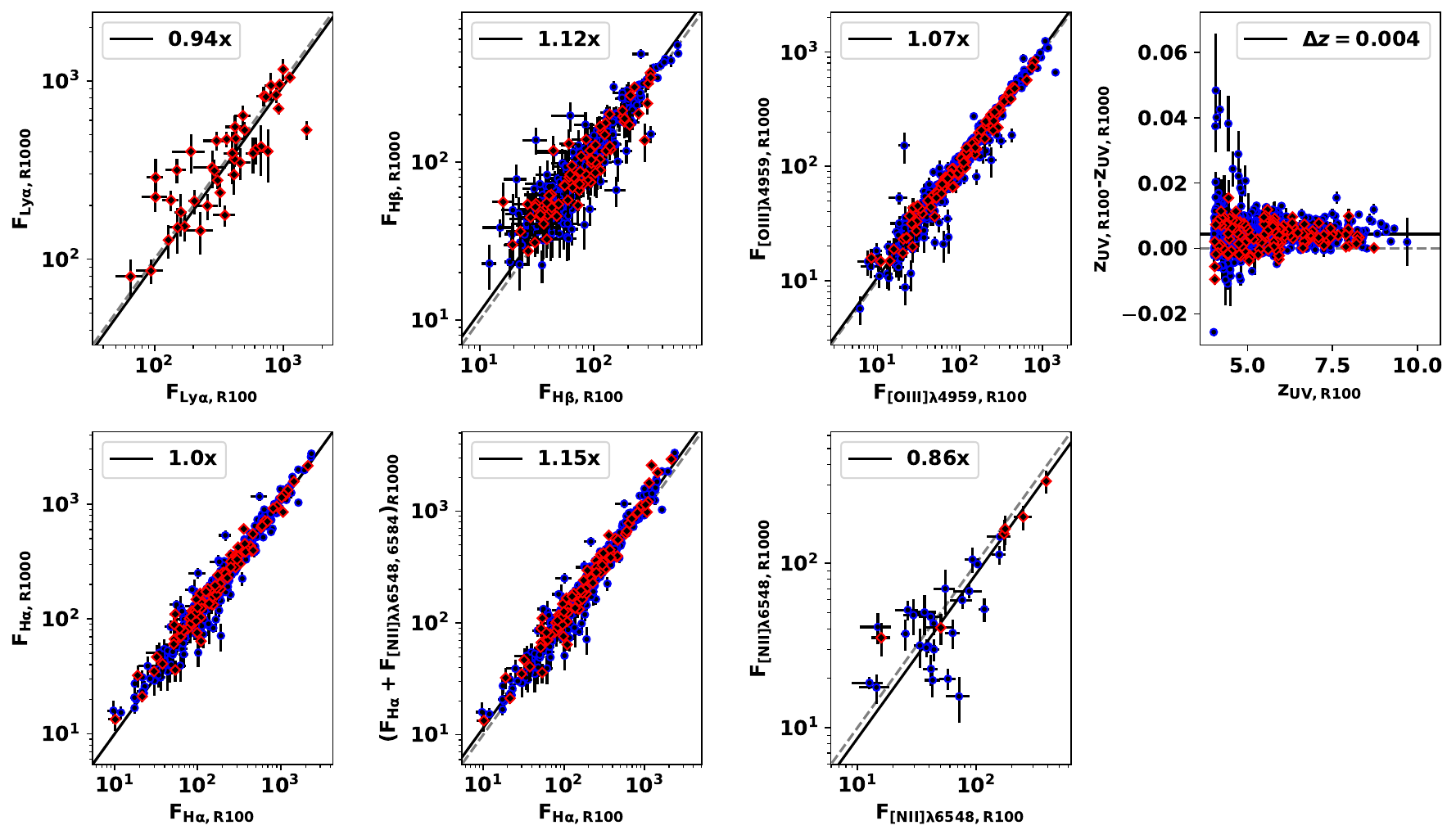}
\caption{Comparison of line fluxes and spectroscopic redshifts for fits to R100 and R1000 data. We compare fluxes for \lya, \hb, \oiiia, \ha, and \niia. In addition, we compare the flux of \ha from the R100 data to the combined [NII]-\ha flux of the R1000 fit. The upper right panel shows the difference in spectroscopic redshift derived from the R100 and R1000 data. LAEs and non-LAEs are shown with red and blue outlines, respectively. In each panel, a best-fit line and its slope (for line flux comparisons) or offset (for redshift comparison) is listed.}
\label{fluxcomp}
\end{figure*}

Because the R100 and R1000 fits were performed separately, we may directly compare the best-fit integrated line flux for multiple strong emission lines (Figure \ref{fluxcomp}). For \oiiia and \hb, the R1000-based fluxes are $\sim$10$\%$ larger. This agrees with the findings of \citet{bunk23b}, who used NIRCam comparisons to suggest that the R100-based fluxes may be more accurate. 

In a curious reversal, we find that the R100-based estimates of \niia flux (of which there are not many $3\sigma$ detections) are lower than the R1000-based estimates. When comparing the R100- and R1000-based fluxes for \ha alone, excellent agreement ($<1\%$ deviation) is found. However, when the combined [NII]-\ha flux of the R1000 fit is compared to the \ha flux of the R100 fit, we find a similar $\sim14\%$ deviation in slope as in the other strong lines. This suggests that \niia and \niib are blended with \ha in the R100 spectra, and our R100-based \ha flux encompasses the full [NII]-\ha complex.

The upper right panel of Figure \ref{fluxcomp} instead presents the difference in spectroscopic redshift as derived from the R100 and R1000 spectra. We find a best-fit offset of $\Delta z\sim0.005$, which is consistent with the median offset presented by \citet[][0.00388]{bunk23b} and \citet[][0.0042]{deug24}. Thus, while the R100 and R1000 results are in approximate agreement, the disagreements in flux and wavelength suggest that they should be analysed separately. 

\subsection{Grating redshift reliability}
For each galaxy, we perform up to four separate fits: the full R100 spectrum, the R1000 data around \lya, the R1000 data around the \oiiiab-\hb complex, and the R1000 data around the \ha-\niiab complex. These fits reveal that the resulting line fluxes and redshifts are in agreement (with the exception of calibration-level offsets, see Section \ref{prisgrat}).

\begin{figure}
\centering
\includegraphics[width=0.5\textwidth]{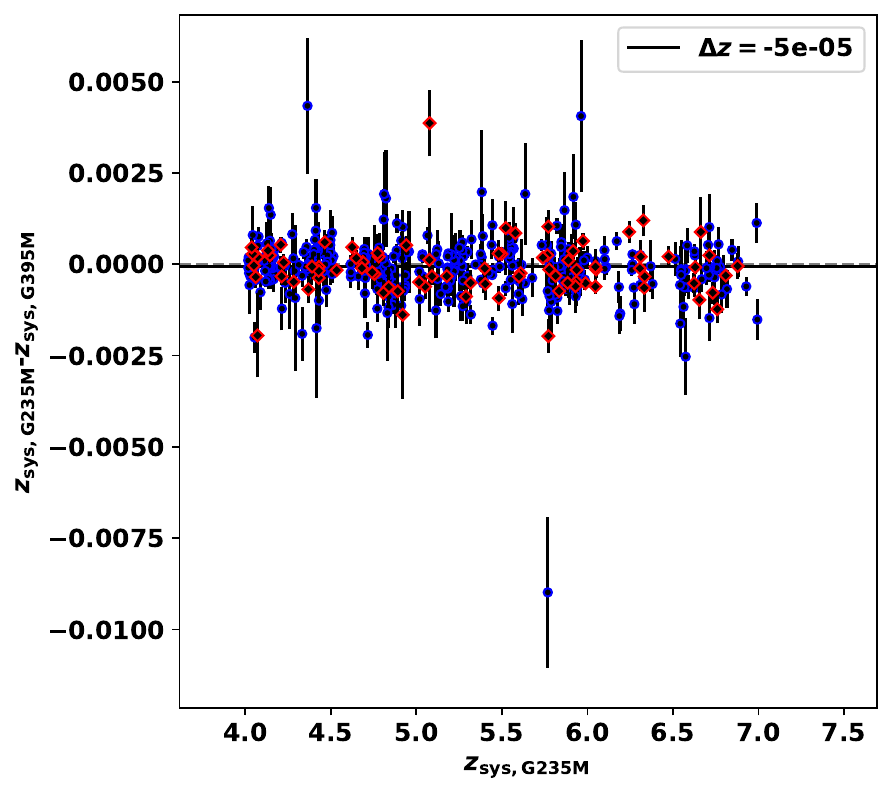}
\caption{Comparison of redshifts derived from the \oiiiab-\hb complex (G235M) and the \ha-\niiab complex (G395M). LAEs and non-LAEs are shown with red and blue outlines, respectively. The average deviation is listed.}
\label{g235395}
\end{figure}

However, it is also possible that the results from each of the three R1000 gratings may yield different results. To inspect this, we consider the redshifts derived from the \oiiiab-\hb complex (G235M) and the \ha-\niiab complex (G395M). As shown in Figure \ref{g235395}, these redshifts are in great agreement, with an average deviation of only $|\delta z|=0.00005$, or $<5$\,km\,s$^{-1}$. Thus, we do not find significant differences in redshifts from different R1000 gratings.

\subsection{Ly$\alpha$ velocity offset measurement}\label{delvapp}

\begin{figure}
\centering
\includegraphics[width=0.5\textwidth]{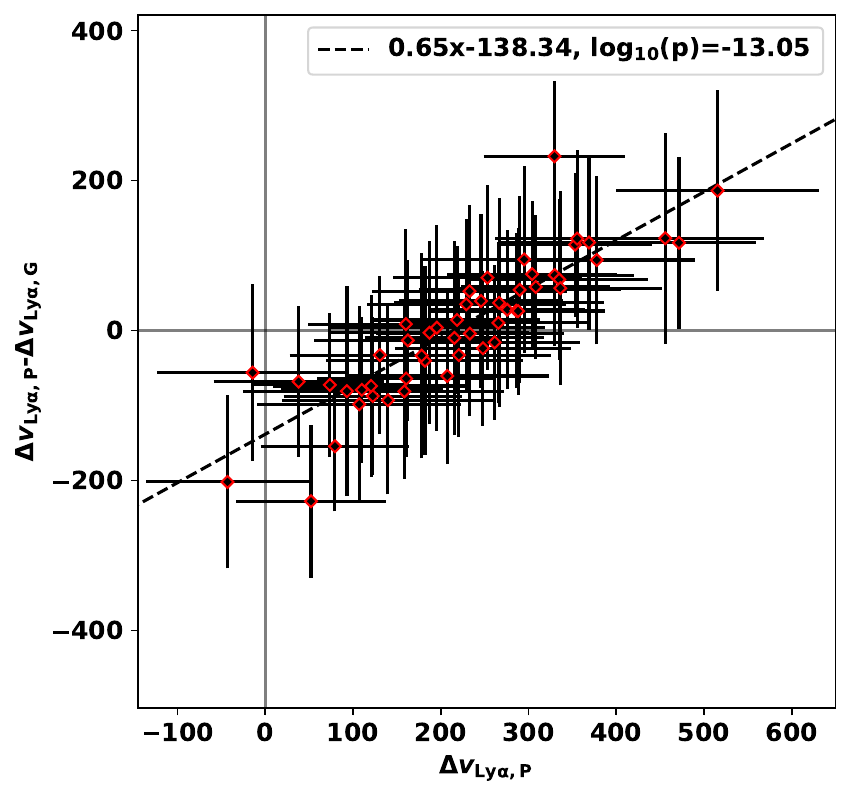}
\caption{Comparison of \lya velocity offset derived using two methods: from the centroid of the best-fit Gaussian model ($\Delta v_{\rm Ly\alpha,G}$), and from the highest-flux wavelength within [-500,+1000]\,km\,s$^{-1}$ of \lya ($\Delta v_{\rm Ly\alpha,P}$). The best-fit correlation is shown by a dashed line.}
\label{veloff}
\end{figure}

As discussed in Section \ref{furobs}, we measure the velocity offset of \lya with respect to the redshift of the rest-optical lines in two ways: the centroid wavelength of a best-fit Gaussian model ($\Delta v_{\rm Ly\alpha,G}$) and the brightest pixel within [-500,1000]\,km\,s$^{-1}$ of \lya ($\Delta v_{\rm Ly\alpha,P}$). In Figure \ref{veloff}, we show the difference between these velocities as a function of $\Delta v_{\rm Ly\alpha,P}$.

Ideally, these two velocities would always agree, resulting in a line of slope 0. But we find that $\Delta v_{\rm Ly\alpha,G}>\Delta v_{\rm Ly\alpha,P}$ for the bulk of the galaxies. This is expected from simulations of how \lya emission profiles are affected by IGM absorption (e.g., \citealt{maso18a}). \lya is intrinsically shifted to the red, and the blue edge is preferably absorbed, resulting in red wings. A symmetric Gaussian fit to these profiles returns a more positive centroid velocity than the peak-finding approach, and this difference correlates with asymmetry. Because lines with lower $\Delta v$ have more absorption and feature higher asymmetry, it is expected that $\Delta v_{\rm Ly\alpha,G}>\Delta v_{\rm Ly\alpha,P}$ for sources with low $\Delta v_{\rm Ly\alpha,P}$, and that this difference decreases with increasing $\Delta v_{\rm Ly\alpha,P}$. This is what we observe.

\subsection{Ly$\alpha$ escape fraction calculation}\label{fesc_calc}

By fitting the R100 and R1000 spectra, we have up to two estimates of the fluxes of \lya, \ha, and \hb (i.e., from the R100 and R1000 fits), as well as a single estimate of the flux of the continuum underlying the \lya line (i.e., from the R100 fit). These may be used to determine two estimates of \rew: $REW_{\rm Ly\alpha,R100}$ and $REW_{\rm Ly\alpha,R1000}$, where both use the same R100-based continuum value. In addition, we may calculate eight versions of \fesc: using the \lya/\ha or \lya/\hb ratio (see Section \ref{furobs}), including a dust correction based on the measured $E(B-V)$ or not (denoted DC or No\_DC, respectively), and using values from the R100 or R1000 fits.

\begin{figure*}
\centering
\includegraphics[width=\textwidth]{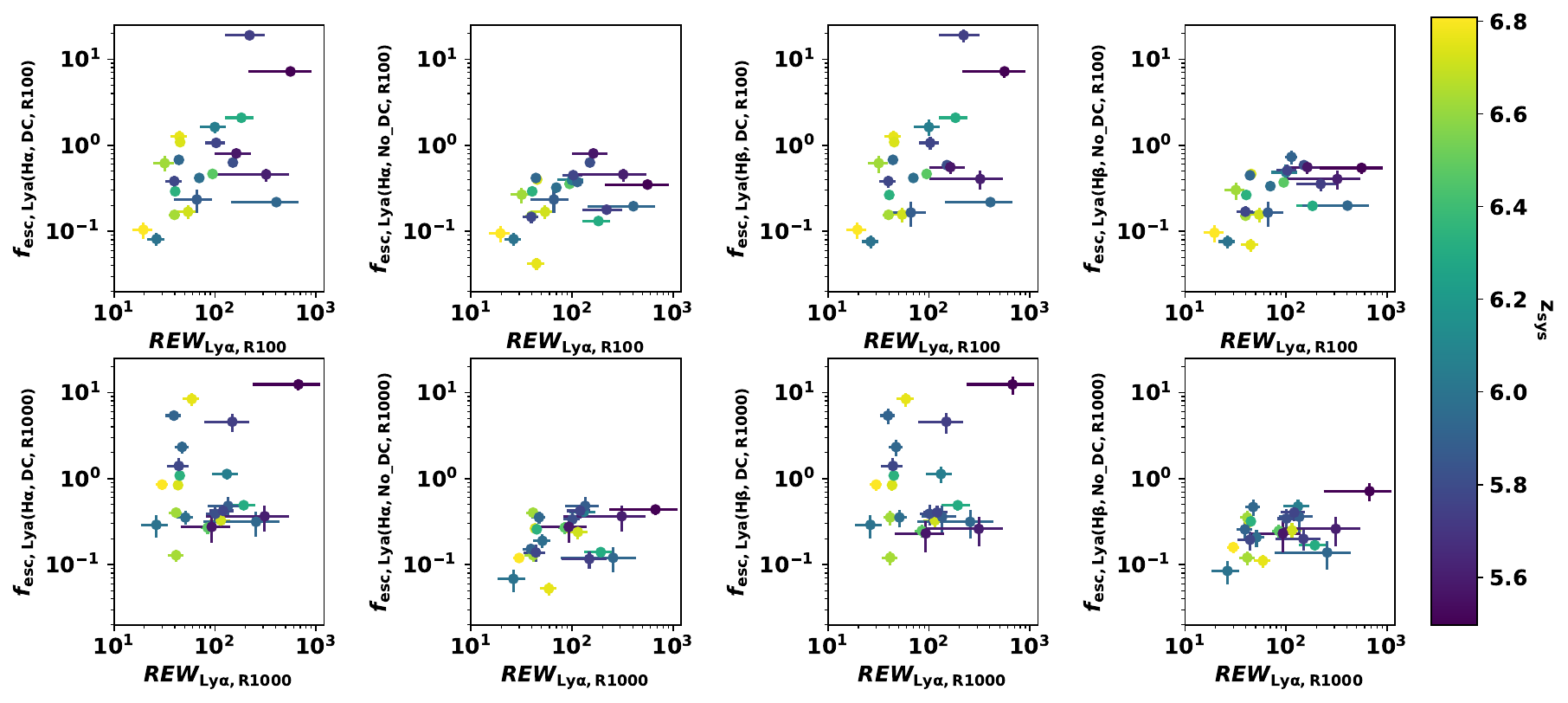}
\caption{\lya escape fraction as a function of \rew for a single sample. The upper row shows results from the R100 data, while the lower row shows the R1000 results. In the left four plots, \fesc is derived by comparing the observed and intrinsic \lya/\ha flux ratio, while the right four plots use the intrinsic \lya/\hb flux ratio. The first and third include dust correction, while the second and fourth do not. Points are coloured by redshift.}
\label{eightfesc}
\end{figure*}

To examine these quantities further, we isolate a subsample of galaxies with both measures of \rew and all eight measures of \fesc (i.e., detections of \lya, \ha, and \hb in R100 and R1000) and plot \fesc as a function of \rew in Figure \ref{eightfesc}. This comparison immediately yields several useful findings. First, the application of a dust correction (which is assumed to be identical for the Balmer lines and \lya) shifts some escape fractions to high values. Some of these fractions are shifted to non-physical values of $>$$100\%$, suggesting that an incorrect dust correction was applied. Some studies have found that \lya and \ha are extincted differently due to the resonant nature of \lya (e.g., \citealt{roy23,begl24,chou24}), implying that different corrections are needed. This may also be an effect of our assumptions of case B recombination rather than case A, or our use of the \citet{calz00} law rather than others (e.g., \citealt{sali18,redd20}).

The most crucial finding here is that all four non-dust corrected escape fractions show the same positive correlation. Throughout the analysis of the main text we consider the \fesc value derived from the R1000 data using the \lya/\hb ratio with no dust correction, and the associated R1000-based \rew. This is driven by our ability to detect \hb in our data out to higher redshifts, uncertainty in the applicability of our applied dust correction, and the higher spectral resolution of the R1000 data.

\section{FitsMap extract}\label{FV}

\begin{figure}
    \centering
    \includegraphics[width=0.5\textwidth]{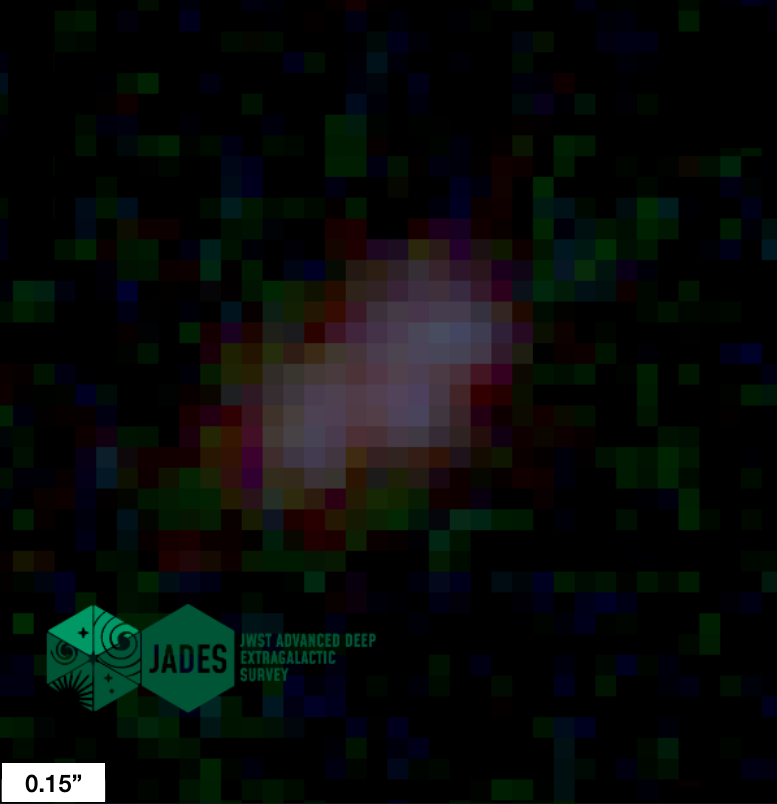}
    \caption{RGB image ($>3\,\mu$m, $2-3\,\mu$m, $<2\,\mu$m, respectively) created using JWST/NIRCam data from JADES observations. The map is centred at RA=53.1374136$^{\circ}$, Dec=-27.7652120$^{\circ}$, and a $0.15''$ scale bar is shown to the lower left corner. Retrieved from FitsMap \citep{haus22}: \url{https://jades.idies.jhu.edu/?ra=53.1374139&dec=-27.7652125&zoom=12}}
    \label{FVfig}
\end{figure}

\section{Alternate $X_{\rm HI}$ estimate}\label{altxhi}
In Section \ref{xhisec}, we combined our observed \rew distribution at $6.5<z<7.5$ with the model outputs presented by \citet{naka23} to place an estimate on $X_{\rm HI}(z\sim7)$. This model was chosen for its assumption of a physically motivated intrinsic \rew distribution with $REW_{\rm Ly\alpha,c}=30\angstrom$. Here, we demonstrate that the use of a model with a more top-heavy \rew distribution results in a higher estimated $X_{\rm HI}(z\sim7)$.

\begin{figure}
    \centering
    \includegraphics[width=0.5\textwidth]{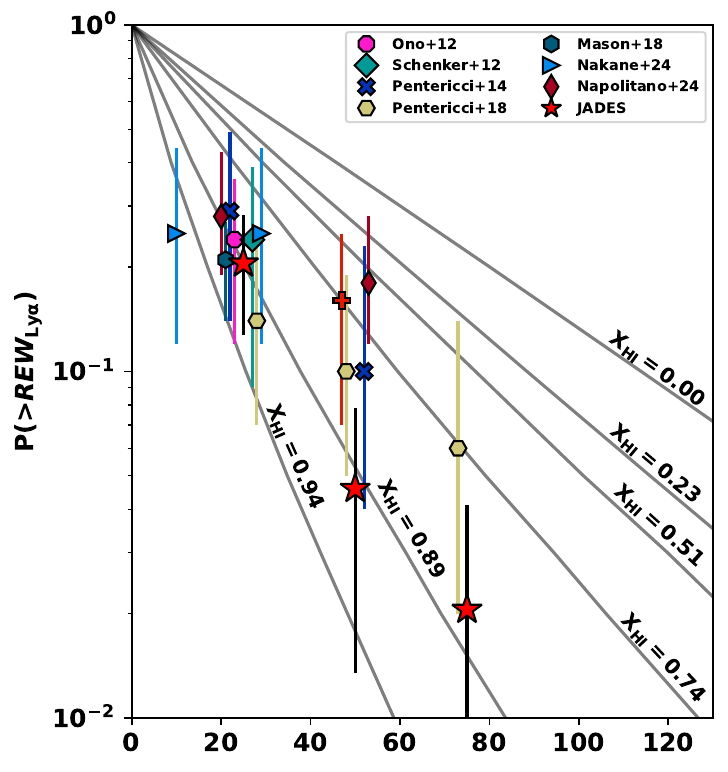}
    \caption{Cumulative distribution for \rew at $z\sim7$ using galaxies with $-20.25<{M_{\rm UV}}<-18.75$, as in Figure \ref{xhifig}. Each solid line shows the expected distribution for a model with $N_{\rm HI}=10^{20}$\,cm$^{-2}$, a wind speed of 200\,km\,s$^{-1}$, and an assumed intrinsic \rew distribution scale length of 50\,$\angstrom$, but with a different neutral fraction \citep{pent14}. Estimates from the literature (\citealt{ono12,sche12,pent14,pent18,maso18a,naka23,napo24}) are shifted by $1\angstrom$ for visibility.}
    \label{xhifig2}
\end{figure}

In Figure \ref{xhifig2}, we plot our \rew CDF at $z\sim7$, but include the model grid of \citet{pent14}. This model is nearly identical to that of \citet{naka23}, but features an intrinsic \rew distribution with $REW_{\rm Ly\alpha,c}=50\angstrom$. This yields a best-fit $X_{\rm HI}=0.89_{-0.06}^{+0.04}$, which is $\sim2\sigma$ higher than the estimate using the \citet{naka23} model grid.

\label{lastpage}
\end{document}